\documentclass[twoside,12pt]{report} 
\usepackage{epsf,graphics}
\usepackage{latexsym,amssymb}
\usepackage{cite}

\usepackage{latexsym,a4wide} 
\usepackage{amsfonts} 
\usepackage{amsbsy} 
\usepackage{amssymb} 

\usepackage{setspace,cite} 
\usepackage{fancyhdr}

\usepackage{vmargin}
\setpapersize{A4}
\setmargins{32mm}{20mm}{154mm}{240mm}{12pt}{20pt}{12pt}{36pt}

\newcommand{\ba}{\begin{array}}
\newcommand{\ea}{\end{array}}
\newcommand{\beq}{\begin{equation}}
\newcommand{\eeq}{\end{equation}}
\newcommand{\bea}{\begin{eqnarray}}
\newcommand{\eea}{\end{eqnarray}}
\newcommand{\ra}{\right>}
\newcommand{\la}{\left<}
\newcommand{\be}{\begin{equation}}
\newcommand{\ee}{\end{equation}}
\newcommand{\lb}{\left(}
\newcommand{\rb}{\right)}

\newcommand{\CF}{{\mathcal F}}
\newcommand{\CP}{{\mathcal P}}
\newcommand{\CQ}{{\mathcal Q}}
\newcommand{\CD}{{\mathcal D}}
\newcommand{\CR}{{\mathcal R}}

\newcommand{\p}{\partial}
\newcommand{\f}{\frac}

\newcommand{\state}[1]{\left| \right. \left. #1 \right>}

\newcommand{\italic}[1]{\begin{it}#1\end{it}}

\newcommand{\half}{{1\over 2}}

\def\half{{1 \over 2}}

\def\KZ{Knizhnik-Zamolodchikov }

\begin{document}


\newpage
\thispagestyle{empty}

\thispagestyle{empty}
\begin{center}
{\Large\bf $SU(2)_k$ Logarithmic Conformal Field Theories\\}
\vspace{1.3cm}

\vspace{1.7cm}

{
Alexander Nichols\\
\vspace{0.8cm}
\begin{it}
Department of Physics, 
University of Oxford\\
Theoretical Physics,\\
1 Keble Road,\\
Oxford OX1 3NP, UK
\end{it}
}

\vspace{0.5cm}
{a.nichols1@physics.ox.ac.uk}
\vspace{1.5cm}
\doublespacing

\bf Abstract
\vspace{0.5cm}
\end{center}
\setlength{\parindent}{0mm}
\setlength{\parskip}{0.5\baselineskip}

We analyse the $SU(2)_k$ WZNW models beyond the integrable representations and in particular the case of $SU(2)_0$. We find that these are good examples of logarithmic conformal field theories as indecomposable representations are naturally produced in the fusion of discrete irreducible representations. We also find extra, chiral and non-chiral, multiplet structure in the theory. The chiral fields, which we construct explicitly in $SU(2)_0$, generate extended algebras within the model. We also study the process of quantum hamiltonian reduction of $SU(2)_0$, giving the $c=-2$ triplet model, in both the free field approach and at the level of correlation functions. For rational level $SU(2)_k$ this gives us a useful technique to study the $h_{1,s}$ correlators of the $c_{p,q}$ models and we find very similar structures to $SU(2)_0$. We also discuss LCFT as a limit of a sequence of ordinary CFTs and some of the subtleties that can occur.

\setlength{\parindent}{6mm}
\setlength{\parskip}{0\baselineskip}

\vspace{3cm}
\begin{center}
Thesis submitted for the Degree of Doctor of Philosophy
in the\\
 University of Oxford\\

\end{center}
\chapter*{}
\thispagestyle{empty}
\begin{center}
\large \bf {Acknowledgments}
\end{center}
\vspace{0.5cm}
I am very grateful for all the support and help I have received throughout my D.Phil. First and foremost I would like to thank my supervisor Dr Ian Kogan for his inspirational insight, guidance and enthusiasm. I would also like to thank the many participants of the past two conferences in LCFT from whom I learnt a great deal. 

I am grateful to my office mates for many useful discussions and for tolerating my mess. I am also grateful to many others in the department who have provided a friendly and intellectually stimulating environment. A big thank you also to all my other friends and house-mates for making my time in Oxford so enjoyable.

\vspace{0.3 in}
\noindent I am grateful to PPARC and Worcester College, Oxford for their financial support.

\vspace{0.3 in}
\noindent Finally special thanks to all my family, and particularly my parents, for all their support.

\pagebreak

\doublespacing

\pagenumbering{roman}
\setcounter{page}{1} \pagestyle{plain}

\onehalfspacing

\tableofcontents
\pagebreak

\doublespacing

\pagestyle{fancy}

\newpage

\pagenumbering{arabic}
\setcounter{page}{1} \pagestyle{fancy}
\renewcommand{\chaptermark}[1]{\markboth{\chaptername%
\ \thechapter:\,\ #1}{}}
\renewcommand{\sectionmark}[1]{\markright{\thesection\,\ #1}}

\addtolength{\headheight}{3pt}
\fancyhead{}
\fancyhead[RE]{\sl\leftmark}
\fancyhead[RO,LE]{\rm\thepage}
\fancyhead[LO]{\sl\rightmark}
\fancyfoot[C,L,E]{}
\pagenumbering{arabic}

\onehalfspacing

\addcontentsline{toc}{chapter}
                 {\protect\numberline{Preface\hspace{-96pt}}} 
\chapter*{Preface}

Quantum field theory is an amazingly rich and beautiful subject. It has been intensely studied in both physics and mathematics for over fifty years and is at the heart of our deepest descriptions of nature.

One recurring theme throughout physics is the special, and very powerful, role played by symmetries. These symmetries may be of a global nature, such as rigid rotation, or of a local nature affecting each point in a different way, such as a local diffeomorphism. The study of these symmetries is central to the modern approach to theoretical physics; gauge theories, relativity, and string theories all being good examples. In almost all physically reasonable quantum field theories there are many divergences and in order to produce sensible results one must first regularise the theory. However in the process one may potentially break some of the classical symmetries (or may even generate non-classical ones). Unfortunately much of the study of physically realistic theories has been confined to the perturbative regimes. As we shall see in two dimensional conformal field theory, where the underlying symmetry is infinite-dimensional, a fully non-perturbative approach can be followed.

Two dimensional conformal field theories (CFT) have been the subject of intense interest since the seminal paper of Belavin, Polyakov and Zamolodichikov \cite{Belavin:1984vu}. They form the basis of the perturbative description of string theory \cite{Polchinski,GSW} and are also used to describe systems at the critical point of a second order phase transition \cite{Polyakov:1970xd}. Many more references and details can be found in any of the standard textbooks \cite{YellowPages,Ketov}.

We shall begin with a brief review of CFT and the concepts that we shall use throughout this thesis. In chapter 2 we begin the study of the discrete representations of $\widehat{SU(2)}$ models beyond the integrable representations with the example of $SU(2)_0$. We shall see that the fusion of $j=\half$ representations in $SU(2)_0$ creates indecomposable representations. From the correlators of the $j=1$ representations we observe a rational solution corresponding to the affine Kac-Moody chiral algebra. We also observe two other logarithmic solutions which come from a non-chiral doublet of operators. In chapter 3 we continue the study of the rational solutions and the corresponding extensions of the chiral algebra and also present a simple free field representation. In chapter 4 we discuss the process of quantum hamiltonian reduction and relations between $\widehat{SU(2)}$ and $c_{p,q}$ theories both in the free field representation and at the level of correlation functions . We show that the generic structures that exist are very similar to those found in $SU(2)_0$. Finally in chapter 5 we discuss some of the subtleties in considering LCFTs as a limit of ordinary CFTs and the appearance of logarithmic partners to the stress tensor in $c=0$ theories.

\vspace{0.2in}
Most of the work presented in this thesis has appeared in the following publications:
\begin{itemize}
%

%
\item \italic{Logarithmic currents in the $SU(2)_0$ WZNW model}

Phys.\ Lett.\ B {\bf 516} (2001) 439, hep-th/0102156.
\item \italic{$SU(2)_0$ and $OSp(2|2)_{-2}$ WZNW models: Two current algebras, one  logarithmic CFT}, with I. I. Kogan. 

Int.\ J.\ Mod.\ Phys.\ A {\bf 17} (2002) 2615, hep-th/0107160.
\item \italic{Stress energy tensor in LCFT and the logarithmic Sugawara construction}, with I. I. Kogan. 

 JHEP {\bf 0201} (2002) 029, hep-th/0112008.
\item \italic{Extended chiral algebras in the $SU(2)_0$ WZNW model}, 

JHEP {\bf 0204} (2002) 056, hep-th/0112094.
\item \italic{Stress energy tensor in c = 0 logarithmic conformal field theory}, with  I. I. Kogan.

Contribution to Michael Marinov Memorial Volume, hep-th/0203207.
\item \italic{Extended multiplet structure in logarithmic conformal field theories}, 

Submitted to JHEP, hep-th/0205170.
\end{itemize}

\chapter{Introduction}
In this chapter we shall introduce some of the basic concepts that we shall use throughout the thesis. In particular we shall explain how the powerful techniques available in two dimensional conformal field theory allow one to exactly determine the operator content and correlation functions. We shall then explain how in a certain subset of theories the irreducible operators do not close under fusion - leading us to the concept of logarithmic conformal field theory.
\section{Conformal Field Theory}
\subsection{General $d$-dimension CFT}
The infinitesimal concept of distance on a manifold can be conveniently expressed through the metric tensor:
\bea
ds^2= g_{\mu \nu}(x) dx^{\mu} dx^{\nu}
\eea
Conformal transformations, a subset of all diffeomorphisms, are defined as those transformations $x \rightarrow x'(x)$ which only change the metric tensor by an overall scale factor:
\bea \label{eqn:conformaltrans}
g_{\mu \nu}(x) \rightarrow g'_{\mu \nu}(x')=\Lambda^2(x) g_{\mu \nu}(x)
\eea
We shall always restrict ourselves to considering conformal transformations in flat space with Euclidean metric $g_{\mu \nu}=\delta_{\mu \nu}$. Clearly the Poincare group, of rigid translations and rotations, is a subgroup of the conformal transformations which exactly preserves the metric. In general however conformal transformations scale lengths in a non-uniform way but always preserve angles.

If we study the effect of an infinitesimal conformal transformation \mbox{$x^{\mu} \rightarrow x'^{\mu}=x^{\mu}+\epsilon^{\mu}(x)$} then one can show:
\bea \label{eqn:infintconftrans}
\p_{\mu} \epsilon_{\nu} + \p_{\nu} \epsilon_{\mu}=f(x) \delta_{\mu \nu}
\eea
It is easily shown that for $d \ne 1,2$ \footnote{For the case $d=1$ there is no concept of angles and thus any smooth map is conformal} the most general solution of (\ref{eqn:infintconftrans}) is: 
\bea 
f(x)=\alpha+\beta_{\mu}x^{\mu}
\eea
\bea \label{eqn:globalconftrans1}
\epsilon_{\mu}=a_{\mu}+b_{\mu \nu} x^{\nu} +c_{\mu \nu \sigma} x^{\nu} x^{\sigma}
\eea
These transformations contain the rigid translations and rotations of the Poincare group. However we have, in addition, the dilatations $x'^{\mu}=\lambda x^{\mu}$ and special conformal transformations\footnote{Although this is a discrete transformation it yields a continuous one by operating with it before and after translation} $x'^{\mu}=x^{\mu}/{x^2}$. Together these transformations form the conformal group.

One can now proceed to construct a theory invariant under this symmetry. An essential role is played by the stress tensor:
\bea \label{eqn:Tdef}
T^{\mu \nu}=-\f{2}{\sqrt{g}} \f{\delta S}{\delta g_{\mu \nu}(x)}
\eea
The importance of this is seen by considering the Ward identity for conformal transformations in flat-space:
\bea \label{eqn:Ward}
\sum_{j=1}^N \left< \Phi_{1}(x_1) \cdots \delta \Phi_{j}(x_j) \cdots \Phi_{N}(x_N) \right>= \int d^2 x \p_{\mu} \epsilon_{\nu}(x) \left< T^{\mu \nu}(x) \Phi_{1}(x_1) \cdots  \Phi_{N}(x_N) \right>
\eea
The Noether current associated to a global conformal transformation is:
\bea
J^{\mu}=\epsilon^{\nu} T_{\mu \nu}
\eea
Requiring that this current be conserved for all the transformations (\ref{eqn:globalconftrans1}) gives us the conditions:
\bea \label{eqn:Tconstraints}
T^{\mu \nu}=T^{\nu \mu} \quad \quad 
\p_{\mu} T^{\mu \nu}=0 \quad \quad 
T^{\mu}_{\mu}=0 
\eea
The first two of these are true in any Poincare invariant theory. Also requiring invariance under dilatations gives us the tracelessness conditions and is sufficient, in a local field theory, to guarantee invariance under the full conformal group.

We shall not discuss generalities further as it is in two dimensions that the subject of conformal invariance gains its real power.

\subsection{Two dimensional CFT}

It is well known that in the Euclidean two dimensional plane one can use complex coordinates $z=x^1+ix^2$ and $\bar{z}=x^1-ix^2$. The flat space metric then becomes:
\bea
ds^2= dz d\bar{z}
\eea
Similarly we introduce the infinitesimal conformal transformations $\epsilon^z(z,\bar{z})$ and $\epsilon^{\bar{z}}(z,\bar{z})$. Then (\ref{eqn:infintconftrans}) takes the form of the Cauchy-Riemann equations and therefore requires that $\epsilon^z$ is in fact a \emph{holomorphic} function; in other words it has no $\bar{z}$ dependence.

One can expand these transformations in a basis:
\bea \label{eqn:localconfbasis}
z \rightarrow z'=z- a_n z^{n+1} \quad \quad \bar{z} \rightarrow \bar{z}'=\bar{z}- \bar{a}_n \bar{z}^{n+1} \quad \quad  n \in Z
\eea
They are generated by the operators:
\bea
l_n = -z^{n+1} \f{d}{dz} \quad \quad \bar{l}_n = -\bar{z}^{n+1} \f{d}{d\bar{z}}
\eea
which satisfy the local algebra of conformal transformations:
\bea
[l_n,l_m]=(n-m)l_{n+m} \quad \quad [\bar{l}_n,\bar{l}_m]=(n-m)\bar{l}_{n+m} \quad \quad [l_n,\bar{l}_m]=0
\eea
We see that the holomorphic and antiholomorphic parts are independent. However this does \emph{not} mean that the theory is a trivial product of the two sectors.

So far we have considered only the local constraints of conformal invariance. The set of invertible, globally defined, conformal transformations are precisely the ones that we found earlier in general dimensions (\ref{eqn:globalconftrans1}). By comparing the transformation (\ref{eqn:localconfbasis}) with the solutions we see that these transformations are generated by the $l_n,\bar{l}_n$ with $n=-1,0,1$. In finite form, in terms of $z$, they are the M\"{o}bius transformations:
\bea
\epsilon(z)=\f{az+b}{cz+d} \quad \quad  \quad \quad ad-bc=1
\eea
The constraints on the stress tensor (\ref{eqn:Tconstraints}) also become much simpler in two dimensions:
\bea
T_{zz}\equiv T(z), \quad \quad T_{\bar{z} \bar{z}} \equiv \bar{T}(\bar{z}), \quad \quad T_{z \bar{z}}=T_{\bar{z} z}=0
\eea
We shall mostly concentrate our attention on $T(z)$ as the behaviour of $\bar{T}(\bar{z})$ is similar. Under a conformal transformation $T$ should behave as a rank $2$ tensor and its variation is, by dimensional reasons, of the form:
\bea \label{eqn:Ttrans}
\delta_{\epsilon}T= \f{c}{12} \epsilon'''(z) + 2 \epsilon'(z)T + \epsilon(z)\p T
\eea
or in finite form:
\bea \label{eqn:finiteTtrans}
T(z) \rightarrow \left( \f{d z'}{dz} \right)^2 T(z') + \f{c}{12}  \{ z',z \}
\eea
where $\{ z',z \}$ denotes the Schwarzian derivative \footnote{ $\{ w,z \}= \f{d^3 w/d^3 z}{dw/dz} - \f{3}{2} \left( \f{d^2 w/d^2 z}{dw/dz}\right)^2$}. The number $c$, known as the central charge, appears in the quantum theory as the anomaly for conformal transformations. As we shall see this plays a crucial role in the structure of the theory. 

In the theory there will be many fields but one may consider a distinguished set, known as primary fields, transforming under all conformal transformations $z \rightarrow z'(z)$ as:
\bea \label{eqn:primary}
\Phi (z,\bar{z}) \rightarrow \Phi' (z',\bar{z}') = \Phi (z,\bar{z}) \left( \f{d z'}{d z} \right)^{-h}  \left( \f{d \bar{z'}}{d \bar{z}} \right)^{-\bar{h}}
\eea
or in infinitesimal form:
\bea
\delta_{\epsilon,\bar{\epsilon}} \Phi (z,\bar{z})=  -\left[ (h \p \epsilon+\epsilon \p) + (\bar{h} \bar{\p} \bar{\epsilon} + \bar{\epsilon} \bar{\p}) \right]  \Phi (z,\bar{z})
\eea
The quantity $h$ is known as the conformal weight, or conformal dimension, of the field. Using the expression for the variation of a field (\ref{eqn:Ward}) and Cauchy's theorem one can show that for correlation functions of primary operators:
\bea \label{eqn:Tinsertion}
\la T(z) \phi_1(z_1) \cdots \phi_n(z_n) \ra= \sum_{i=1}^n \left( \f{h_i}{(z-z_i)^2}+ \f{\p_i}{z-z_i} \right)  \la \phi_1(z_1) \cdots \phi_n(z_n) \ra
\eea
If $z_i\ne \infty$ then the above expression must be regular at infinity. Using the transformation law for the stress tensor (\ref{eqn:finiteTtrans}) we find that we must have the asymptotic behaviour:
\bea
T(z) \sim \f{1}{z^4} \quad \quad {\rm as }~~  z \rightarrow \infty
\eea
Now examining the behaviour of the correlator (\ref{eqn:Tinsertion}) as $z \rightarrow \infty$ gives us immediately the differential equations, or Ward identities:
\bea \label{eqn:VirWard}
\sum_i \p_i \la \phi_1(z_1) \cdots \phi_n(z_n) \ra&=&0 \nonumber \\
\sum_i \left( h_i + z_i \p_i \right) \la \phi_1(z_1) \cdots \phi_n(z_n) \ra&=&0\\
\sum_i \left( 2 h_i z_i + z_i^2 \p_i \right) \la \phi_1(z_1) \cdots \phi_n(z_n) \ra&=&0 \nonumber
\eea
These constrain the correlators and effectively allow one to use the global M\"{o}bius symmetries to fix the positions of three of the fields. Therefore the two and three point functions of primary fields are determined up to constants:
\bea \label{eqn:twopointfn}
\la \phi_{h_1}(z_1) \phi_{h_2}(z_2) \ra &=& A ~~\delta(h_1,h_2) z_{12}^{-2h_1} \\
\la \phi_{h_1}(z_1) \phi_{h_2}(z_2) \phi_{h_3}(z_3) \ra &=& C(h_1,h_2,
h_3) ~~z_{12}^{-h_1-h_2+h_3} z_{13}^{-h_1-h_3+h_2} z_{23}^{-h_2-h_3+h_1} \nonumber
\eea
Normally, in a unitary theory, one normalises the two point functions so that $A=1$ but as we shall be considering non-unitary theories, where this is not always possible, we shall leave this unfixed. The numbers $C(h_1,h_2,h_3)$, known as the structure constants of the theory, contain all the information that one needs to reconstruct the correlators of the theory. An important aspect, emphasized by Polyakov \cite{Bootstrap}, is that the associativity of the operator algebra is a dynamical constraint on the theory. In other words not all possible three point functions lead to a consistent quantum field theory with well defined correlation functions. Solving these constraints is called the conformal bootstrap. In general this is an extremely hard problem. However it becomes tractable if we have a finite set of basic operators. As the three point functions are in general rather hard to find directly it is much easier in practice to deal with the four point functions. The four point functions are only determined up to a functional form. There is some obvious freedom in defining this and here we shall use the convention:
\bea \label{eqn:fourpointfn}
\hspace{-0.3cm}\langle \phi_{h_1}(z_1) \phi_{h_2}(z_2) \phi_{h_3}(z_3) \phi_{h_4}(z_4) \rangle
=z_{43}^{h_2+h_1-h_4-h_3}z_{42}^{-2h_2}z_{41}^{h_3+h_2-h_4-h_1} 
 z_{31}^{h_4-h_1-h_2-h_3}F(z)
\eea 
where the cross ratio $z$, invariant under M\"{o}bius transformations, is given by:
\bea
z=\frac{z_{12}z_{34}}{z_{13}z_{24}} 
\eea
As we shall make extensive use of the crossing symmetry transformations it is useful to note that under $1 \leftrightarrow 3$ we have $z \leftrightarrow 1-z$ and under $1 \leftrightarrow 4$ we have $z \leftrightarrow \f{1}{z}$. As the correlator is a physical object it must certainly be a single-valued function of all variables. The associativity of the operator algebra can equivalently be stated in terms of crossing symmetry of identical fields in the correlator. Clearly any other quantum numbers that the operators possess, in particular their Bose or Fermi nature, must be  properly taken into account.

As we commented earlier the other local conformal transformations are not globally invertible and therefore relate correlation functions in different topologies. We shall see later how they can be used to constrain the correlators further.

In Euclidean QFT the evolution of an operator is given by the \italic{equal-time} commutator:
\bea
\f{d \Phi}{d t}=[H,\Phi]
\eea
For an arbitrary conformal transformation this generalises to:
\bea \label{eqn:changeinphi}
\delta_{\epsilon} \Phi= [T_{\epsilon},\Phi]
\eea
with:
\bea
T_{\epsilon}= \oint \f{d z}{2 \pi i} \epsilon(z) T(z)
\eea
In order to make (\ref{eqn:changeinphi}) well defined one first needs to define a notation of time. This is conveniently provided by the technique of radial quantisation. One parameterises the plane by:
\bea \label{eqn:zplane}
z=e^{\tau + i \sigma} \quad \quad \bar{z}=e^{\tau - i \sigma} \quad \quad \tau,\sigma \in R
\eea
and takes the variable $\tau$ to be the \italic{time} coordinate.

However there still remains an operator ordering ambiguity in the commutator (\ref{eqn:changeinphi}). To render this well defined one makes use of Schwinger's time splitting technique \cite{ItzyksonZuberQFTBook} to rewrite this in the form:
\bea \label{eqn:timesplit}
[T_{\epsilon},\Phi(w,\bar{w})] = \lim_{|z| \rightarrow |w|} \oint \f{d z}{2 \pi i} \epsilon(z) T(z) \Phi( w, \bar{w})
\eea
To evaluate this we need to understand the behaviour of the integrand in the limit as $z \rightarrow w$. We use the Wilsonian operator product expansion \cite{Wilson:1969zs} of the form:
\bea
O_i(z) O_j(w) \sim \sum_{k} C_{i j}^k(z-w) O_k (w)
\eea
The $\{O_i\}$ are a complete set of local operators and the coefficients $C_{i j}^k(z-w)$ are functions and are fixed in form by conformal invariance. This expression must always be understood as being valid when inserted into a correlation function and we shall assume this without further comment.

From the correlator with an insertion of $T$ (\ref{eqn:Tinsertion}) one can see that the primary fields have the following simple OPE with the stress tensor:
\bea \label{eqn:OPETprimary}
T(z) \Phi(w,\bar{w}) \sim \f{h \Phi(w,\bar{w}) }{(z-w)^2} + \f{ \p_w \Phi(w,\bar{w}) }{z-w} + \cdots
\eea
The non-singular terms are called the descendent fields. Although these have a more complicated transformation law under conformal transformations their behaviour is completely determined from that of the primary fields. If the operator algebra of the primary fields closes then the OPE takes the general form:
\be
\label{eqn:generalOPE}
\phi_{h_1}(x_1,z_1) \phi_{h_2}(x_2,z_2) = \sum_h {C(h_1,h_2,h) z_{12}^{-h_1-h_2+h}  [ \phi_h(z_2) ] } 
\ee
where we have denoted by $[ \phi_h ]$ all descendent fields that can be produced from the given primary field $ \phi_h $.

Furthermore one can show using the transformation law for the stress tensor (\ref{eqn:Ttrans}) in the regularised expression (\ref{eqn:timesplit}) that the stress tensor must obey the following OPE:
\bea \label{eqn:TTOPE}
T(z) T(w) \sim \f{c}{2(z-w)^4} + \f{2 T(w)}{(z-w)^2} + \f{\p T(w)}{z-w} + \cdots 
\eea
As $T(z)$ is a holomorphic field one may perform a Laurent expansion:
\bea
T(z)= \sum_n L_n z^{-n-2}
\eea
One may use the expression for the commutator (\ref{eqn:timesplit}) to re-express the OPE  (\ref{eqn:TTOPE}) in terms of the modes and one gets the famous Virasoro algebra:
\bea \label{eqn:Virasoroalg}
\left[ L_n, L_m \right] = (n-m) L_{n+m} + \f{c}{12} n(n^2-1) \delta_{n+m,0}
\eea
The most fundamental state in any theory is the vacuum $\state{0}$. In CFT this is defined by the insertion of a unit operator at the origin of the $z$-plane (\ref{eqn:zplane}). It is $SL(2,C)$ invariant and therefore satisfies:
\bea \label{eqn:vacuumdefn}
L_n \state{0} &=& 0 \quad n \ge -1 \\
\bar{L}_n \state{0} &=& 0 \nonumber
\eea
The primary states are then defined by:
\bea
\state{h} = \lim_{z \rightarrow 0} \Phi_h (z) \state{0}
\eea
where $\Phi_h(z)$ is a primary field. From the OPE of the stress tensor with the primary fields (\ref{eqn:OPETprimary}) we get:
\bea
\left[ L_n, \Phi_h(z) \right] &=& \oint \f{dw}{2 \pi i} w^{n+1} T(w) \Phi_h(z) \\
&=& \left( z^{n+1} \f{d}{dz} +(n+1) z^n h \right) \Phi_h(z) \nonumber
\eea
we find:
\bea
L_{n} \state{h}&=&0 \quad \quad n \ge 1 \\
L_{0} \state{h}&=& h \state{h} \nonumber
\eea
Descendent fields are written using the Virasoro modes acting on a primary state $\state{h}$:
\bea
L_{-n_1} L_{-n_2} \cdots L_{-n_k} \state{h} \quad \quad N= \sum_{i=1}^k n_{i}
\eea
where the number $N$ is called the level. From a given primary state we may construct many different secondary, or descendent, states. However there may be certain combinations which are orthogonal to every other primary state. Such combinations are known as \emph{null-vectors}. From the two-point function (\ref{eqn:twopointfn}) we know that primary states of different conformal weight are orthogonal and so it is enough to consider the orthogonality of the null vector with states at the same level.

For instance at level two we find the combination:
\bea \label{eqn:leveltwonull}
( L_{-2} -\f{3}{2(2h+1)} L_{-1}^2 ) \left.| h \right>
\eea
is orthogonal to both the states $L_{-2} \left.| h \right>$ and $L_{-1}^2  \left.| h \right>$. If we are considering correlation functions of irreducible representations then such null vectors must be set to zero. Using these one can find, and solve, differential equations for the four-point functions of the theory. The study of null vectors \cite{KacBook,Feigin:1983tg} of the Virasoro algebra leads to the `minimal' $c_{p,q}$ models \cite{Belavin:1984vu} which have:
\bea
c_{p,q}&=&1-\f{6(p-q)^2}{pq} \\
h_{r,s}&=&\f{(pr-qs)^2-(p-q)^2}{4pq}
\eea
and the identifications $h_{r,s}=h_{q-r,p-s}$. These models have a \emph{finite} number of primary operators which form a closed algebra under fusion. Such models are called \emph{rational} CFTs as it can be shown that a finite closed operator algebra implies that the conformal weights and central charge are all rational numbers.
\section{WZNW models}
For much of conformal field theory the algebraic approach is central. In contrast to much of physics in higher dimensions no reference is made to any underlying Lagrangian, indeed several different Lagrangians may yield the same quantum theory. The WZNW model is one of the few cases in which we do have a well defined action:
\bea \label{eqn:WZaction}
{\mathcal{S}}_{WZNW}= \f{k}{8 \pi} \int_M d^2 x Tr(\p g ~\p g^{-1} ) + \f{k}{12 \pi} \int_{\p M} d^3 y Tr( g^{-1} \p g \wedge g^{-1} \p g \wedge g^{-1} \p g)
\eea
The number $k$ is known as the level. The field $g$ takes its value in a Lie group $G$. The integrand of the second term, the famous Wess-Zumino topological term, is a total derivative. This action is conformally invariant (the boundary conditions on $g$ are such that complex coordinates are well defined on $M$) but also has a much larger affine Kac-Moody symmetry:
\bea
g(z,\bar{z}) \rightarrow \bar{\Omega}(\bar{z}) g(z,\bar{z}) \Omega(z)
\eea
This symmetry is characterised by the currents:
\bea
J(z)&=&J^a(z) t^a = k g^{-1} \p g \\
\bar{J}(\bar{z})&=& \bar{J}^a(\bar{z}) t^a = -k(\bar{\p} g) g^{-1} \nonumber
\eea
where $t^a$ is a generator of the Lie group $G$: $\left[t^a,t^b \right]=if^{ab}_c t^c$ and $f^{ab}_c$ are the structure constants.

The OPE of these currents is given by:
\bea \label{eqn:KMOPE}
J^a(z) J^b(w) \sim \f{k \delta^{ab}}{(z-w)^2}+ \f{i f^{ab}_c J^c(w)}{z-w} + \cdots
\eea
with a similar behaviour for $\bar{J}$. As $J^a(z)$ is a holomorphic field one may also express it as a mode expansion:
\bea
J^a(z)=\sum_n J^a_n z^{-n-1}
\eea
Then, using the same techniques as with Virasoro case, from (\ref{eqn:KMOPE}) we get the affine Kac-Moody algebra:
\bea \label{eqn:KMalg}
\left[ J^a_n,J^b_m \right] = i f^{ab}_c J^c_{m+n} + k n \delta^{ab} \delta_{m+n,0}
\eea
On a compact group, due to a quantisation condition, the WZNW action (\ref{eqn:WZaction}) is only well defined for integer values of $k$. However from the algebraic point of view one can consider arbitrary values of $k$ in the algebra (\ref{eqn:KMalg}).

Now specialising to the case of $SU(2)$ which we shall consider in this thesis one finds that for positive integer values of $k$ one can construct a vacuum null vector using just the affine Kac-Moody generators:
\bea \label{eqn:introvacnullvect}
{\mathcal N}= \left( J^+_{-1} \right)^{k+1} \state{0}
\eea
It can be verified using (\ref{eqn:KMalg}) that this state is indeed an affine Kac-Moody primary field. Insisting that this vector vanishes in all correlation functions gives us the closed set of representations $0 \le j \le \f{k}{2}$, with $2j \in Z$ and also all the fusion rules amongst these. This is known as the \emph{integrable}, and in this case also unitary, sector of the theory. We shall consider representations beyond these and so we shall not decouple this field.

To get the classical stress tensor for the WZNW model we use the standard definition (\ref{eqn:Tdef}). This leads to an expression that is quadratic in the currents $J^a$. In the quantum theory one also assumes a quadratic expression but fixes the overall normalisation of $T$ so that the affine currents $J^a$ are $h=1$ fields as one would want for conserved currents. With this choice, known as the Sugawara construction, we get: \footnote{The normal ordering in defined using the time-splitting procedure (\ref{eqn:timesplit}) as before:
\bea
:J^a J^a :(w) \equiv \lim_{z \rightarrow w} \f{1}{2 \pi i} \oint_w \f{d z}{(z-w)} J^a(z) J^a(w) \nonumber
\eea}
\bea \label{eqn:Sugawara}
T(z)=\f{1}{2(k+g)} :J^a J^a:(z)
\eea
where $g$ is the dual Coxeter number of the algebra $G$ (for $SU(N) \quad g=N$). It is amazing that such a simple modification to $T$ is only possible when we include the effects of all quantum corrections. The central charge for this stress tensor is:
\bea
c=\f{k ~\rm{dim~} g}{k+g}
\eea
where $\rm{dim~} g$ is the dimension, or number of generators, of the group $G$.
\subsection{Free field representation of $\widehat{SU(2)}$}
Using the Lagrangian approach for $SU(2)_k$ we can obtain a free field, or Wakimoto, representation \cite{Wakimoto:1986gf}. We shall follow the presentation given in \cite{Gerasimov:1990fi}. The classical action is as given previously (\ref{eqn:WZaction}). We shall use the Gauss decomposition of a group element given by:
\be
g= \left( \begin{array}{ll} 1 & \Psi  \\ 0 & 1 \end{array} \right)
\left( \ba{ll} e^{\Phi} & 0 \\ 0 & e^{-\Phi} \ea \right)
\left( \ba{ll} 1 & 0 \\ \chi & 1 \ea \right)
\ee
In the following, unless otherwise stated, we restrict attention to the holomorphic sector as the anti-holomorphic one behaves in a similar manner.
The classical conserved currents are:
\bea \label{eqn:classicalcurrents}
k g^{-1}dg=k \left( \ba{ll} e^{-2\Phi}d \Psi \chi + d \Phi & e^{-2\Phi}d \Psi \\ -e^{-2\Phi}d \Psi \chi^2 -2\chi d\Phi + d\chi & -e^{-2\Phi}d \Psi\chi -d\Phi\ea \right)= J^a {\sigma}^a
\eea
where ${\sigma}^a$ are the Pauli matrices. We also re-define:
\bea \label{eqn:W}
W=ke^{-2\Phi}d \Psi
\eea
Therefore the Lagrangian becomes:
\be
{\mathcal{L}}=-\f{1}{4\pi} (W \bar{\p}\chi + k \p{\Phi}\bar{\p}\Phi )
\ee
So far the results are purely classical. The transformation (\ref{eqn:W}) is anomalous and one must also take into account the change in the measure. When this is done the full quantum action becomes:
\be
{\mathcal{L}}_q= -\f{1}{4 \pi} (W \bar{\p}\chi+ (k+2) \p \Phi \bar{\p} \Phi + {\mathcal{R}} \Phi )
\ee
The curvature term ${\mathcal{R}} \Phi$ can be concentrated at a point by appropriate choice of metric. This can be taken to infinity and is known as the background charge \cite{Dotsenko:1984nm}. It causes a modification of the central charge to exactly reproduce the anomaly.

To get the standard normalisation we rescale $ -(k+2) \p \Phi \bar{\p} \Phi = \f{1}{2} \p \phi \bar{\p} \phi $ and define $\beta=-W~,~\gamma=\chi$. Then:
\be
{\mathcal{L}}_q= -\f{1}{4 \pi} (-\beta \bar{\p} \gamma -\half \p \phi \bar{\p} \phi + {\mathcal{R}} \Phi )
\ee
The stress tensor now becomes:
\be \label{eqn:stressT}
T=-\beta \p \gamma -\half \p \phi \p \phi - \f{i}{\sqrt{2(k+2)}} \p^2 \phi
\ee
There is now no guarantee that the currents in the quantum theory can be written in such a simple form as before (\ref{eqn:classicalcurrents}). However, with slight modification, they can indeed be and are known as the Wakimoto representation \cite{Wakimoto:1986gf}:
\bea \label{eqn:Wakimoto}
J^+&=&\beta \nonumber \\
J^3&=&i \sqrt{\f{k+2}{2}} \p \phi +  \gamma \beta\\ 
J^-&=&-i \sqrt{2(k+2)} \p \phi \gamma - k \p \gamma - \beta\gamma^2 \nonumber
\eea
where the bosonic ghost system $(\beta,\gamma)$ (see for example \cite{Polchinski}) obeys the standard free field relations:
\be
\beta(z) \beta(w) \sim 0 \sim \gamma(z) \gamma(w) ~~~~ \beta(z) \gamma(w) \sim \f{-1}{z-w} \sim -\gamma(z) \beta(w)
\ee
The symbol $\sim$ denotes the singular terms in the OPE. As these are chiral fields this is sufficient information to calculate all correlators. The field $\phi$ is a standard free boson:
\bea
\phi(z) \phi(w) \sim -\ln(z-w) 
\eea
We define normal ordered products using the time-splitting procedure:
\bea
(AB) = \f{1}{2 \pi i} \oint_w \f{dz}{z-w} A(z) B(w)
\eea
This does \emph{not} define an associative product and we shall always use the convention of right normal ordering $ABC=(A(BC))$. The currents obey the $\widehat{SU(2)}$ algebra:
\bea \label{eqn:SU2KM}
J^3(z) J^{\pm}(w) &\sim& \pm \f{J^{\pm}(w)}{z-w} \nonumber \\
J^+(z)J^-(w) &\sim& \f{k}{(z-w)^2}+\f{2 J^3(w)}{z-w} \\
J^3(z)J^3(w) &\sim& \f{k}{2(z-w)^2} \nonumber
\eea
The Sugawara stress tensor is:
\bea \label{eqn:sugawara}
T&=&\f{1}{2(k+2)}\left( J^+J^- + J^-J^+ + 2 J^3J^3 \right)  \nonumber \\
&=& -\beta \p \gamma - \half \p \phi \p \phi - \f{i}{\sqrt{2(k+2)}} \p^2 \phi
\eea
which coincides with the expression found before (\ref{eqn:stressT}). The central charge is:
\bea
c=\f{3k}{k+2}
\eea
and is made up by: $2+\left(1- \f{6}{k+2} \right)=\f{3k}{k+2}$. The bosonic ghost system contributes $c=2$ and the remainder comes from the $\phi$ part.
\subsection{The \KZ equation}
The free field formulation gives a very powerful approach to study the $\widehat{SU(2)}$ theories. However in many cases, and certainly in the study of logarithmic conformal field theory, it is much easier initially to follow an algebraic approach. We shall use the powerful consistency relations known as the \KZ equations \cite{Knizhnik:1984nr} based on our knowledge of the connection between the Virasoro and affine Kac-Moody algebras (\ref{eqn:Sugawara}).

In a similar manner to the Virasoro primaries we consider affine Kac-Moody primary fields having the simple OPEs:
\bea \label{eqn:KMprim}
J^a(z) \Phi(w) \sim \f{-t^a \Phi(w)}{z-w} + \cdots
\eea
where the matrix $t^a$ is a matrix representation \footnote{The minus sign arises as we define $[t^a,t^b]=if^{ab}_c t^c$ following \cite{YellowPages}} of $\Phi$ in the group $G$.

As before the pole structure (\ref{eqn:KMprim}) gives us the identity:
\bea \label{eqn:KMWardID}
\la J^a(z) \phi_1(z_1,\bar{z}_1) ... \phi_n(z_n,\bar{z}_n) \ra = - \sum_{i=1}^n \f{t^a_i}{z-z_i} \la  \phi_1(z_1,\bar{z}_1) ... \phi_n(z_n,\bar{z}_n) \ra 
\eea
However as $J^a(z)$ is an $h=1$ field we must have the asymptotic behaviour:
\bea 
J^a(z) \sim \f{1}{z^2} \quad \quad {\rm as} ~~ z \rightarrow \infty
\eea
Using this in (\ref{eqn:KMWardID}) we find the Ward identity:
\bea \label{eqn:KMWard}
\sum_{i=1}^n t^a_i  \la  \phi_1(z_1,\bar{z}_1) ... \phi_n(z_n,\bar{z}_n) \ra =0
\eea
The physical interpretation of this is that the correlator must transform as a singlet under the group $G$.

We now consider inserting the expression for the stress tensor (\ref{eqn:Sugawara}) into correlation functions of affine Kac-Moody primary fields:
\bea
\la T(z) \phi_1(z_1) \cdots \phi_n (z_n) \ra = \la \f{:J^a J^a:}{2(k+g)} (z) \phi_1(z_1) \cdots \phi_n (z_n) \ra
\eea
Now we find from examining the poles as $z$ approaches each of the points  $z_i$:
\bea \label{eqn:confweight}
h_i=\f{t^a_i t^a_i}{2(k+g)}
\eea
and
\bea \label{eqn:KZgeneral}
\left[(k+g) \frac{\p}{\p z_i}-\sum_{j\neq i}\frac{ t^a_i \otimes t^a_j}
{z_i-z_j} \right] \left<\phi_{1}(z_1) \cdots \phi_n (z_n) \right> =0 
\eea
This set of partial differential equations are known as the \KZ equations \cite{Knizhnik:1984nr}. For two and three point functions they give no new information beyond what may be obtained using the expressions for the conformal weights (\ref{eqn:confweight}) and the Virasoro Ward identities (\ref{eqn:VirWard}). However for the four point function (\ref{eqn:fourpointfn}) they become a series of coupled differential equations.\footnote{Essentially there is an equation for each appearance of a singlet in the tensor product $\phi_1 \otimes \cdots \otimes \phi_n$} For correlation functions involving the fundamental representations of a compact Lie group this equation can be easily solved \cite{Knizhnik:1984nr}. The procedure for general finite dimensional representations is, as far as we are aware, not known. For the infinite dimensional representations the situation is even more complicated and has not been completed even for the simplest case of $SU(2)$. 

In this thesis we shall restrict ourselves to the finite dimensional representations of $G=SU(2)$ as the tensor structure is rather simple and the KZ equation reduces to a set of coupled ordinary differential equations. 
\subsubsection{Auxiliary variables}
It will extremely useful in much of this thesis to introduce the following representation for the  $SU(2)$ generators \cite{Zamolodchikov:1986bd}:
\bea \label{eqn:repn}
t^+=x^2\frac{\p}{\p x}-2jx, ~~~
t^-=-\frac{\p}{\p x}, ~~~
t^3=x\frac{\p}{\p x}-j \eea
There is also a similar algebra in terms of $\bar{x}$ for the antiholomorphic part. It is easily verified that these obey the global $SU(2)$ algebra:
\bea
[t^+,t^-]=2t^3 \quad \quad [t^3,t^{\pm}]=\pm t^{\pm}
\eea
With our conventions the Casimir is given by:
\bea
t^a t^a=  t^+ t^- +  t^- t^+ + 2 t^3 t^3 
\eea
Using these auxiliary variables we can write a discrete representation with \mbox{$m=-j,\cdots, j$} as a single field:
\bea
\phi_j(x,z)=\sum_{m=-j}^{j} \phi_{j,m}x^{m+j}
\eea
Using the action of the generators $J^a(z)$ one can show they satisfy:
\bea \label{eqn:JphiOPEs}
J^-(z) \phi_j(x,w) &\sim&  -\f{ -\frac{\p}{\p x} \phi_j(x,w) }{z-w} \nonumber\\
J^3(z) \phi_j(x,w) &\sim&  -\f{\left( x\frac{\p}{\p x}-j \right)\phi_j(x,w) }{z-w} \\
J^+(z) \phi_j(x,w) &\sim&  -\f{\left(x^2\frac{\p}{\p x}-2jx \right)\phi_j(x,w) }{z-w} \nonumber
\eea
The fields $\phi_j(x,z)$ are also primary with respect to the Virasoro algebra with $L_0$ eigenvalue:
\be
h=\frac{j(j+1)}{k+2}
\ee
Using (\ref{eqn:repn}) in the affine Kac-Moody Ward identity (\ref{eqn:KMWard}) gives us a similar set of conditions\footnote{This is because the global Virasoro algebra generated by $L_{0},L_{\pm 1}$ is isomorphic to $SL(2,R)$} to that obtained from global Virasoro invariance (\ref{eqn:VirWard}). Therefore there are analogous expressions for the correlators.

The two and three point functions are fully determined using global $SU(2)$ and conformal transformations:
\be \label{eqn:2pt}
\la \phi_{j_1}(x_1,z_1) \phi_{j_2}(x_2,z_2) \ra = A(j_1) \delta_{j_1 j_2} x_{12}^{2j_1}z_{12}^{-2h}
\ee
\bea \label{eqn:3pt}
\la \phi_{j_1}(x_1,z_1) \phi_{j_2}(x_2,z_2) \phi_{j_3}(x_3,z_3) \ra = C(j_1,j_2,j_3)~~ x_{12}^{j_1+j_2-j_3} x_{13}^{j_1+j_3-j_2} x_{23}^{j_2+j_3-j_1} \\
z_{12}^{-h_1-h_2+h_3} z_{13}^{-h_1-h_3+h_2} z_{23}^{-h_2-h_3+h_1} \nonumber
\eea
For the case of the four point correlation functions of $SU(2)$ primaries the form is only determined up to a function of the cross ratios. Our convention is:
\bea \label{eqn:correl}
\langle \phi_{j_1}(x_1,z_1) \phi_{j_2}(x_2,z_2) \phi_{j_3}(x_3,z_3) \phi_{j_4}(x_4,z_4) \rangle
&=&z_{43}^{h_2+h_1-h_4-h_3}z_{42}^{-2h_2}z_{41}^{h_3+h_2-h_4-h_1} \nonumber \\
& & z_{31}^{h_4-h_1-h_2-h_3}x_{43}^{-j_2-j_1+j_4+j_3}x_{42}^{2j_2}  \\
& & x_{41}^{-j_3-j_2+j_4+j_1}x_{31}^{-j_4+j_1+j_2+j_3}~F(x,z) \nonumber
\eea 
Here the invariant cross ratios are:
\be 
x=\frac{x_{12}x_{34}}{x_{13}x_{24}} \quad \quad z=\frac{z_{12}z_{34}}{z_{13}z_{24}} 
\ee
We can now use the representation of the $SU(2)$ generators (\ref{eqn:repn}) in the \KZ equation (\ref{eqn:KZgeneral}). For the case of the four point functions it becomes a partial differential equation for $F(x,z)$:
\be \label{eqn:KZ}
(k+2) \frac{\p}{\p z} F(x,z)=\left[ \frac{\CP}{z}+\frac{\CQ}{z-1} \right] F(x,z)
\ee
The operators $\CP,\CQ$ are explicitly given by:
\bea
\CP&=&x^2(1-x)\frac{\p^2}{\p x^2}+((j_1+j_2+j_3-j_4-1)x^2-2j_1x-2j_2x(1-x))\frac{\p}{\p x} \nonumber \\
& & -2j_2(j_1+j_2+j_3-j_4)x+2j_1j_2 \\
\CQ&=&(1-x)^2 x\frac{\p^2}{\p x^2}-((j_1+j_2+j_3-j_4-1)(1-x)^2-2j_3(1-x)
-2j_2x(1-x))\frac{\p}{\p x} \nonumber \\
& & -2j_2(j_1+j_2+j_3-j_4)(1-x)+2j_2j_3 
\eea
As we shall always be dealing with the finite dimensional representations of spin $j$, with $m=-j,-j+1,\cdots,j$, we must have:
\bea
\left( J^-_0 \right)^{2j+1} \Phi_j(x,z)=0 \Rightarrow \f{\p^{2j+1}}{\p x^{2j+1}} \Phi_j(x,z)=0
\eea
Using this constraint for all the fields in the correlator, where we assume for simplicity all spins are equal i.e. $j_i=j$, we find:
\bea \label{eqn:discreterepns}
F (x,z)= \sum_{i=0}^{2j} x^i F_i(z) 
\eea
Inserting this into the \KZ equation (\ref{eqn:KZ}) allows us to reduce it to an \emph{ordinary} differential equation, of degree  $2j+1$, in terms of the lowest component $F_0(z)$.
\section{Logarithmic CFT}
During the last ten years an interesting class of conformal field theories (CFTs) has emerged called logarithmic conformal field theories (LCFTs). In normal conformal field theory the primary operators (which are in irreducible representations of the Virasoro algebra) and their descendents form a complete set of operators and all others can be expressed as linear combinations of these. However in LCFT when we compute correlation functions we find that this is not true and one must add further operators to ensure that a complete set is obtained.

In \cite{Gurarie:1993xq} Gurarie introduced the concept of LCFT and the presence of logarithmic structure in the operator product expansion was explained by the indecomposable representations that can occur in the fusion of primary operators. These can occur when there are fields with conformal dimensions differing by integers allowing operators to have a more general Jordan block structure.

We shall use as an example the case first studied in \cite{Gurarie:1993xq} as it will also be important in this thesis. The null vector for the $h_{1,2}=-\f{1}{8}$ fields in the $c=-2$ model is given by (\ref{eqn:leveltwonull}):
\bea 
( L_{-2} - 2 L_{-1}^2 )  \left|  -\f{1}{8} \right>
\eea
Requiring that this vanishes and using the definitions of mode expansion and normal ordering one can form a differential equation for the four point function:
\bea \label{cminustwocorrel}
\la \mu(z_1,\bar{z}_1) \mu(z_2,\bar{z}_2) \mu(z_3,\bar{z}_3) \mu(z_4,\bar{z}_4) \ra = |z_{13} z_{24}|^{-1/2} |z(1-z)|^{1/2} G(z,\bar{z})
\eea
\bea
z(z-1) \f{d^2 G}{d z^2} + (2z-1) \f{d G}{d z} + \f{1}{4} G =0
\eea
The solutions $F_i(z)$, known as conformal blocks, are easily found:
\bea \label{eqn:cminus2sols}
F_1(z)&=&K(z)\\
F_2(z)&=&\tilde{K}(z) \equiv K(1-z) \nonumber
\eea
where $K(z)$ given by:
\bea
K(z)=\f{\pi}{2} ~{}_2F_1 \left( \f{1}{2} ,\f{1}{2} ; 1;z \right)
\eea
This is the complete elliptic integral of the first kind (see \cite{Erdelyi} Section 13.8 but note, as has become standard in the LCFT literature, the conventions for the argument):
\bea
K(z)= \int_0^1 \f{1}{\sqrt{(1-x^2)(1-zx^2)}}  ~ dx
\eea
Now expanding $K(z)$ we find although it has a regular series expansion around $z=0$ it has \italic{logarithmic} singularities at $z=1,\infty$. Obviously $\tilde{K}(z)$ is therefore regular around $z=1$ but has logarithmic singularities at $z=0,\infty$. Therefore we cannot avoid logarithmic terms in the correlator.

As always one must combine the holomorphic and anti-holomorphic conformal blocks to get a single-valued correlator:
\bea
G(z,\bar{z}) = \sum_{i,j} U_{i,j} F_i(z) \overline{F_j(z)}
\eea
Normally for a solution behaving as $z^a$ for $z \rightarrow 0$ one has the single-valued diagonal combination:
\bea
|z|^{2a}=z^a \bar{z}^a
\eea
However if there is a logarithm present then we must instead form the combination:
\bea
\ln |z|^2 = \ln z + \ln \bar{z}
\eea
It is this difference which makes LCFT more complicated as the fields are not just products of left and right moving parts. 

The only single-valued solution with $h=\bar{h}=-\f{1}{8}$ is:
\bea \label{eqn:cminus2singval}
G(z,\bar{z}) = K(z) \overline{ \tilde{K}(z)} + \tilde{K}(z) \overline{K(z)}
\eea
This clearly leads to a correlator invariant under the exchange of fields $1 \leftrightarrow 3$, i.e. $z \leftrightarrow 1-z$, and slightly less obviously \footnote{To see this use the identities $z^{-1/2} K \left( \f{1}{z} \right) =K(z)+ i \tilde{K}(z) $ and $z^{-1/2} \tilde{K} \left( \f{1}{z} \right)=\tilde{K}(z)$} also under  $1 \leftrightarrow 4$.

Now by expanding (\ref{cminustwocorrel}) we get the OPE:
\bea
\mu(z_1,\bar{z}_1) \mu(z_2,\bar{z}_2)  \sim |z_{12}|^{1/2}  \Bigr[ D(z_2,\bar{z}_2) + \ln|z_{12}|^2 C(z_2,\bar{z}_2) + \cdots \Bigl]
\eea
and the correlators:
\bea
\la C(z_1,\bar{z}_1) C(z_2,\bar{z}_2) \ra &=& 0 \nonumber\\
\la C(z_1,\bar{z}_1) D(z_2,\bar{z}_2) \ra &=& 1 \\
\la D(z_1,\bar{z}_1) D(z_2,\bar{z}_2) \ra &=& 1-2 \ln |z|^2 \nonumber
\eea
These correlators are explained by the fact that the fields $C,D$ transform as:
\bea 
T(z) C(w,\bar{w}) &\sim& \f{h C(w,\bar{w}) }{(z-w)^2} + \f{ \p_w C(w,\bar{w}) }{z-w} + \cdots \\
T(z) D(w,\bar{w}) &\sim& \f{h D(w,\bar{w}) + C(w,\bar{w}) }{(z-w)^2} + \f{ \p_w D(w,\bar{w}) }{z-w} + \cdots \nonumber
\eea
The generator $L_0$ has a Jordan block structure:
\bea
L_0 \left( \begin{array}{c}
C \\ 
D
\end{array} \right) = \left( \begin{array}{cc} h ~~~ 0 \\ 1 ~~~ h \end{array} \right) \left( \begin{array}{c}
C \\ 
D
\end{array} \right)
\eea
where in this case we have $h=0$. The field $D$ is known as the \italic{logarithmic partner} of $C$. Now proceeding exactly as before we can form differential equations for the correlation functions of these fields \cite{Caux:1996nm}. They lead to the general form:
\bea
\la C(z_1,\bar{z_1})  C(z_2,\bar{z_2}) \ra&=&0 \nonumber\\
\la C(z_1,\bar{z_1})  D(z_2,\bar{z_2}) \ra&=&\f{A}{z^{2h}} \\
\la D(z_1,\bar{z_1})  D(z_2,\bar{z_2}) \ra&=&\f{-2 A \ln|z|^2+B}{z^{2h}} \nonumber
\eea 
In particular, the irreducible sub-representation, $C(z,\bar{z})$ is always a zero norm state. It is immediately obvious that such theories must be non-unitary as the zero-norm state is not decoupled. It is the fact that there are inevitably indecomposable representations that give rise to the logarithmic terms.

LCFTs have now been studied for over ten years now and have emerged in many different areas such as: WZNW models \cite{Caux:1997kq,Kogan:1997nd,Bernard:1997bp,Kogan:1998vx,Kogan:1999hz,Nichols:2000mk,Nichols:2001du,Giribet:2001qq,Kogan:2001nj,Gaberdiel:2001ny,Nichols:2001cv,Hadjiivanov:2001kr,Nichols:2002dk,Lesage:2002ch}, gravitational dressing \cite{Bilal:1994nx,Bilal:1995rc}, polymers and percolation \cite{Saleur:1992hk,Cardy,Gurarie:1999yx,Cardy:1999zp},  2d turbulence \cite{RahimiTabar:1996dh,RahimiTabar:1997nc,Flohr:1996ik,RahimiTabar:1997ki,RahimiTabar:1996si}, certain limits of QCD \cite{Korchemsky:2001nx,Derkachov:2002wz,Gorsky:2002ju}, the Seiberg-Witten solution of ${\mathcal N}=2$ supersymmetric Yang-Mills\cite{Cappelli:1997qf,Flohr:1998ew}, and the Abelian sand-pile model \cite{Mahieu:2001iv,Ruelle:2002jy}. One of the most important applications has been to disordered systems and the quantum hall effect \cite{Caux:1996nm,Caux:1998sm,Gurarie:1999bp,Bhaseen:1999nm,Bhaseen:2000mi,Bhaseen:2000gr,Maassarani:1997jn,Guruswamy:1999hi,Ludwig:2000em,Bhaseen:2000bm,Read:2001pz}. There has also been much work on the application of LCFT to string theory \cite{Kogan:1996df,Periwal:1996pw,Mavromatos:1998nz,Ellis:1999mj,Kogan:2000nw,Gravanis:2001zv,Mavromatos:2001iz,Gravanis:2001cy,Bakas:2002qh,Sfetsos:2002cn} and in the AdS/CFT correspondence \cite{Ghezelbash:1998rj,Kogan:1999bn,Myung:1999nd,Sanjay:1999uw,Lewis:1999qv,Lewis:2000tn,Moghimi-Araghi:2001fg,Jabbari-Faruji:2002xz}. The holographic relation between logarithmic operators and vacuum instability was considered in \cite{Kogan:1998xm,Lewis:1998fg}. An approach to LCFT using nilpotent dimensions was given in \cite{Moghimi-Araghi:2000qn,Moghimi-Araghi:2002gk}. An example which we shall return to later is the appearance of a logarithmic partner of the stress tensor in $c=0$ LCFTs \cite{Gurarie:1999yx,CardyTalk,Kogan:2002mg}. This has also been discussed in more general settings \cite{Moghimi-Araghi:2000dt,Kogan:2001ku}. A recent area of particular interest has been the analysis of LCFTs in the presence of a boundary \cite{Moghimi-Araghi:2000cx,Kogan:2000fa,Ishimoto:2001jv,Kawai:2001ur,Bredthauer:2002ct}.

There has also been a lot of important work on analysing the general structure and consistency of such models in particular the $c_{p,q}$ models and the special case of $c=-2$ which is by far the best understood \cite{Kausch:1995py,Flohr:1996ea,Gaberdiel:1996kx,Gaberdiel:1996np,Gaberdiel:1998ps,Kausch:2000fu}. For more about the general structure of LCFT see \cite{RahimiTabar:1997ub,Rohsiepe:1996qj,Kogan:1997fd,Flohr:2001tj} and references therein. Recently a very general construction of LCFT has been proposed via deformations of ordinary CFT \cite{Fjelstad:2002ei}. Many of the concepts of standard CFT, for example null vectors, characters and fusion rules, also have analogues in LCFT although there are important differences \cite{Flohr:1996ea,Flohr:1997vc,Flohr:1998wm,Flohr:2000mc,Flohr:2001tj}. Introductory lecture notes on LCFT and more references can be found in \cite{Tabar:2001et,Flohr:2001zs,Gaberdiel:2001tr,Kawai:2002fu}. 
\subsection{Rational models}
In conformal field theory a fundamental role is played by the chiral algebra of the theory and all fields transform in representations of this algebra. The Virasoro algebra (\ref{eqn:Virasoroalg}) is one of the simplest and most universal examples. The models which are mathematically best defined are those in which there exist only a finite number of basic fields. In these cases it is sufficient to describe the fusion rules and correlation functions of these fields. In non-logarithmic CFT these fields are all in irreducible representations and such models are called rational CFTs, the minimal models being an excellent example \cite{Belavin:1984vu}.

In the $c_{p,1}$ models, and in particular the $c=-2$ case that we will discuss in detail, it was found that although these models involve an infinite number of Virasoro representations they can be rearranged into a finite number of representations of a larger chiral algebra. Therefore they are not rational with respect to the Virasoro algebra but \emph{are} rational with respect to a larger chiral algebra. The use of the term rational here differs slightly from the previous one, and also the usual mathematical definition, as although there are only a finite number of basic fields they are not all in irreducible representations. This extended chiral algebra is generated by a triplet of $h_{3,1}=2p-1$ fields. In the case of $c_{2,1}=-2$ the algebra can be explicitly calculated with relative ease:
\bea \label{eqn:tripletalgebra}
T(z) T(w) &\sim& \f{-2}{2 (z-w)^4} + \f{2 T(w)}{(z-w)^2} + \f{\p T(w)}{z-w} \nonumber \\
T(z) W^a(w) &\sim& \f{3 W^a(w)}{(z-w)^2}+ \f{\p W^a(w)}{z-w} \\
W^a(z) W^b(w) &\sim&  g^{ab} \left( \f{1}{(z-w)^6}- 3 \f{T(w)}{(z-w)^4}-\f{3}{2} \f{\p T(w)}{(z-w)^3} + \f{3}{2} \f{\p^2 T(w)}{(z-w)^2} \right. \nonumber  \\
&&~~~~ \left.  -4 \f{(T^2)(w)}{(z-w)^2}  + \f{1}{6} \f{\p^3 T(w)}{z-w} -4 \f{\p(T^2)(w)}{z-w} \right) \nonumber \\ 
&& - 5 f^{ab}_c \left( \f{W^c(w)}{(z-w)^3} + \f{1}{2}\f{\p W^a(w)}{(z-w)^2} + \f{1}{25} \f{\p^2 W^c(w)}{z-w} + \f{1}{25} \f{(TW^c)(w)}{z-w} \right) \nonumber
\eea
where $g^{ab}$ and $f^{ab}_c$ are the metric and structure constants of $SU(2)$. This can also be written in terms of modes:
\bea
\left[L_m,L_n \right] &=& (m-n)L_{m+n} - \f{1}{6} m(m^2-1) \delta_{m+n,0} \nonumber \\
\left[L_m,W^a_n \right] &=& (2m-n)W^a_{m+n} \\
\left[W^a_m,W^b_n \right] &=& g^{ab} \Bigl( 2(m-n) \Lambda_{m+n} + \f{1}{20}(m-n)(2m^2+2n^2-mn-8) L_{m+n} \nonumber \\
 &&-\f{1}{120}m(m^2-1)(m^2-4)\delta_{m+n,0} \Bigr)  \nonumber\\
&&+f^{ab}_c \Bigl( \f{5}{14}(2m^2+2n^2-3mn-4)W^c_{m+n} +\f{12}{5} V^a_{m+n} \Bigr)\nonumber
\eea
where the normal ordered fields are:
\bea
\Lambda= (LL)-\f{3}{10}\p^2 L \quad \quad
V^a= (LW^a)-\f{3}{14}\p^2 W^a 
\eea
This triplet algebra is only associative if certain vacuum null vectors decouple \cite{Gaberdiel:1996np}. The vanishing of these null vectors in all correlators determines a consistent set of representations that close under fusion. The operator content is given by six fundemental representations: the irreducible ones : $\nu_0,\nu_1,\nu_{-1/8},\nu_{3/8}$ and the two indecomposable ones ${\mathcal R}_0,{\mathcal R}_1$.
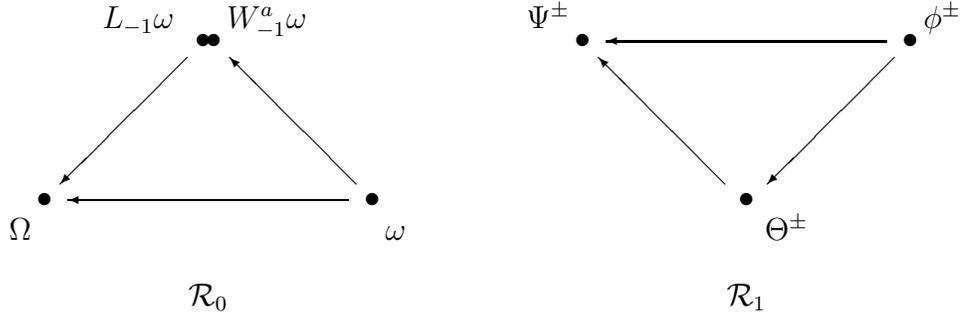
\begin{figure}
\begin{displaymath}
\begin{picture}(180,140)(-90,-50)
\put(62,0){\vbox to 0pt{\vss\hbox to 0pt{\hss$\bullet$\hss}\vss}}
\put(-62,0){\vbox to 0pt{\vss\hbox to 0pt{\hss$\bullet$\hss}\vss}}
\put(2,60){\vbox to 0pt{\vss\hbox to 0pt{\hss$\bullet$\hss}\vss}}
\put(-2,60){\vbox to 0pt{\vss\hbox to 0pt{\hss$\bullet$\hss}\vss}}
\put(56,6){\vector(-1,1){48}}
\put(-8,54){\vector(-1,-1){48}}
\put(53,0){\vector(-1,0){106}}
\put(-67,-15){\hbox to 0pt{\hss$\Omega$}}

\put(67,-15){\hbox to 0pt{$\omega$\hss}}
\put(7,65){\hbox to 0pt{$W^a_{-1}\omega$\hss}}
\put(-40,65){\hbox to 0pt{$L_{-1}\omega$\hss}}
\put(0,-40){\hbox to 0pt{\hss${\cal R}_0$\hss}}
\end{picture}
\qquad
\begin{picture}(180,140)(-90,-50)
\put(62,60){\vbox to 0pt{\vss\hbox to 0pt{\hss$\bullet$\hss}\vss}}
\put(-62,60){\vbox to 0pt{\vss\hbox to 0pt{\hss$\bullet$\hss}\vss}}
\put(0,0){\vbox to 0pt{\vss\hbox to 0pt{\hss$\bullet$\hss}\vss}}
\put(56,54){\vector(-1,-1){48}}
\put(-8,6){\vector(-1,1){48}}
\put(53,60){\vector(-1,0){106}}
\put(-67,65){\hbox to 0pt{\hss$\Psi^\pm$}}
\put(67,65){\hbox to 0pt{$\phi^\pm$\hss}}
\put(7,-15){\hbox to 0pt{$\Theta^\pm$\hss}}
\put(0,-40){\hbox to 0pt{\hss${\cal R}_1$\hss}}
\end{picture}
\end{displaymath}
\caption{Indecomposable representations in the $c=-2$ triplet model.} \label{fig:indecomprepn}
\end{figure}
The vertices in Figure \ref{fig:indecomprepn} correspond to representations and the arrows indicate the action of the chiral algebra. The representation ${\cal R}_0$ is generated from the cyclic state $\omega$ whereas ${\cal R}_1$ is generated from $\phi^{\pm}$. The defining relations of ${\cal R}_0$ and ${\cal R}_1$ are:
\bea
\begin{array}{ll}
L_0 \omega = \Omega \quad \quad &  W^a_0 \omega =0 \\
\\
L_0 \Psi^{\alpha}=\Psi^{\alpha}\quad \quad  &  W^a_0 \Psi^{\alpha}=t^{a \alpha}_{\beta} \Psi^{\beta} \\
L_0 \Theta^{\alpha}=0 \quad \quad & W^a_0 \Theta^{\alpha}=0 \\
L_{-1} \Theta^{\alpha}=\Psi^{\alpha} \quad \quad & W^a_{-1} \Theta^{\alpha}=t^{a \alpha}_{\beta} \Psi^{\beta} \\
L_1 \phi^{\alpha}=-\Theta^{\alpha} \quad \quad & W^a_1 \phi^{\alpha}=-t^{a \alpha}_{\beta} \Theta^{\beta} \\
L_0 \phi^{\alpha}=\phi^{\alpha}+ \Psi^{\alpha} \quad \quad & W^a_0 \phi^{\alpha}=2 t^{a \alpha}_{\beta} \phi^{\beta} \\
\end{array}
\eea
where $t^{a \alpha}_{\beta}$ is a $2 \times 2$ matrix representation of $SU(2)$. One can also see how the irreducible representations $\nu_0$ and $\nu_1$, generated from the states $\Omega$ and $\Psi^{\pm}$ respectively, are embedded within ${\cal R}_0$ and ${\cal R}_1$.

 The full set of fusion rules have also been calculated \cite{Gaberdiel:1996np}:
\bea \label{eqn:oldfusionrules}
\nu_0 \otimes X &=&  X \quad \quad  {\rm for~ all ~ X }\nonumber \\
\nu_{-1/8} \otimes \nu_{-1/8} &=& \CR_0 \nonumber\\
\nu_{-1/8} \otimes \nu_{3/8} &=& \CR_1 \nonumber\\
\nu_{-1/8} \otimes \nu_1 &=& \nu_{3/8} \nonumber\\
\nu_{-1/8} \otimes \CR_0 &=& 2 \nu_{-1/8} \oplus 2 \nu_{3/8} \nonumber\\
\nu_{-1/8} \otimes \CR_1 &=& 2 \nu_{-1/8} \oplus 2 \nu_{3/8} \nonumber\\
\nu_{3/8} \otimes \nu_{3/8} &=& \CR_0 \nonumber\\
\nu_{3/8} \otimes \nu_1 &=& \nu_{-1/8}  \\
\nu_{3/8} \otimes \CR_0 &=& 2 \nu_{-1/8} \oplus 2 \nu_{3/8} \nonumber\\
\nu_{3/8} \otimes \CR_1 &=& 2 \nu_{-1/8} \oplus 2 \nu_{3/8} \nonumber\\
\nu_1 \otimes \nu_1 &=& \nu_0 \nonumber\\ 
\nu_1 \otimes \CR_0 &=& \CR_1 \nonumber\\
\nu_1 \otimes \CR_1 &=& \CR_0 \nonumber\\
\CR_0 \otimes \CR_0 &=& 2\CR_0 \oplus 2\CR_1  \nonumber\\
\CR_0 \otimes \CR_1 &=& 2\CR_0 \oplus \CR_1 \nonumber\\
\CR_1 \otimes \CR_1 &=& 2\CR_0 \oplus 2\CR_1\nonumber
\eea
In order to get a well defined theory one must combine the holomorphic and anti-holomorphic sectors together and ensure that all correlators are single-valued. In LCFT this is always non-trivial due to the appearance of logarithmic terms. For the case of the rational $c=-2$ triplet model all the fundemental correlators of the local theory have been calculated explicitly to verify the overall consistency \cite{Gaberdiel:1998ps}.
\chapter{The $SU(2)_0$ WZNW model: Introduction}
We shall now begin to analyse in some detail a particular $\widehat{SU(2)}$ theory, namely $SU(2)_0$, beyond the integrable representations. The correlators of the $j=\half$ representation were previously discussed in \cite{{Caux:1997kq}} but here we shall use the \KZ equation with auxiliary variables to analyse many of the other representations of the theory.

We shall study the correlation functions of discrete irreducible representations of $SU(2)_0$, with $2j \in Z$, because in these cases one can easily solve the \KZ equation. However as we shall see not all the operators produced in fusion are in irreducible representations of the full affine Kac-Moody algebra.  We also create indecomposable representations which are characterised by logarithmic singularities in the correlation functions. It is important to note that the indecomposable nature is due to the affine algebra and \emph{not} the global $SU(2)$. Indeed it is known that as far as the global $SU(2)$ symmetry is concerned the discrete representations must be irreducible and must fuse to give other discrete representations \cite{SL2RBook}. We shall further find that by careful consideration of certain correlators one can deduce extra fermionic operators in the theory.

Our discussion of the theory, and indeed this entire thesis, will be limited to considering correlators on the sphere. It is quite possible that on the torus modular transformations may force one to introduce other representations. However it seems reasonable if one wanted to produce a rational theory the chiral algebras would be generated by discrete representations.

In addition to the question of other representations there is also considerable doubt as to the precise correspondence between the conformal blocks that we shall calculate and the actual correlators that realise them. It is quite possible that, as with many other LCFTs, one needs to insert extra logarithmic partners to produce the given conformal blocks. We shall also make the simplifying assumption when deducing OPEs that the vacuum state is normalisable i.e. $\left< \Omega \right>=1$. These are all questions of global consistency of the theory which we are not in a position to answer at this stage. However we believe that the general results will not be affected by such discussions. 
\section{Doublet solutions}
We shall now discuss the simplest, non-trivial, irreducible fields in the model namely the $j=\half$ representations. These were previously discussed in \cite{Caux:1997kq} but we shall rederive them here partly for completeness but also to demonstrate the auxiliary variable approach.

Putting $j_i=\half$ we find:
\bea
\CP &=& x^2(x-1)\frac{\p^2}{\p x^2}+(2x-x^2)\frac{\p}{\p x}+(x-\half) \\
\CQ &=& -(1-x)^2x\frac{\p^2}{\p x^2}+(x^2-1)\frac{\p}{\p x}+(\half-x) \nonumber 
\eea
As explained earlier, using the fact that the representations are discrete, we have:
\be
F(x,z)=A(z)+xB(z)
\ee
Substituting this into the \KZ equation (\ref{eqn:KZ}) and separating powers of $x$ leads to:
\bea
2z(k+2)(z-1) \f{d A}{d z} + A -2zB &=& 0 \\
2z(k+2)(z-1) \f{d B}{d z} -2A + (4z-3)B  &=& 0 \nonumber 
\eea
Now eliminating $B(z)$ we get:
\bea
&&4z^2(k+2)^2(z-1)^2 \f{d^2 A}{d z^2} \\
&& \quad 4(k+2)z(z-1)(4z+kz-1)\f{d A}{d z} - (2kz+4z-2k-1)A=0 \nonumber 
\eea
For generic $k$ this has two linearly independent hypergeometric solutions:
\bea
A^{(1)}(z)&=&z^{\f{1}{2(k+2)}} (z-1)^{\f{-3}{2(k+2)}} {}_2F_1 \lb \f{1}{k+2}, \f{-1}{k+2} ; \f{2}{k+2};z \rb \\
A^{(2)}(z)&=&z^{\f{2k+1}{2(k+2)}} (z-1)^{\f{-3}{2(k+2)}} {}_2F_1 \lb  \f{k+1}{k+2}, \f{k-1}{k+2} ; \f{2(k+1)}{k+2};z \rb \nonumber
\eea
However for certain values of $k$ the solutions develop logarithmic singularities. A necessary, but not sufficient, condition is that $\f{2}{k+2} \in Z$. In these cases in the tensor product $\f{1}{2} \otimes \f{1}{2} \rightarrow 0 \oplus 1$ the $0$ and $1$ fields have conformal weights differing by integers and may therefore be contained in an indecomposable representation.

At $k=0$ the solutions of the \KZ equation are found to be:
\bea \label{eqn:SU20doublet}
F_{\half \half \half \half}^{(1)}(x,z)&=& z^{\f{1}{4}}(1-z)^{\f{1}{4}} \left\{ \left( -\f{E}{z(1-z)}+\f{K}{z} \right) x + \f{E}{1-z} \right\} \\
F_{\half \half \half \half}^{(2)}(x,z)&=& z^{\f{1}{4}}(1-z)^{\f{1}{4}} \left\{ \left( \f{\tilde{E}}{z(1-z)}-\f{\tilde{K}}{1-z} \right) x + \f{\tilde{K}}{1-z} -\f{\tilde{E}}{1-z} \right\} \nonumber
\eea
In the case of $\widehat{SU(2)}$ we shall always use the subscript labels to denote the spin $j$ of the fields invloved. The complete elliptic integrals $E(z),K(z)$ (see \cite{Erdelyi} Section 13.8 but note again the conventions for the argument) are:
\bea
E(z)&=& \int_0^1 \f{\sqrt{1-z x^2}}{\sqrt{1-x^2}} dx = \f{\pi}{2} ~{}_2F_1\left( -\half,\half;1;z \right)  \nonumber \\
K(z)&=& \int_0^1 \f{1}{\sqrt{(1-x^2)(1-z x^2)}} dx = \f{\pi}{2} ~{}_2F_1\left( \half,\half;1;z \right) \\
\tilde{E}(z)&=&E(1-z) \quad \quad \quad  \tilde{K}(z)=K(1-z)\nonumber 
\eea
Imposing single-valuedness leads to the unique result for the correlation function:
\bea
G(x,\bar{x},z,\bar{z})= F_{\half \half \half \half}^{(1)}(x,z) \overline{F_{\half \half \half \half}^{(2)}(x,z)} + F_{\half \half \half \half}^{(2)}(x,z) \overline{F_{\half \half \half \half}^{(1)}(x,z)}
\eea
%
%
%
%
%
We can write the OPEs derived from this correlator:
\bea \label{eqn:OPEhalfhalf}
g(x_1,z_1;\bar{x}_1,\bar{z}_1) g(x_2,z_2;\bar{x}_2,\bar{z}_2)\sim |z_{12}|^{-3/2} |x_{12}|^{2} \left\{ \f{z_{12}}{x_{12}}  K(x_2,z_2) + \f{\bar{z}_{12}}{\bar{x}_{12}}\bar{K} (\bar{x}_2, \bar{z}_2) \right. \nonumber \\
 \left. + \f{|z_{12}|^2}{|x_{12}|^2} \Bigl[ D(x_2,z_2;\bar{x}_{2},\bar{z}_2) + \ln |z_{12}|^2 C(x_2,z_2;\bar{x}_{2},\bar{z}_2)\Bigr] + \cdots \right\}
\eea
where $g(x,z)$ is the $j=\half$ field. Now we find non-trivial correlators:
\bea \label{eqn:KKOPE}
\la K(x_1,z_1) K(x_2,z_2) \ra &=& \f{x_{12}^2}{z_{12}^2} \nonumber \\
\la C(x_1,z_1;\bar{x}_{1},\bar{z}_1) C(x_2,z_2;\bar{x}_{2},\bar{z}_2) \ra &=& 0 \\
\la C(x_1,z_1;\bar{x}_{1},\bar{z}_1) D(x_2,z_2;\bar{x}_{2},\bar{z}_2) \ra &=& \f{2 |x_{12}|^4}{|z_{12}|^4} \nonumber \\
\la D(x_1,z_1;\bar{x}_{1},\bar{z}_1) D(x_2,z_2;\bar{x}_{2},\bar{z}_2) \ra &=& \f{-4 |x_{12}|^4 \ln |z_{12}|^2 }{|z_{12}|^4} \nonumber
\eea
We also have:
\bea
\la g(x_1,z_1;\bar{x}_{1},\bar{z}_1) g(x_2,z_2;\bar{x}_{2},\bar{z}_2) \ra &=& 0
\eea
These are exactly the same as those given in \cite{Caux:1997kq} written using auxilary variables. In this simple case there is little value in using the auxiliary variable approach. However for higher spin representations it allows one to very quickly derive the differential equations to be solved.
\section{Triplet solutions}
\label{sec:Tripletsolutions}
We now describe the structure of the $j=1$ operators of the theory. We shall see that this gives the first real hint of a deeper underlying structure.

For $j_i=1$ we get:
\bea
\CP &=& x^2(x-1)\frac{\p^2}{\p x^2}+(-3x^2+4x)\frac{\p}{\p x}+(4x-2) \\
\CQ &=& -(1-x)^2x\frac{\p^2}{\p x^2}+(3x^2-2x-1)\frac{\p}{\p x}+(2-4x) \nonumber
\eea
We now have:
\be
\CF_{1111}(x,z)=F(z)+xG(z)+x^2H(z)
\ee
Proceeding as before we now get three equations:
\bea \label{eq:3eqns}
(k+2)z(z-1) \f{d F}{d z} + 2F-zG&=&0 \nonumber\\
(k+2)z(z-1) \f{d G}{d z} -4F-2G+2zG-4zH &=&0 \\
(k+2)z(z-1) \f{d H}{d z} -G -4H+6zH &=& 0\nonumber
\eea
These lead to third order ODEs for $F,G$ and $H$ which are actually sufficiently simple that one may write down the general solution to them. First observe that under $1 \leftrightarrow 3$ we have $x,z \leftrightarrow 1-x,1-z$ and so $H(1-z)=H(z)$. Anticipating such a symmetry one may introduce the variable $u=z(1-z)$ and rewrite the equation for $H(z)$ in terms of $u$. One can then find the following solutions:
\bea
H^{(1)}(u)&=&u^{-2/(k-2)} {}_3F_2 \left(\f{1}{2}-\f{3}{k-2},1-\f{2}{k-2},-\f{5}{k-2}; 1-\f{6}{k-2},1-\f{4}{k-2};4u \right) \nonumber \\
H^{(2)}(u)&=&u^{2/(k-2)} {}_3F_2 \left(\f{1}{2}+\f{1}{k-2},1+\f{2}{k-2},-\f{1}{k-2}; 1-\f{2}{k-2},1+\f{4}{k-2};4u \right) \nonumber\\
H^{(3)}(u)&=&u^{4/(k-2)} {}_3F_2 \left(\f{1}{2}+\f{3}{k-2},1+\f{4}{k-2},\f{1}{k-2}; 1+\f{2}{k-2},1+\f{6}{k-2};4u \right)\nonumber
\eea
These are similar to those found in the context of $OSp(2|2)_k$ \cite{Maassarani:1997jn}. However they are of limited use as the transformation $u=z(1-z)$ is not globally invertible. We shall therefore proceed in a similar manner to before. The equation for $F$ is:
\bea \label{eq:Feqn}
(k+2)^3 z^3 (z-1)^3 \f{d^3 F}{d z^3} + (k+2)^2 z^2 (z-1)^2  \left( 14z +3kz-4 \right) \f{d^2 F}{d z^2} \nonumber \\
+(k+2) z (z-1) \left( k^2 z^2+12k z^2+32 z^2-6kz-24z+2k \right) \f{d F}{d z} \\
+ 2 \left( 2k^2 -8z^2 -2kz^2-3k^2 z+k^2 z^2 +2 kz \right) F =0 \nonumber
\eea
In \cite{Fuchs:1987ew} the four point function of vector fields in the  $O(N)$ model was given by an integral representation for general $k$. Our calculation is very similar, as $O(3) \sim SU(2)$, and has the integral representation:
\be
F(z)= \int_{C_1} {ds \int_{C_2} {dt ~ (s t)^{\alpha} \left[(s-1)(t-1) \right]^{\beta} \left[(s-z)(t-z)\right]^{\gamma} (s-t)^{\delta} }}
\ee
where $\alpha=-\f{2}{k+2}, ~~ \beta=-\f{2}{k+2}-1, ~~ \gamma=-\f{2}{k+2}, ~~ \delta=\f{2}{k+2}$. However for certain values of $k$, or choices of contours, this fails to converge and is not a particularly useful form. 
At $k=0$ one finds the following factorisation of (\ref{eq:Feqn}):
\bea \label{eqn:Feqnfact3}
\left( (z-1)\f{d}{d z}+2 \right)
\left( z(z-1)\f{d}{d z}-1 \right)
\left( (z-1)\f{d}{d z}+1 \right) F(z)=0
\eea
We can easily solve this and hence find solutions for $F(x,z)$. These can be conveniently written in a basis:
\bea \label{eqn:spinone}
F_{1111}^{(1)}(x,z)&=&\f{1}{1-z}+\f{2x}{z}-\f{x^2}{z(1-z)} \nonumber \\
F_{1111}^{(2)}(x,z)&=& \lb 1-\f{x^2}{z^2} \rb + \ln z \CF_{1111}^{(1)} \\ 
F_{1111}^{(3)}(x,z)&=&\CF_{1111}^{(2)}(1-x,1-z)\nonumber
\eea
Now we must again combine left and right moving parts to get a single valued correlator. However $F_{1111}^{(1)}(x,z)$ and its complex conjugate already give well defined correlators for fields with $j_i=1,\bar{j}_i=0$ and $j_i=0,\bar{j}_i=1$ respectively and so these are good candidates for local chiral fields. A crucial point, which we shall return to later, is the appearance of rational solutions for these correlators.

Let us briefly revert to the index notation by expanding:
\bea \label{eqn:indexnotationexpanded}
\phi_{1}(x,z)= \phi_{1}^{-}(z) -2 x  \phi^{0}_{1}(z) - x^2 \phi^{+}_{1}(z)
\eea
Inserting this into the general form of the two and three point functions (\ref{eqn:2pt},~\ref{eqn:3pt}) we get:
\be 
\la \phi^{a}_{1}(x_1,z_1) \phi^{b}_{1}(x_2,z_2) \ra = A \frac{\eta_{ab}}{z_{12}^{2}}
\ee
\be
\la \phi^{a}_{1}(x_1,z_1) \phi^{b}_{1}(x_2,z_2) \phi^{c}_{1}(x_3,z_3) \ra = \frac{C~\epsilon^{abc}}{z_{12} z_{13} z_{23}} \nonumber
\ee
We can now do the same for the four point function (\ref{eqn:correl}) with the solution $F_{1111}^{(1)}$:
\bea  \label{eq:allsame}
\langle \phi^{a}_{1}(z_1) \phi^{b}_{1}(z_2) \phi^{c}_{1}(z_3) \phi^{d}_{1}(z_4) \rangle =  \frac {\delta^{ab}\delta^{cd}}{z_{13}z_{14}z_{23}z_{24}} + \frac {\delta^{ac}\delta^{bd}}{z_{12}z_{34}z_{14}z_{32}} + \frac {\delta^{ad}\delta^{bc}}{z_{13}z_{12}z_{43}z_{42}} 
\eea
One might initially think that since the Kac-Moody current $J^a$ in $SU(2)_0$ is itself a primary $j=1$ field with $h=1$ one should equate this with the correlator:
\bea 
\la J^a(z_1) J^b(z_2) J^c(z_3) J^d(z_4) \ra
\eea
However use of the $SU(2)_0$ Kac-Moody algebra (\ref{eqn:SU2KM}) shows that this must vanish\footnote{In fact it is equal to $\lim_{k \rightarrow 0} \la J^a(z_1) J^b(z_2) J^c(z_3) J^d(z_4) \ra$. This suggests that such a rational solution must also exist for correlators of adjoint fields in $SU(N)_0$.}. The resolution, as is common in LCFT, see for example \cite{Flohr:2000mc}, is that in order to have a non-vanishing correlator, and realise the conformal block of the irreducible operators, one must have an insertion of a logarithmic partner:
\bea 
\la J^a(z_1) J^b(z_2) J^c(z_3) N^d(z_4,\bar{z}_4) \ra =  \frac {\delta^{ab}\delta^{cd}}{z_{13}z_{14}z_{23}z_{24}} + \frac {\delta^{ac}\delta^{bd}}{z_{12}z_{34}z_{14}z_{32}} + \frac {\delta^{ad}\delta^{bc}}{z_{13}z_{12}z_{43}z_{42}}
\eea
Such a field $N^a$ was found in \cite{Caux:1997kq} by considering the fusion of the $K(x,z)$ field (\ref{eqn:OPEhalfhalf}) with itself and has the behaviour:
\bea \label{eqn:JNOPE}
J^a(z) N^b(w) \sim \f{\delta^{ab}}{(z-w)^2} + \f{i f^{abc} N^c(w)}{z-w}
\eea
As it is a logarithmic partner it must actually be non-chiral although this is not seen in the above correlator. We still have full crossing symmetry of the correlator as we can we consider it to be the lowest component of a correlator involving the full Jordan block containing both $J^a$ and $N^a$ (see for example \cite{Moghimi-Araghi:2000qn}). All the fields behave bosonically exactly as one would expect.

There is another way to produce a well behaved correlator if we have \emph{fermionic} fields. To see this consider the most general expression involving \mbox{$j=1$, $\bar{j}=1$} fields:
\bea
G(x,\bar{x},z,\bar{z})=\sum_{i,j=1}^{3}{U_{i,j}  F_{1111}^{(i)}(x,z) \overline{ F_{1111}^{(j)}(x,z)}}
\eea
To make this single-valued everywhere we must have:
\bea \label{eq:Gfull}
G(x,\bar{x},z,\bar{z})&=&U_{1,1}  F_{1111}^{(1)}(x,z) \overline{ F_{1111}^{(1)}(x,z)} \nonumber\\
&+& U_{1,2} \Bigl[  F_{1111}^{(1)}(x,z) \overline{ F_{1111}^{(2)}(x,z)} +  F_{1111}^{(2)}(x,z) \overline{ F_{1111}^{(1)}(x,z)} \Bigr]  \\
&+& U_{1,3} \Bigl[  F_{1111}^{(1)}(x,z) \overline{ F_{1111}^{(3)}(x,z)} +  F_{1111}^{(3)}(x,z) \overline{ F_{1111}^{(1)}(x,z)} \Bigr] \nonumber
\eea
The first of these is similar to the case just discussed and therefore we shall only consider the additional solutions. In contrast to normal CFT these do not have just a simple diagonal form. This is one well known difference in logarithmic CFT.

Under $1 \rightarrow 3$ we have $x,z \rightarrow 1-x,1-z$ and the blocks transform as:
\bea
 F_{1111}^{(1)} &\rightarrow&  F_{1111}^{(1)} \nonumber \\
 F_{1111}^{(2)} &\rightarrow&  F_{1111}^{(3)} \\
 F_{1111}^{(3)} &\rightarrow&  F_{1111}^{(2)}\nonumber
\eea
Under $1 \rightarrow 4$ we have $x,z \rightarrow \f{1}{x},\f{1}{z}$:
\bea
 F_{1111}^{(1)} &\rightarrow& \f{z^2}{x^2}  F_{1111}^{(1)} \nonumber\\
 F_{1111}^{(2)} &\rightarrow& - \f{z^2}{x^2}  F_{1111}^{(2)} \\
 F_{1111}^{(3)} &\rightarrow& \f{z^2}{x^2} \left( - i \pi  F_{1111}^{(1)} -  F_{1111}^{(2)} +  F_{1111}^{(3)} \right) \nonumber
\eea
Immediately we see that the single-valued combination:
\bea
 F_{1111}^{(1)}(x,z) \overline{ F_{1111}^{(2)}(x,z)} +  F_{1111}^{(2)}(x,z) \overline{ F_{1111}^{(1)}(x,z)}
\eea
behaves fermionically under $1 \leftrightarrow 4$ but has no symmetry under $1 \leftrightarrow 2$ or $1 \leftrightarrow 3$. To indicate these statistics we shall write the correlator as:
\bea
\la \Theta^+(0,0) \Theta^-(x,z) \Theta^-(1,1) \Theta^+(\infty,\infty) \ra =  F_{1111}^{(1)}(x,z) \overline{ F_{1111}^{(2)}(x,z)} +  F_{1111}^{(2)}(x,z) \overline{ F_{1111}^{(1)}(x,z)} \nonumber
\eea
where the minus sign arises from the exchange of identical fermionic fields. We emphasize that these extra labels have nothing to do with the Kac-Moody algebra. All the $\widehat{SU(2)}$ structure is encoded, as before, in the $x$-variables. Therefore by careful consideration of the solutions we have deduced an extra structure underlying the correlators. This is one of the main results of this thesis. The fields $\Theta^{\pm}(x,z)$ are clearly non-chiral. We can again easily revert to the index notation and write:
\bea
&&\la  \lb \Theta^+\rb^{a \bar{a}}(z_1,\bar{z}_1) \lb \Theta^- \rb^{b \bar{b}}(z_2,\bar{z}_2) \lb \Theta^- \rb^{c \bar{c}}(z_3,\bar{z}_3) \lb \Theta^+ \rb^{d \bar{d}}(z_4,\bar{z}_4) \ra \\
&=& \ln \left| \f{z_{12} z_{34} }{z_{13} z_{24} } \right|^2  F_{1111}^{(1)} \overline{ F_{1111}^{(1)}} 
+   \Bigl[ \f{\delta^{ab}\delta^{cd}}{z_{12}^2 z_{34}^2} - \f{\delta^{ac}\delta^{bd}}{z_{13}^2 z_{24}^2} \Bigr] \overline{ F_{1111}^{(1)}} 
+  F_{1111}^{(1)} \overline{\Bigl[ \f{\delta^{ab}\delta^{cd}}{z_{12}^2 z_{34}^2} - \f{\delta^{ac}\delta^{bd}}{z_{13}^2 z_{24}^2} \Bigr]} \nonumber
\eea
where $ F_{1111}^{(1)}$ is given by the RHS of equation (\ref{eq:allsame}). 

Clearly as $ F_{1111}^{(3)}(x,z)$ is related to $ F_{1111}^{(2)}(x,z)$ under $1 \leftrightarrow 3$ other solutions correspond to:
\bea
\la \Theta^+(0,0) \Theta^+(x,z) \Theta^-(1,1) \Theta^-(\infty,\infty) \ra &=&  F_{1111}^{(1)}(x,z) \overline{ F_{1111}^{(3)}(x,z)} +  F_{1111}^{(3)}(x,z) \overline{ F_{1111}^{(1)}(x,z)} \nonumber \\
\la \Theta^+(0,0) \Theta^-(x,z) \Theta^+(1,1) \Theta^-(\infty,\infty) \ra &=&  F_{1111}^{(1)}(x,z) \overline{\left(  F_{1111}^{(3)}(x,z) -  F_{1111}^{(2)}(x,z) \right) } \\
&& + \left(  F_{1111}^{(3)}(x,z) -  F_{1111}^{(2)}(x,z) \right) \overline{ F_{1111}^{(1)}(x,z)} \nonumber 
\eea
The OPEs deduced from these correlators are:
\bea \label{eqn:OPEthetatheta}
\Theta^{\alpha}(x_1,z_1;\bar{x}_1,\bar{z}_1) \Theta^{\beta}(x_2,z_2;\bar{x}_2,\bar{z}_2)&\sim& d^{\alpha \beta} |z_{12}|^{-4} |x_{12}|^{4} \left\{ \f{z_{12}}{x_{12}}  K(x_2,z_2) + \f{\bar{z}_{12}}{\bar{x}_{12}}\bar{K} (\bar{x}_2, \bar{z}_2) \right. \nonumber \\
&& \hspace{-1.3in}\left. + \f{|z_{12}|^2}{|x_{12}|^2} \left[ D(x_2,z_2;\bar{x}_{12},\bar{z}_2) + \ln |z_{12} |^2 C(x_2,z_2;\bar{x}_{12},\bar{z}_2)\right] +  \cdots \right\}~
\eea
%
%
This looks very similar to the OPE deduced for two $j=\f{1}{2}$ fields (\ref{eqn:OPEhalfhalf}) and indeed we also find the same non-vanishing two point functions. We also see directly from the OPE (\ref{eqn:OPEthetatheta}) that we again have a vanishing two point function:
\bea \label{eqn:ThetaThetaequalszero}
\la \Theta^{\alpha}(x_1,z_1;\bar{x}_1,\bar{z}_1) \Theta^{\beta}(x_2,z_2;\bar{x}_2,\bar{z}_2) \ra =0
\eea
From examination of the correlators it appeared in \cite{Caux:1997kq} that these fields $K(x,z)$, or equivalently written $K^a(z)$, were also $j=1$ affine Kac-Moody primaries but with a non-zero two-point function  (\ref{eqn:KKOPE}). However if this is the case one would expect the four point function $\left<KKKK \right>$ to be given by the conformal blocks (\ref{eqn:spinone}). We have seen the only bosonic solution is that involving $F_{1111}^{(1)}(z)$ and it leads to a correlator having vanishing two point function. Of course this is dependent on our assumption of the existence of a normalisable vacuum state. There clearly remains some serious questions regarding the overall consistency of the correlators which actually produce these conformal blocks and the indecomposable representations that one deduces from them. It seems likely that other, more complicated, representations may have to be introduced into correlators to remedy this. For the arguments presented in this thesis this is a subtlety and does not affect the main results.

We can also find explicit solutions for the case of the correlators with $j_1=j_3=1$, $j_2=j_4=\half$. In this case we have only two conformal blocks. We obtain two solutions, one of which has logarithmic terms in its expansion as $1 \rightarrow 3$, the other being well behaved:
\bea \label{eqn:halfhalf11}
F_{\half \half 1 1}^{(1)}(x,z) &=& (1-z)^{-1} z^{1/4} ( z-2 +x ) \nonumber \\
F_{\half \half 1 1}^{(2)}(x,z) &=& (1-z)^{-3/2} z^{1/4} \Bigr[ \left( (z-2) \sqrt{z-1} \arctan{\sqrt{z-1}} +z -1 \right)   \\
 && \hspace{1.5in} \left( 1-z + z \arctan{\sqrt{z-1}} \right) \f{x}{z} \Bigl] \nonumber
\eea
\section{Factorisation of KZ equations}
\label{sec:factorisation}
The fact that the third order equation (\ref{eqn:Feqnfact3}) factorised is by itself not remarkable as any linear ordinary differential equation can always be written as a product of linear factors. What is significant however is that there is a subset of solutions, namely $F_{1111}^{(1)}(x,z)$, on which one can perform the conformal bootstrap; i.e. construct a well behaved correlator. The factorisation which reflects this is:
\bea
\CD_3 F(z) = \CD_2 \CD_1 F(z)
\eea
where:
\bea
\CD_2&=& z(1-z)^2 \f{d^2}{dz^2} +2(1-z)(1-2z) \f{d}{dz} -2 \\
\CD_1&=&(z-1)\f{d}{dz} +1 \nonumber
\eea
The solutions satisfying $\CD_1 F=0$ lead to:
\bea
F(z)=\f{1}{1-z} \Rightarrow F(x,z)=F_{1111}^{(1)}(x,z)
\eea
It is only such factorisations that have any invariant meaning in the theory.

In the work of \cite{Zamolodchikov:1986bd,Gepner:1986wi} selection rules were derived for the unitary $\widehat{SU(2)}$ theories as a consequence of pure current algebra null vectors. For instance for $k \in N$, following from the existence of the vacuum null vector (\ref{eqn:introvacnullvect}), we have a set of null vectors in the model:
\bea
\lb J^+_{-1} \rb ^{k-2j+1} \left. | j; m=j \ra,
\eea
Decoupling these leads to a theory with the spin restricted to the integrable sector: $0 \le j \le \f{k}{2}$. It was later shown by Christe and Flume \cite{Christe:1987yf} that such null vectors can also be deduced from the corresponding factorisation of the KZ equation for $j \ge \f{k}{4}$:
\bea
\CD_{2j+1} = \CD_{2j+1-s} \CD_{s}
\eea
where $\CD_n$ denotes a differential operator of order $n$ and $s=k-2j+1$. We may therefore expect that the factorisation of the \KZ equation we observe in $SU(2)_0$ is also related to the existence of null vectors. However as this is a non-unitary CFT the decoupling of these is much more subtle. At $j=1$ there \emph{is} indeed a null vector:
\bea \label{eqn:nullvector}
{\mathcal N}\state{(j=1)^+}  = \lb  J^+_{-1} L_{-1} - J^+_{-2} \rb \state{(j=1)^+}
\eea
where $(j=1)^+$ denotes the highest component of a $j=1$ operator. Clearly this is satisfied by the affine Kac-Moody current $\state{(j=1)^+} =J^+_{-1} \state{0}$ suggesting that this is indeed the correct null vector if we wish to keep only $F_{1111}^{(1)}(x,z)$.

To see this more explicitly we consider inserting the null vector (\ref{eqn:nullvector}) into a correlation function. In order to do so it is convenient to first express it in terms of auxiliary variables. Now $J^-_0=-\f{\p}{\p x}$ acts as a translation operator in the $x$ space so one can introduce:
\bea
J(x,z)&=&e^{x J^-_0} J^+(z) e^{-x J^-_0} \\
&=& J^+(z) - 2xJ^3(z) -x^2 J^-(z) \nonumber 
\eea
which is similar to the expansion of a $j=1$ primary (\ref{eqn:indexnotationexpanded}). One can then show using (\ref{eqn:JphiOPEs}) that primary fields have the following OPE with $J(x,z)$:
\bea
J(x,z) \phi(y,w) \sim - \f{\left( (x-y)^2 \f{\p}{\p y} +2j (x-y) \right) \phi(y,w)}{z-w}
\eea
The states must transform as:
\bea
\state{\phi(x,z)} = e^{x J^-_0} \state{\phi(z)^+}
\eea
Using this notation one may rewrite the null vector (\ref{eqn:nullvector}) using the auxiliary variables as:
\bea \label{eqn:nullvectxvars}
{\mathcal N} \phi(x,z) = J(x,z) \f{\p \phi(x,z)}{\p z} - \f{\p J(x,z)}{\p z} \phi(x,z) 
\eea
Now inserting this into a correlation function we get:
\bea
0=&& \left< \phi(x_1,z_1) \p_{z_2} J(x_2,z_2) \phi(x_2,z_2)  \cdots \phi(x_n,z_n) \right> \nonumber\\
&&- \left< \phi(x_1,z_1) J(x_2,z_2) \p_{z_2} \phi(x_2,z_2) \cdots \phi(x_n,z_n) \right> \nonumber\\
&&=\f{1}{2 \pi i} \oint_{z=z_2} \f{dz}{z-z_2} \left< \phi(x_1,z_1) \p_{z} J(x_2,z) \phi(x_2,z_2) \cdots \phi(x_n,z_n) \right> \nonumber\\
&& - \f{1}{2 \pi i} \oint_{z=z_2} \f{dz}{z-z_2} \left< \phi(x_1,z_1)  J(x_2,z) \p_{z_2} \phi(x_2,z_2) \cdots \phi(x_n,z_n) \right> \\
&&=- \sum_{i \ne 2} \f{1}{(z_i-z_2)^2} \left\{ (x_2-x_i)^2 \f{\p}{\p x_i} + 2j_i(x_2-x_i) \right\} \left< \phi(x_1,z_1)  \cdots \phi(x_n,z_n) \right> \nonumber \\
&&+ \sum_{i \ne 2} \f{1}{z_i-z_2} \left\{ (x_2-x_i)^2 \f{\p}{\p x_i} + 2j_i(x_2-x_i) \right\} \p_{z_2} \left< \phi(x_1,z_1)  \cdots \phi(x_n,z_n) \right> \nonumber
\eea
We must of course have $j_2=1$ as it is a spin $1$ null vector. For the case of the four point function (\ref{eqn:correl}) we get:
\bea \label{eqn:correlwithnullvect}
&&\lb \f{1}{z^2} + \f{1}{z} \f{\p}{\p z} \rb \left[ \lb x^2 (j_4-j_1-1-j_3) F(x,z)+(x-1) \f{\p F(x,z) }{\p x} \rb +2 j_1 x F(x,z) \right] \nonumber \\
&&+ \lb \f{1}{(z-1)^2} + \f{1}{z-1} \f{\p}{\p z} \rb \left[  \lb (1-x)^2 (-j_4+j_1+1+j_3) F(x,z)-x \f{\p F(x,z) }{\p x} \rb \right. \nonumber \\
&&\left. \phantom{\f{\p}{\p z}}\hspace{2in}
-2 j_3 (1-x) F(x,z) \right]=0 
\eea
As $j_2=1$ we must have a solution $F_{j_1,1,j_3,j_4}(x,z)=F(z)+xG(z)+x^2H(z)$. Inserting this into the above equation we find the unique solution:
\bea
F_{j_1,1,j_3,j_4}(x,z)=\f{j_1-j_3-j_4}{z-1}+2\f{j_4z-j_1}{z(z-1)}x +\f{j_1-j_4+j_3}{z(z-1)} x^2 
\eea
In particular for $j_i=1$ we recover the previous expression $F_{1111}^{(1)}(x,z)$.

One can also consider what the effect of the null vector (\ref{eqn:nullvector}) is on the other two solutions, which involved logarithms, obtained at $j=1$. By operating with ${\mathcal N}$, given by the LHS of (\ref{eqn:correlwithnullvect}), we find they become:
\bea \label{eqn:nullvectoraction}
{\mathcal N} F_{1111}^{(1)} &=& 0 \nonumber\\
{\mathcal N} F_{1111}^{(2)} &=& \f{6(xz-z-x+x^2)x}{z^3(z-1)} \\
{\mathcal N} F_{1111}^{(3)} &=& \f{6(xz-z+x^2-3x+2)x}{z(z-1)^3}\nonumber
\eea
The RHS of these equations exactly satisfy the KZ equation with \mbox{$j_1=1$},\mbox{$j_2=2$},\mbox{$j_3=1$},\mbox{$j_4=1$}. Therefore we see that the action of the null vector is to map the non-chiral fermionic operators at $j=1$ into a $j=2$ field. We shall see in the next chapter that there are indeed a doublet of \emph{chiral} fields at $j=2$. Although we have not discussed these operators we can already see how they arise by the action of the null vectors.
\subsection{Operator identification}
Several of the issues which we have discussed in this chapter are also pertinent in the well known $c=-2$ model. In the normal `minimal models' one identifies fields that have null vectors at different levels leading us to the standard identifications in the Kac-table \cite{Belavin:1984vu}. In the $c=-2$ model we do not have any such minimal theory and this identification of operators is \emph{not} correct. 

Our starting point is the Kac table for the $c_{2,1}=-2$ model by which we mean the table that is related to Virasoro null vectors and not some kind of formal extension involving half-integer entries \cite{Saleur:1992hk}:
\begin{center}
\begin{tabular}{|c||ccc|} \hline
7 & 3 & 1 & 0 \\
6 & $\f{15}{8}$ & $\f{3}{8}$ & $-\f{1}{8}$ \\
5 & 1 & 0 &  0  \\
4 & $\f{3}{8}$ & $-\f{1}{8}$ & $\f{3}{8}$ \\
3 & 0 & 0 & 1 \\
2 & $-\f{1}{8}$ & $\f{3}{8}$ &  $\f{15}{8}$\\
1 & 0 & 1 & 3 \\ \hline \hline
$h_{r,s}$ & 1 & 2 & 3  \\ \hline
\end{tabular} 
\end{center}
Consider for example the conformal blocks of correlators with four $h_{1,3}=0$ operators. They are easily found to be:
\bea \label{eqn:h13cminus2}
F_{0000}^{(1)}&=&1 \nonumber\\
F_{0000}^{(2)}&=&\ln (z) \\
F_{0000}^{(3)}&=&\ln (1-z)\nonumber
\eea
where the subscript labels now refer to the \italic{conformal weight} of the fields $\phi_i(z_i)$. In this theory there is, of course, another operator with the same conformal weight namely $h_{1,1}=0$. For the correlator involving four of these fields the conformal block is a constant $F_{0000}^{(1)}$. If we were to identify operators having null vectors at different levels then we must discard the solutions $F^{(2)},F^{(3)}$. However these solutions are an integral part of the triplet model as they correspond to conformal blocks of the fermionic operators\footnote{We trust no confusion will result between these and the $\Theta^{\pm}(x,z)$ operators that we discussed in $SU(2)_0$. } $\Theta^{\pm}(z,\bar{z})$ \cite{Kausch:2000fu}:
\bea
\la \Theta^+(0,0) \Theta^-(z,\bar{z}) \Theta^-(1,1) \Theta^+(\infty,\infty) \rb = -\ln | z |^2
\eea
These operators are actually GSO projected out in the local theory \cite{Gaberdiel:1998ps} but this is irrelevant from the point of view of the Kac-table which is only concerned with the holomorphic or anti-holomorphic sectors separately. Under the action of $L_{-1}$ the identity is annihilated whereas the other states are mapped into chiral fermionic states at $h=1$.

The correct identifications can only properly be made in the context of a particular rational model. For instance in the triplet model we do not identify the $h_{2,1}=1$ and $h_{1,5}=1$ operators. The former are the chiral states in the doublet representation $\nu_1$ and the latter are these together with the additional states $W^a_{-1} \left. | \omega \right>$. Using $W^a_{-1} \left. | \Omega \right>=0$ and $W^a_{0} \left. | \omega \right>=0$ one can check that $W^a_{-1} \left. | \omega \right>$ is indeed a Virasoro primary state. However the operators at $h_{1,2}$ and $h_{2,4}$ both of which have dimension $-\f{1}{8}$ must be identified as there is only one state in the triplet model with this dimension. In this same sense the Coulomb gas methods of Dotsenko and Fateev \cite{Dotsenko:1984nm} cannot hope to reproduce all results beyond the minimal sector as they imply a certain, in general \emph{different}, identification of operators.

In this thesis, in the absence of a rational model, we shall not discard any of the conformal blocks. Clearly one hopes that a consistent model could be constructed and we shall comment on this as we proceed. We have already seen that the role of null vectors is a subtle issue in non-unitary CFT and we shall be conservative and not decouple any of them - except of course the one corresponding to the KZ equation which is the very definition of our model.
\chapter{Extended chiral algebras in $SU(2)_0$ model}
As we have already emphasized chiral algebras play a central role in conformal field theory. In the case of the $c_{p,1}$ models the $W(2,2p-1^3)$ algebra allowed the construction of rational models; that is models having a finite number of irreducible and indecomposable representations.

We already know, since we have gone beyond the integrable sector, that $SU(2)_0$ is not rational with respect to the $\widehat{SU(2)}$ affine Kac-Moody algebra alone. We shall be able to show that there are \emph{extra} chiral algebras in the $SU(2)_0$ model. We leave the question of their role in producing a rational theory to future work.

We shall find convincing evidence for the existence of chiral algebras by studying the four point functions of certain operators. A chiral operator algebra is completely characterised by its singular behaviour. This fact implies the existence of chiral \emph{rational} solutions, that is analytic funtions with a finite number of poles, for the correlators. We shall also present a free field description of these fields which confirms their chiral nature.

Before we discuss the case of the $SU(2)_0$ model it is instructive to first return to the case of the $c=-2$ model. As this model is much better understood it provides a good example of the methods we shall use later.
\section{Structure of chiral algebras in $c=-2$}
The $c=-2$ triplet model is one of the best known examples of a rational LCFT. In this case the structure of the fields is determined from studying the representation theory of the $W(2,3^3)$ algebra. In more complicated orbifolds it is not clear if one can also form rational theories by using other chiral algebras.

As we have already explained knowledge of the null vectors allows one to compute correlation functions between fields in irreducible representations of the Virasoro algebra. In the $c_{2,1}=-2$ model we have seen that, by considering the $h_{1,2}=-\f{1}{8}$ fields, these irreducible representations do not close on themselves and one must introduce some indecomposable representations. 

Here we shall restrict ourselves to considering extensions of the Virasoro algebra by integer dimension fields with dimensions found \emph{within} the standard Kac-table; for example $h_{r,1}=1,3,6,10,\cdots$ in $c=-2$. This is natural if one wishes to construct theories that have some connection with the operators present in the Kac-table. We can use the Virasoro null vectors to find the conformal blocks for correlators involving these integer dimension fields. The correlators of these fields in principle encode all the information about their possible structure and fusion. We always assume the existence of a unique translation invariant vacuum state to avoid the obvious structure that would result from simply having several identical copies of every operator.

It is interesting to see how the possible structure of the integer dimension fields emerges from the correlation functions of the model. The conformal blocks for a correlator with four $h_{2,1}=1$ operators are easily found:
\bea \label{eqn:cminus2doublet}
F_1(z)&=&1- \f{1}{z^2}\\
F_2(z)&=&1- \f{1}{(1-z)^2} \nonumber
\eea
Immediately we see the striking fact that these two solutions are both rational functions. This is in stark contrast to the generic situation in which one would expect to obtain a hypergeometric function from solving a second order differential equation. The appearance of a rational solution indicates that one may consistently consider such operators as chiral fields\footnote{Actually it is only a necessary condition. To be sufficient they must be consistently chiral in \emph{all} correlation functions of the theory. However the appearance of a rational solution is so restrictive that we do not know of examples where it is insufficient.}. It does not force us to do so in a particular model but merely gives us the possibility. The fact that a chiral OPE is completely determined by its singular terms is in coincidence with the fact that rational functions are fully determined by their pole structure at the singular points. In this way from the rational conformal blocks of a theory one may deduce the existence of chiral algebras.

It is very simple to deduce the well known extended chiral algebras in certain minimal models. For example in the $c_{5,6}=\f{4}{5}$ model the correlation function of four $h_{1,4}=3$ fields is given by:
\bea \label{eqn:ZamW23}
\la W(0) W(z) W(1) W(\infty) \ra &=& \f{c^2}{9} \left[ \f{1}{z^6} + \f{1}{(1-z)^6} +1 \right] \nonumber \\
&&+2c\left[ \f{1}{z^4} + \f{1}{(1-z)^4} + \f{1}{z^3} + \f{1}{(1-z)^3} - \f{1}{z} - \f{1}{1-z} \right] \\
&& + c \lb \f{9}{5} + \f{32}{5} \f{16}{(22+5c)} \rb \left[ \f{1}{z^2} + \f{1}{(1-z)^2} + \f{2}{z} + \f{2}{(1-z)} \right] \nonumber 
\eea
In this example, the famous Zamolodchikov $W(2,3)$ algebra \cite{Zamolodchikov:1985wn}, we actually have associativity for arbitrary values of central charge as can be seen from crossing symmetry of (\ref{eqn:ZamW23}). However it is only at the very special case $c=\f{4}{5}$ that it has the interpretation of a correlator of a degenerate $h_{1,4}$ field in the Kac-table.

In the $c=-2$ example there is not one but two rational correlation functions. They are related under crossing symmetry in the following way:
\bea
F_1(1-z)&=&F_2(z) \\
z^{-2} F_1 \left(\f{1}{z} \right) &=& -F_1(z) \nonumber
\eea
We immediately see that we cannot produce a non-vanishing chiral correlator that is invariant under all crossing symmetries. In other words it is not consistent to regard these as identical $h=1$ chiral operators in a theory. One therefore has two choices: either there are no chiral operators at $h=1$; or such operators must have some extra quantum numbers allowing one to relax the constraints of crossing symmetry.

As $F_1(z)$ acquires a negative sign under $z \leftrightarrow \f{1}{z}$ ( i.e exchange of $2 \leftrightarrow 3$, or $1 \leftrightarrow 4$) this could come from the exchange of identical fermionic operators. Clearly there must be more than one species of fermion as we cannot construct a chiral correlator invariant under fermionic symmetry in all exchanges. If we were to have more than two species of fermions then there would be several conformal blocks behaving in the same manner under $1 \leftrightarrow 4$ in order to accomodate this. Therefore we deduce that there are only two types of fermionic operators. These are precisely realised by the symplectic fermions:
\bea \label{eqn:symplecticfermion}
\Psi^{\alpha}(z) \Psi^{\beta}(w) \sim \f{d^{\alpha \beta}}{(z-w)^2} \quad \quad \alpha,\beta=\pm
\eea
with $d^{\alpha \beta}$ antisymmetric. One can similarly examine the conformal blocks of the $h_{3,1}=3$. One finds that there are exactly three rational solutions, given later in (\ref{eqn:cminus2triplet}), and by similar arguments to those just presented one can conclude that these can be a triplet of bosonic fields. These are again exactly realised in the symplectic fermion model. In more complicated orbifold models not all of these chiral fields would be present and one would take the chiral fields to be some sub-sector of the ones described here. For example the $Z_4$ twisted $c=-2$ theory has an invariant $W(2,3,10^2)$ algebra \cite{Eholzer:1998se,Kausch:2000fu}. The conformal blocks give us the maximal consistent extensions of the chiral algebra by degenerate fields in the Kac-table and we are not required to have all of these in any specific theory.
\section{Rational Solutions to the $SU(2)_0$ KZ equation}
For the fields in $SU(2)_0$ with $j \in N$ we found, for all cases studied, that we have either rational or logarithmic solutions. As we are discussing chiral algebras we shall concentrate on the rational solutions. 

For a four point correlator involving a discrete representation of spin $j$ we have already seen how we can write the general solution to the KZ equation (\ref{eqn:discreterepns}):
\be \label{eqn:xexp}
F(x,z)=F_0(z)+xF_1(z)+x^2F_2(z)+ \cdots +x^{2j}F_{2j}(z)
\ee
This allows one to reduce the KZ equation to a linear ordinary differential equation of order $2j+1$. In \cite{Kogan:2001nj} we found an ansatz for the four point correlator in $SU(2)_0$ with $j_i=j$. The lowest component was:
\bea
F_0(z)=z^{-j(j-1)}(1-z)^{-j(j+1)}{}_2F_1(j,-j;1;z)
\eea
Given this all other components are fully determined by substitution into the KZ equation. For $j \in N$ these become rational solutions and are thus everywhere single-valued (i.e. they have trivial monodromy properties). Now using the crossing symmetries one can easily find a full set of rational solutions on which one can perform the conformal bootstrap.
\subsection{Spin $j=1$}
This is the case that was discussed in the previous chapter and will be given again here for completeness. For $j_1=j_2=j_3=j_4=1$ we found one rational solution:
\bea \label{eqnjone}
F_{1111}(x,z)=-\f{1}{2(z-1)}+\f{x}{z}+\f{x^2}{2z(z-1)}
\eea
This obeys:
\bea
F_{1111}(1-x,1-z)&=&F_{1111}(x,z) \\
z^{-2h}x^{2j}F_{1111}\left(\f{1}{x},\f{1}{z}\right)&=&F_{1111}(x,z) \nonumber
\eea
Under the crossing symmetry operation we must exchange both $z$ \emph{and} $x$ simultaneously as all fields have dependence on both variables. In terms of the original component notation this ensures we  exchange not only the positions but also all the $SU(2)$ labels as well.

As there is just one solution and it is invariant under all exchanges of fields it must correspond to a single bosonic set of fields transforming in the $j=1$ representation. As we discussed before this is essentially the four point function of the Kac-Moody current $J(x,z)$ but with one of the fields replaced by the logarithmic partner $N(x,z)$ to make the correlator non-vanishing. In this and the following correlators we shall, for convenience, be slightly sloppy and write this as:
\bea
\la J(0,0) J(x,z) J(1,1) J(\infty,\infty) \ra = F_{1111}(x,z)
\eea
\subsection{Spin $j=2$}
For the four point correlator with $j_1=j_2=j_3=j_4=2$ we find two rational solutions:
\bea \label{eqn:jtwo}
F_{2222}^{(1)}(x,z)&=&\f{1}{z^5}\left( (-z^3-2z^4)+(-12z^2-16z^3+4z^4)x+(-18z+18z^3)x^2 \right. \nonumber  \\
&&\left.+(-4+16z+12z^2)x^3+(2+z)x^4 \right) \\
F_{2222}^{(2)}(x,z)&=&\f{1}{(z-1)^5} \left( (10z-15z^2+9z^3-2z^4)+(60-140z+108z^2-36z^3+4z^4)x \right. \nonumber\\
&&\left.+(-90+162z-90z^2+18z^3)x^2+(36-44z+12z^2)x^3+(-3+z)x^4 \right) \nonumber
\eea
These obey:
\bea \label{eqn:jtwoconfblocks}
F_{2222}^{(1)}(1-x,1-z)&=&F_{2222}^{(2)}(x,z)\nonumber \\
F_{2222}^{(2)}(1-x,1-z)&=&F_{2222}^{(1)}(x,z)\\
x^4z^{-6}F_{2222}^{(1)}\left(\f{1}{x},\f{1}{z} \right)&=&-F_{2222}^{(1)}(x,z) \nonumber\\
x^4z^{-6}F_{2222}^{(2)}\left(\f{1}{x},\f{1}{z} \right)&=&-F_{2222}^{(1)}(x,z)+F_{2222}^{(2)}(x,z) \nonumber
\eea
We see that from the leading forms of these solutions that they do not go as $z^{-6}$ as one would expect from an $h=3$ field. Therefore there is no contribution from the identity field in the OPE and the two point functions of these fields must vanish. It is clear that a similar approach to the $j=1$ fields, with logarithmic insertions, must be carried out here but we shall not discuss this in detail. 

As we commented earlier in section \ref{sec:factorisation} the $j=2$ operators are actually produced by acting on the fermionic $j=1$ operators with a null vector. We therefore suspect that they should be a fermionic doublet of fields and indeed the above behaviour of the conformal blocks (\ref{eqn:jtwoconfblocks}) under crossing symmetry confirms this.

It is easily seen that all this is correctly reproduced by adding extra labels in the following way:
\bea
\la \Psi^+ (0,0) \Psi^-(x,z) \Psi^-(1,1) \Psi^+(\infty,\infty) \ra = F_{2222}^{(1)}(x,z) \\
\la \Psi^+ (0,0) \Psi^+(x,z) \Psi^-(1,1) \Psi^-(\infty,\infty) \ra = F_{2222}^{(2)}(x,z) \nonumber
\eea
\subsection{Spin $j=3$}
For $j=3$ we have three rational solutions given in (\ref{eqn:jthree}). Analysing their behaviour under crossing symmetry, in exactly the same manner as before, we deduce that they are in fact a bosonic triplet of operators which we shall denote as $W^{\pm,3}(x,z)$. We can write the correlators as:
\bea
\la W^+ (0,0) W^+(x,z) W^-(1,1) W^-(\infty,\infty) \ra &=& F_{3333}^{(1)}(x,z) \nonumber\\
\la W^3 (0,0) W^3(x,z) W^3(1,1) W^3(\infty,\infty) \ra &=& F_{3333}^{(2)}(x,z)\\
\la W^+ (0,0) W^-(x,z) W^-(1,1) W^+(\infty,\infty) \ra &=& F_{3333}^{(3)}(x,z)\nonumber
\eea
The first of these is the unique solution that has no singular terms as $z \rightarrow 0$. The third is similar but with no singular terms as $z \rightarrow 1$. The second is the unique solution invariant under all crossing symmetries.
\subsection{Mixed correlators}
We can of course consider correlators where not all four fields have the same spin. For instance the correlator with $j_1=1,~j_2=2,~j_3=1,~j_4=1$ yields two rational solutions given already, up to numerical factor, in (\ref{eqn:nullvectoraction}):
\bea
F_{1211}^{(1)}(x,z)&=& \f{(xz-z-x+x^2)x}{z^3(z-1)} \\
F_{1211}^{(2)}(x,z)&=& \f{(xz-z+x^2-3x+2)x}{z(z-1)^3} \nonumber
\eea
We already know that there is only one chiral $j=1$ operator so we should have full crossing symmetry under exchange of $1,3$ and $4$. However no combination of the above conformal blocks respects this and therefore, if all operators are chiral, the correlator must vanish. This was to be expected from the fermionic nature of the chiral $j=2$ operator:
\bea
\la J (0,0) \Psi^{\pm} (x,z) J(1,1) J (\infty,\infty) \ra = 0
\eea
\section{Free field representation}
As we have already explained the appearance of rational four point functions is a neccessary, and highly restrictive, condition for the existence of a chiral algebra. However, due to the fact that all the operators involved have integer dimensions, it is in general extremely hard to study if the operator algebra is closed or not solely from the correlation functions.

In general the most convincing method to verify that fields are really chiral and that they obey a closed $W$-algebra is to construct them in the free field approach and then directly compute the OPEs.

The standard Sugawara stress tensor for $SU(2)_k$ is (\ref{eqn:sugawara}):
\bea \label{eqn:sugawara1}
T&=&\f{1}{2(k+2)}\left( J^+J^- + J^-J^+ + 2 J^3J^3 \right) \nonumber \\
&=& -\beta \p \gamma - \half \p \phi \p \phi - \f{i}{\sqrt{2(k+2)}} \p^2 \phi
\eea
It is composed of two commuting parts: the $\beta \gamma$ system with $c=2$ and the $\phi$ system with $c=\f{3k}{k+2}-2$. 

The affine Lie algebra primary fields $\Phi_{j,m}$ (with $m=-j,\cdots,j$) obey (\ref{eqn:KMprim}):
\bea \label{eqn:KMprimary}
J^a(z) \Phi_{j,m}(w) \sim \f{-t^a \Phi_{j,m}(w)}{z-w}
\eea
where $t^a$ is a matrix in the spin $j$ representation of $SU(2)$. They have conformal dimension:
\bea
h=\f{j(j+1)}{k+2}
\eea
In the standard free field construction of $\widehat{SU(2)}$ the vertex operators of primary affine Lie algebra fields are taken to be:
\bea
\Phi_{j,m}=(-\gamma)^{j-m}e^{ij \sqrt{\f{2}{k+2}} \phi}
\eea
In order to calculate correlators with such operators, vacuum charges and screening operators must be inserted \cite{Dotsenko:1990ui,Dotsenko:1991zb}. This is  quite a complicated procedure and moreover the result is, in general, only an integral representation. For $SU(2)_0$ however the $j \in N$ operators have integer conformal weight and we have a subset of rational solutions for these correlators. On may therefore suspect that there is a simpler way of evaluating correlators for such fields. 

We now observe that for $k=0$ the $\phi$ part of the stress tensor (\ref{eqn:sugawara1}) is none other than a $c=-2$ system. We therefore consider using a fermionic ghost representation for this system. This is not essential but makes it easier to see the extended multiplet nature in the free field representations. We therefore write:
\bea
\xi \sim e^{i \phi} ~~~~~~~ \eta \sim e^{-i \phi} 
\eea
where $\xi,\eta$ are a fermionic ghosts with $h=(1,0)$ respectively:
\bea \label{eqn:cminus2}
\xi(z)\eta(w) \sim \f{1}{z-w} \sim \eta(z)\xi(w) ~~~~~~~ i \p \phi =  \xi \eta 
\eea
These fields are related to the symplectic fermions in (\ref{eqn:symplecticfermion}) via: $\Psi^+=\xi, \Psi^-=\p \eta$. The currents and stress tensor now become:
\bea \label{eqn:SU20currents}
J^+&=&\beta ~~~~~~
J^3=\xi \eta + \beta \gamma  ~~~~~~
J^-=-2 \xi \eta \gamma - \beta\gamma^2 
\eea
\bea \label{eqn:stressTsu2level0}
T= -\beta \p \gamma - \xi \p \eta 
\eea
\section{Extended algebras in $SU(2)_0$}
\label{sec:extendedalg}
As we are interested in finding free field descriptions of the discrete representations it is obvious that the highest component, annihilated by $J^+_0$, cannot have any $\gamma$ dependence. For each value of $h \le 6$ we formed the most general linear combination of the three fields $\beta,\xi,\eta$ and their derivatives having conformal weight $h$. Using the expressions for the affine currents (\ref{eqn:SU20currents}) we then imposed the very restrictive condition that these fields are actually affine Kac-Moody primary fields (\ref{eqn:KMprimary}). As they are in discrete representations they must also be eventually annihilated by continued application of $J^-_0$.

We found the following expressions for the fields\footnote{We are extremely grateful to K. Thielemans for giving us a copy of his OPEDEFS Mathematica program \cite{Thielemans:1991uw} which was invaluable in finding many of the free field expressions.}. Throughout we use the convention of right normal ordering: $ABC=(A(BC))$. For brevity we shall only write the highest component, that is the one annihilated by $J^+_0$ . All others in the multiplet may be obtained by the action of $J^-_0$:
\begin{itemize}
\item{$j=1$}
\bea \label{eqn:jequalone}
J^+=\beta
\eea
\item{$j=2$}
\bea \label{eqn:jequaltwo}
 \lb\Psi^+ \rb ^{++}&=&-\beta \p \xi + \p \beta \xi \\
\lb \Psi^- \rb ^{++} &=&\beta \beta \beta \eta \nonumber
\eea
\item{$j=3$}
\bea
\lb W^+ \rb ^{+++}&=&-\beta \p^2 \xi \p \xi + \p \beta \p^2 \xi \xi - \p^2 \beta \p \xi \xi  \nonumber  \\ 
\lb W^3 \rb ^{+++}&=&\f{1}{2} \beta \beta \beta \eta \p^2 \xi - \f{1}{2} \beta \beta \beta \p \eta \p \xi - \f{3}{2} \p \beta \beta \beta \eta \p \xi  +\f{1}{2} \p \beta \beta \beta \p \eta \xi + \f{3}{2} \p \beta \p \beta \beta \eta \xi \nonumber \\
&& \quad + \p \beta \p \beta \p \beta - \f{1}{2} \p^2 \beta \beta \beta \eta \xi -\f{3}{4} \p^2 \beta \p \beta \beta + \f{1}{12} \p^3 \beta \beta \beta \\
\lb W^- \rb ^{+++}&=&\beta \beta \beta \beta \beta \p \eta \eta \nonumber
\eea
\end{itemize}
We have also included the $j=1$ field (\ref{eqn:jequalone}) which is of course just the affine current $J^+(z)$ itself. On action by $J^-_0$ this clearly produces the other Kac-Moody generators $J^3(z)$ and $J^-(z)$. It is a non-trivial fact that the $j=2$ and $j=3$ operators given here actually also lead to multiplets belonging to the \emph{discrete} representations of $SU(2)$.

In the case of the $j=2$ fields $\Psi^{\pm}$ the full multiplets are given by:
\bea \label{eqn:Psiplusmultiplet}
\lb \Psi^+ \rb^{++}&=&- \beta \p \xi+ \p \beta \xi \nonumber\\
\lb \Psi^+ \rb^{+}&=&4 \beta \gamma \p \xi+4 \eta \p \xi \xi-4 \p \beta \gamma \xi \nonumber\\
\lb \Psi^+ \rb^0&=&-12  \beta \gamma \gamma \p \xi-24  \gamma \eta \p \xi \xi +12 \p \beta \gamma \gamma \xi \\
\lb \Psi^+ \rb^{-}&=& 24  \beta \gamma \gamma \gamma \p \xi+72 \gamma \gamma \eta \p \xi \xi-24 \p \beta \gamma \gamma \gamma \xi \nonumber\\
\lb \Psi^+ \rb^{--}&=&-24  \beta \gamma \gamma \gamma \gamma \p \xi -96  \gamma \gamma \gamma \eta \p \xi \xi + 24 \p \beta \gamma \gamma \gamma \gamma \xi \nonumber \\
\nonumber \\
%
\lb \Psi^- \rb^{++}&=&\beta \beta \beta \eta \nonumber \\
\lb \Psi^- \rb^{+}&=& -4  \beta \beta \beta \gamma \eta+6 \beta \beta \p \eta-6 \p \beta \beta \eta \nonumber\\
\lb \Psi^- \rb^{0}&=&12 \beta \beta \beta \gamma \gamma \eta-36 \beta \beta \gamma \p \eta+36 \beta \p \eta \eta \xi+6 \beta \p^2 \eta+36 \p \beta \beta \gamma \eta-24 \p \beta \p \eta+6 \p^2 \beta \eta \nonumber\\
\lb \Psi^- \rb^{-}&=&-24  \beta \beta \beta \gamma \gamma \gamma \eta+108 \beta \beta \gamma \gamma \p \eta-216 \beta \gamma \p \eta \eta \xi-36 \beta \gamma \p^2 \eta-108 \p \beta \beta \gamma \gamma \eta \nonumber\\
&& \quad +144 \p \beta \gamma \p \eta-144 \p \eta \eta \p \xi-36 \p^2\beta \gamma \eta+36 \p^2 \eta \eta \xi \\
\lb \Psi^- \rb^{--}&=&24  \beta \beta \beta \gamma \gamma \gamma \gamma \eta-144  \beta \beta \gamma \gamma \gamma \p \eta+432 \beta \gamma \gamma \p \eta \eta \xi+72 \beta \gamma \gamma \p^2 \eta+576 \gamma \p \eta \eta \p \xi \nonumber\\
&& \quad -144 \gamma \p^2 \eta \eta \xi+144 \p \beta \beta \gamma \gamma \gamma \eta-288 \p \beta \gamma \gamma \p \eta+72 \p^2 \beta \gamma \gamma \eta \nonumber
\eea
It can be verified by explicit calculation that the lowest component of both these multiplets is indeed annihilated by $J^-_0$.

We also found other solutions $\xi,\p \xi \xi, \p^2 \xi \p \xi \xi$ for $j=1,2,3$ respectively and we shall comment later on these operators.

It is indeed striking that the operator content obtained through the free field representations, with the simple use of the symplectic fermion system (\ref{eqn:cminus2}), exactly produces the correct chiral field content of the $SU(2)_0$ model. In particular we see that the bosonic and fermionic nature of the operators is in exact agreement with that which we expected from the previous analysis based on the crossing symmetry of correlators.
\subsection{Operator Product Expansions}
\label{sec:OperatorProductExpansions}
We have seen that the chiral $j=1,2,3$ operators are alternately bosonic, fermionic, bosonic and that there are respectively a singlet, doublet and a triplet of possible fields. Using these symmetries and the fact that a chiral algebra is completely determined by its singular terms we can already deduce almost all the structure of the OPEs. We know:
\bea
[j=0] \otimes [j=J]=[j=J]
\eea
where $[~]$ denotes operators and all their affine Lie algebra descendents. In particular the $j=1$ field, $J^a$, is of course just an affine Kac-Moody descendent of the identity $j=0$. A special case of this OPE is the definition of an affine Kac-Moody primary field (\ref{eqn:KMprimary}).

We shall use indices $\alpha,\beta$ for extra doublet index structure and $a,b,c$ for the triplet. We now want to analyse the OPE of $[j=2]^{\alpha}$ with itself. Clearly as these are both fermionic the result must be bosonic. Furthermore the $j=3$ operators have $h=6$ and so cannot occur in the singular terms. Thus we must have:
\bea \label{eqn:jtwoalgebra}
[j=2]^{\alpha} \otimes [j=2]^{\beta}=d^{\alpha \beta} [j=0]
\eea
where $d^{\alpha \beta}$ is the anti-symmetric tensor $d^{+-}=-d^{-+}=1$. 

By similar arguments we find:
\bea
[j=3]^{a} \otimes [j=2]^{\alpha}=t^{a \alpha}_{\beta} [j=2]^{\beta}
\eea
where $t^{a \alpha}_{\beta}$ are $2 \times 2$ matrices which generate an $SU(2)$ Lie algebra. In this case we can only produce fermionic operators in the singular terms with dimension at most  $3+6-1=8$. In particular this excludes a possible fermionic $j=4$ operator with $h=10$. 

Similarly:
\bea 
[j=3]^{a} \otimes [j=3]^{b}=g^{ab} [j=0] + f^{ab}_c [j=3]^c
\eea
where now $g^{ab}$ and $f^{ab}_c$ are respectively the metric and structure constants of $SU(2)$. We thus expect $j=1,2,3$ to yield a closed $W$-algebra. We also expect that in the absence of other symmetries, such as orbifolding, we cannot get a finite $W$-algebra involving $j > 3$. This is a similar situation to $c=-2$ where more complicated $W$-algebras play a role in the orbifolded models.

Using the free field representation of the the affine Lie algebra generators and those of the $j=2$ representations (\ref{eqn:Psiplusmultiplet}) we can explicitly find the full algebra. One can verify, again non-trivially, that:
\bea
\lb \Psi^+ \rb^{A}(z) \lb \Psi^+ \rb^{B}(w) &\sim& 0 \quad \quad \quad A,B=++,+,0,-,-- \\
\lb \Psi^- \rb^{A}(z) \lb \Psi^- \rb^{B}(w) &\sim& 0 \nonumber
\eea
so the only non-trivial OPEs are between $\lb \Psi^+ \rb$ and $\lb \Psi^- \rb$ multiplets. Here we shall just give one example to show the typical structure:
\bea
\lb \Psi^+ \rb^{++}(z) \lb \Psi^- \rb^{+}(w) &\sim&\f{12  J^+ J^+ J^+}{(z-w)^3}+ \f{-4 J^+ J^+ J^+ J^3+12 \p J^+ J^+ J^+ }{(z-w)^2} \\
&&+ \f{1}{z-w} \left( -4 J^+ J^+ J^+ J^3 J^3-4 J^+ J^+ J^+ J^+ J^- \right. \nonumber\\
&&  \left. \quad  -4 J^+ J^+ J^+ \p J^3  -12 \p J^+ \p J^+ J^++12 \p^2 J^+ J^+ J^+ \right)\nonumber
\eea
The full (very tedious!) results were given in \cite{Nichols:2001cv}. In particular one finds that it is indeed of the form stated before from general arguments (\ref{eqn:jtwoalgebra}). Unfortunately using the free field representations it was prohibitively difficult to go beyond the $j=2$ algebra. In particular we could not examine the associativity of the triplet algebra, based on the $j=3$ fields. However we shall see later that these are closely related to the triplet fields in $c=-2$ and therefore one might conjecture that, in analogue to the $c=-2$ case, these are enough to rationalise the $SU(2)_0$ model.
\subsection{Fermionic $j=1$ fields}
We already know from the study of the correlation functions that there are extra primary operators at $j=1$ that have a fermionic behaviour under crossing symmetry. We also found expressions for these in the free field representation and they are in exact agreement with the solutions found from the \KZ equation (\ref{eqn:spinone}). Here we shall restrict our attention to the $j=1$ fields but we also found similar operators in the $j=2,3$ fields. As we commented earlier we found the extra highest component of a $j=1$ multiplet $\xi$ and we shall now show that this leads to the $\lb \Theta^+ \rb^a$ multiplet:
\bea \label{eqn:FFthetaplus}
\lb \Theta^+ \rb^+ &=& -\xi \nonumber\\
\lb \Theta^+ \rb^3 &=& 2 \xi \gamma \\
\lb \Theta^+ \rb^- &=& -2 \xi \gamma^2\nonumber
\eea
The last field is annihilated by $J^-_0$ and so this is indeed a discrete representation. To verify that this is in fact the $\lb \Theta^+ \rb^a$ field let us apply the null vector (\ref{eqn:nullvector}) in the free field expression:
\bea
{\mathcal N} \state{\lb \Theta^+ \rb^+} = \lb J^+ L_{-1} - J^+_{-2} \rb \state{\lb \Theta^+ \rb^+} &=& \lb \beta \p  - \p \beta \rb \lb \Theta^+ \rb^+ \nonumber\\
&=& -\lb \beta \p  - \p \beta \rb \xi  \\
&=& -\beta \p \xi + \p \beta \xi =   \lb \Psi^+ \rb^{++}\nonumber
\eea
This is exactly what we found previously in section \ref{sec:factorisation} from the action of this null vector in the correlation functions of $\Theta^+(x,z)$. The other field $\lb \Psi^- \rb^{++}$, given in (\ref{eqn:jequaltwo}), should be obtained in a similar way:
\bea 
\lb \beta \p  - \p \beta \rb \lb \Theta^- \rb^+ = \lb \Psi^- \rb^{++}= \beta \beta \beta \eta
\eea
A formal solution to this is:
\bea \label{eqn:FFthetaminus}
\lb \Theta^- \rb^+ = \beta \p^{-1} \lb \beta \eta \rb = J^+_{-1} \state{\p^{-1} Q}
\eea
In the standard free field representation the integral over the field $Q=\beta \eta$ occurs as a screening charge. The formal operator $\p^{-1}$, which is similar to integration, produces extra zero modes which one must treat with care. The appearance of such an operator is in agreement with the recent general description of LCFT given in \cite{Fjelstad:2002ei}. 

We do not attempt here to give a full description of this operator but we can verify formally that it is indeed a affine Kac-Moody primary field in a $j=1$ representation. Firstly note:
\bea
J^+(z) Q(w) &\sim& 0 \nonumber \\
J^3(z) Q(w) &\sim& 0 \\
J^-(z) Q(w) &\sim& \p_w \lb \f{2 \eta(w)}{z-w} \rb \nonumber
\eea
Now acting with $\p^{-1}_w$ on these we deduce:
\bea
J^a_n \state{\p^{-1} Q} &=& 0 \quad n \ge 1  \nonumber\\
J^+_0 \state{\p^{-1} Q} &=& 0 \\
J^3_0 \state{\p^{-1} Q} &=& 0  \nonumber\\
J^-_0 \state{\p^{-1} Q} &=& 2 \state{\eta} \nonumber
\eea
Given the form of $\lb \Theta^- \rb^+$ in (\ref{eqn:FFthetaminus}) it is a simple task to verify $J^a_n \state{\lb \Theta^- \rb^+}=0$ for $n \ge 1$. Also:
\bea
\lb J^-_0 \rb^3 \state{\lb \Theta^- \rb^+}&=& \lb J^-_0 \rb^3 J^+_{-1} \state{\p^{-1} Q} \nonumber\\
&=&\left\{ \left[ \lb J^-_0 \rb^3,J^+_{-1} \right]  + J^+_{-1} \lb J^-_0 \rb^3  \right\} \state{\p^{-1} Q} \\
&=& \left\{ -6 J^3_{-1} \lb J^-_0 \rb^2  - 6 J^-_{-1} J^-_0 +  J^+_{-1} \lb J^-_0 \rb^3 \right\} \state{\p^{-1} Q} \nonumber\\
&=& 2 \left\{ -6 J^3_{-1} J^-_0 - 6 J^-_{-1} +  J^+_{-1} \lb J^-_0 \rb^2 \right\} \state{\eta}\nonumber
\eea
Now using the fact that $J^-_0 \state{\eta}= 2 \state{\eta \gamma}$ and $\lb J^-_0 \rb^2 \state{\eta}= 6 \state{\eta \gamma^2}$ one can easily verify that the above expression vanishes in the free field representation. The field $\eta$ belongs to an infinite dimensional $j=-1$ representation and the above expression is precisely a null vector for this field. Therefore the discrete affine Kac-Moody primary operators  $\lb \Theta^{\pm} \rb^a$ can also be written, somewhat formally, in the free field representation. It remains an interesting problem to correctly introduce the zero modes and compute the OPEs. This should lead one to free field expressions for the indecomposable fields (\ref{eqn:OPEthetatheta}) produced in the fusion of $\lb \Theta^{\pm} \rb^a$.

\chapter{Hamiltonian reduction}
So far we have discussed an example of $\widehat{SU(2)}$ beyond the integrable representations and observed the appearance of indecomposable representations and extended chiral structure. We shall now show that in another large class of models, namely the $c_{p,q}$ models beyond the `minimal sector', very similar structures exist. Rather than discussing this separately we shall make use of a powerful relation between $\widehat{SU(2)}$ and $c_{p,q}$ models known as hamiltonian reduction. In particular the hamiltonian reduction of the $SU(2)_0$ theory gives a $c=-2$ model. The discrete representations of $SU(2)$ are related to the $h_{1,s}$ fields in the $c_{p,q}$ case. We shall first discuss the free field approach to the reduction and then an approach for studying the correlation functions.
\section{Free field approach}
\label{sec:Freefieldapproach}
Quantum Hamiltonian reduction of $\widehat{SU(2)}$ theories is achieved by imposing the constraint  $J^+ \sim 1$. Such a reduction not only removes the original group structure from the theory but also affects all the correlation functions of the theory. It has been well studied in the literature and it is known that such a reduction leads to a theory with central charge given by $c_{k+2,1}$ \cite{Drinfeld:1984qv,Polyakov:1988qz,Alekseev:1989ce,Bershadsky:1989mf,Feigin:1990pn}. 

We shall firstly discuss the reduction in the free field approach. We start with the Sugawara stress tensor for $SU(2)_k$:
\bea
T_{Sugawara}=-\beta \p \gamma -\half \p \phi \p \phi - i \sqrt{2} \f{1}{2 \sqrt{k+2}} \p^2 \phi 
\eea
with central charge $c=2+1-24 \lb \f{1}{2 \sqrt{k+2}} \rb^2=\f{3k}{k+2}$.
If we wish to impose $J^+=1$ in the quantum theory then, in order for this to be a sensible constraint, we must first modify the stress tensor so that $J^+$ becomes a dimension zero field. We define:
\bea \label{eqn:improvedT}
T_{modified}=T_{Sugawara}+\p J^3
\eea
then we have:
\bea
T_{modified}(z) J^+(w) \sim \f{\p J^+(w)}{z-w}
\eea
This modified stress tensor also satisfies the Virasoro algebra but with a different central charge:
\bea
c_{modified}=c_{Sugawara}-6k=\f{3k}{k+2}-6k
\eea
To impose the constraint $J^+=1$ in the quantum theory we first introduce a fermionic $(b,c)$ ghost system with $c=-2$ and conformal weights $(1,0)$. We can then form the operator:
\bea
Q=\oint b(z) (J^+(z) -1) dz
\eea
This can be shown to satisfy $Q^2=0$ and so we identify the physical states of the reduced theory with the cohomology:
\bea \label{eqn:physicalstate}
Q \state{\rm phys}=0
\quad \quad \quad \state{\rm phys} \ne Q \state{\chi}  \quad {\rm ~for ~any~ state~} \chi
\eea
The full stress tensor is the modified one (\ref{eqn:improvedT}) with the kinetic term for the $(b,c)$ system:
\bea \label{eqn:improvedstressT}
T&=&T_{improved} - b\p c \nonumber \\
&=& \gamma \p \beta -b \p c - \half \p \phi \p \phi - i \sqrt{2} \left[ \f{1}{2 \sqrt{k+2}} - \f{\sqrt{k+2}}{2} \right] \p^2 \phi \nonumber \\
&=& \left[ Q, \p c \gamma \right] - \half \p \phi \p \phi - i \sqrt{2} \left[ \f{1}{2 \sqrt{k+2}} - \f{\sqrt{k+2}}{2} \right] \p^2 \phi \\
&=& - \half \p \phi \p \phi - i \sqrt{2} \left[ \f{1}{2 \sqrt{k+2}} - \f{\sqrt{k+2}}{2} \right] \p^2 \phi \nonumber 
\eea
where in going from the second to the third line we make use of the fact that: \mbox{$[Q,\p c \gamma]=\gamma \p \beta -b \p c$} and so such terms vanish in matrix elements between physical states.

The central charge of the reduced theory is precisely that of the $c_{p,q}$ model with $k+2=\f{p}{q}$ (we will always take $\gcd(p,q)=1$):
\bea \label{eqn:CpqKactable}
c_{p,q}&=&1 - \f{6 (p-q)^2}{pq} \nonumber\\
&=& 13-6 \left( k+2+\f{1}{k+2} \right)
\eea
\bea \label{eqn:CpqKactableweights}
h_{r,s}=\f{(pr-qs)^2-(p-q)^2}{4pq}
\eea

If we perform hamiltonian reduction of the discrete representations of $SU(2)$ with $2j \in Z$ we get the conformal weights of the $h_{1,2j+1}$ fields in the $c_{p,q}$ model: 
\bea \label{eqn:reduced}
h_{1,2j+1}&=&\f{j(j+1)}{k+2}-j
\eea
Clearly $c_{p,q}$ is unchanged under exchange of $p$ and $q$ but our convention of $k+2=\f{p}{q}$ fixes the reduced field to be $h_{1,2j+1}$ rather than $h_{2j+1,1}$. 
For the case of $SU(2)_0$ we find it reduces to a model with $c_{2,1}=-2$ model. We note that this $c=-2$ theory is that of the reduced stress tensor (\ref{eqn:improvedstressT}) and is \emph{not} related to the other $c=-2$ $(b,c)$ system that was introduced.
\subsection{$SU(2)_0$ example}
In the previous chapter we saw how to use the symplectic fermions to realise the extra structure in the $SU(2)_0$ theory. We shall now show that at the free field level the $j=1,2,3$ multiplets given in Section \ref{sec:extendedalg} reduce to the well known extra chiral fields in the $c=-2$ model.

A crucial point to first notice is that in the $SU(2)_0$ theory, \emph{before} performing the reduction, the stress tensor is (\ref{eqn:stressTsu2level0}):
\bea
T_{Sugawara}= -\beta \p \gamma - \xi \p \eta \nonumber
\eea
and $\xi, \eta$ have conformal weights $h=1,0$ respectively.

However after hamiltonian reduction the stress tensor becomes (\ref{eqn:improvedstressT}):
\bea
T&=& - \xi \p \eta + \p \lb \xi \eta \rb = -\eta \p \xi
\eea
and now $\xi,\eta$ have weights $h=0,1$ respectively. This fact is essential to obtain the correct weights for the fields.

When we impose the physical state conditions (\ref{eqn:physicalstate}) on the previous multiplets of operators found in $SU(2)_0$ we find that only the highest component of the multiplet is a physical operator in the reduced theory. When we set $J^+=\beta=1$ in these we get the physical states:
\begin{itemize}
\item{$j=1$}
\bea 
J^+=1
\eea
\item{$j=2$}
\bea 
\Psi^+ &=&- \p \xi \\
\Psi^-  &=& \eta \nonumber 
\eea
\item{$j=3$}
\bea
W^+ &=&- \p^2 \xi \p \xi \nonumber  \\
W^3 &=&\f{1}{2} \eta \p^2 \xi - \f{1}{2} \p \eta \p \xi \\
W^- &=& \p \eta \eta \nonumber
\eea
\end{itemize}
These are precisely the doublet and triplet fields that are well known to exist in the $c=-2$ theory \cite{Kausch:1995py}. The extra indicial structure of the fields is not affected by the hamiltonian reduction and we see that there is an exact match between the extended multiplet structures of the two theories.
One can also perform the reduction on the $j=1$ fields $\Theta^{\pm}(x,z)$ that we found earlier in the free field representation (\ref{eqn:FFthetaplus},~\ref{eqn:FFthetaminus}). These reduce to the following operators:
\bea
\Theta^+&=& - \xi \\
\Theta^-&=& \p^{-1} \eta \nonumber
\eea
which are precisely the formal expressions for the symplectic fermions \cite{Kausch:1995py}. When the zero modes are correctly introduced \cite{Gurarie:1997dw} then all OPEs can be calculated. 
\section{Correlation functions}
One might also wonder if the reduction can also be performed at the level of correlation functions and indeed this is possible. There is a very elegant procedure first analysed in a series of papers \cite{Furlan:1991by,Furlan:1993mm,Ganchev:1992af}. Here we shall briefly outline the procedure and do not in any way attempt to justify its origin. 

The rather surprising equivalence suggested is that if one takes the limit as $x \rightarrow z$ in the $SU(2)_k$ correlators one obtains those of the reduced $c_{k+2,1}$ model. This is indeed very strange as $z$ is a physical coordinate in the plane and $x$ is an artificial isotopic coordinate to describe the $SU(2)$ structure. A quick check of the two and three point functions of primary fields of $SU(2)_k$ (\ref{eqn:2pt},\ref{eqn:3pt}) does indeed yield the correct form for the two and three point functions in the reduced theory with the correct conformal weights (\ref{eqn:reduced}). However a much more non-trivial statement is that such a procedure gives the correct four-point functions. If, rather than expanding $F(x,z)$ as a power series in $x$ as in (\ref{eqn:discreterepns}), one expands in the alternative basis:
\bea
F(x,z)=\sum_{n=0}^{2j} (x-z)^n \tilde{F}_n (z)
\eea
then it was shown \cite{Furlan:1991by,Furlan:1993mm,Ganchev:1992af} that the lowest component $\tilde{F}_0(z)$, which is the only term surviving in the limit $x \rightarrow z$, obeys the correct differential equation for the primary field $h_{1,2j+1}$ in the $c_{k+2,1}$ model. We have explicitly checked by hand that this works in a few of the lowest order cases.

As we shall only be computing conformal blocks we shall disregard overall numerical factors which only become important when considering the consistency of the entire theory. However there are sometimes subtleties in this reduction process \cite{Petersen:1996xn} when correlators vanish as $x$ approaches $z$. This can be also be seen as an obstacle in reducing the \KZ equation to an expression in terms of the lowest component $\tilde{F}_0(z)$. We did not find any problems in the examples studied in this thesis.

Here will shall explicitly calculate some of the correlation functions to demonstrate how this works in these cases. The conformal weights of the $c=-2$ fields are given by $h=\f{j(j+1)}{2}-j$, where $j$ is the spin of the $SU(2)_0$ operator. The lowest few are:
\bea 
\begin{tabular}{ccccccc}
$j$~~~~ $0$ & $\f{1}{2}$ & $1$ & $\f{3}{2}$ & $2$ & $\f{5}{2}$ & $3$  \\
$h$~~~~ $0$ & $-\f{1}{8}$ & $0$ & $\f{3}{8}$ & $1$ & $\f{15}{8}$ & $3$ \nonumber
\end{tabular}
\eea
We have already seen in the free field approach that the doublet $j=2$ and triplet $j=3$ fields reduce to the fermions $\Psi^{\pm}$ and $W$-algebra fields $W^{a}$ in the triplet model. We also see that the vacuum logarithmic pair at $h=0$ in the $c=-2$ theory is a direct reduction of the indecomposable representation with $j=0,1$ in the $SU(2)_0$ theory.

In the $c=-2$ theory the Kac table is empty and we are forced to extend the representations in order to get a non-trivial theory. In an exactly analogous way we have extended the $SU(2)_0$ beyond the highest weight vector $j=0$ to get a non-trivial theory.

The $h_{r,1}=0,1,3,\cdots$ fields from $c_{2,1}=-2$ theory can also be obtained by reduction of another $SU(2)_k$ theory; with $k=-\f{3}{2}$. In \cite{Kausch:1995py} it was found that these integer dimension fields in $c=-2$ did not have nice properties under modular transformations. In a similar way one suspects that the same is true of the $2j \in Z$ fields in $SU(2)_{-3/2}$ and therefore presumably there must be other sectors of the theory.
\subsection{Logarithmic correlators}
We have explicitly verified that this simple reduction procedure exactly reproduces all the chiral and non-chiral correlators of \cite{Gaberdiel:1998ps}. Several of these were given in \cite{Kogan:2001nj} and here we shall give a few examples.

If we consider the four point correlator of $j=\half$ operators with conformal blocks (\ref{eqn:SU20doublet}):
\bea
F_{\half \half \half \half}^{(1)}(x,z)&=& z^{\f{1}{4}}(1-z)^{\f{1}{4}} \left\{ \left( -\f{E}{z(1-z)}+\f{K}{z} \right) x + \f{E}{1-z} \right\} \\
F_{\half \half \half \half}^{(2)}(x,z)&=& z^{\f{1}{4}}(1-z)^{\f{1}{4}} \left\{ \left( \f{\tilde{E}}{z(1-z)}-\f{\tilde{K}}{1-z} \right) x + \f{\tilde{K}}{1-z} -\f{\tilde{E}}{1-z} \right\} \nonumber
\eea
When we set $x=z$ in these we get the solutions:
\bea
F_{\half \half \half \half}^{(1)}(x,z)&\rightarrow& z^{\f{1}{4}}(1-z)^{\f{1}{4}} K \\
F_{\half \half \half \half}^{(2)}(x,z)&\rightarrow& z^{\f{1}{4}}(1-z)^{\f{1}{4}} \tilde{K} \nonumber
\eea
These are precisely the two conformal blocks of the $\left< \mu \mu \mu \mu \right>$ correlator (\ref{eqn:cminus2sols}) where $\mu$ is the $h=-\f{1}{8}$ operator in $c=-2$. Moreover in $SU(2)_0$ the non-chiral correlator has the structure:
\be \label{eqn:nonchiralpart}
G(x,z,\bar{x},\bar{z})=F_{\half \half \half \half}^{(1)}(x,z) \overline{F_{\half \half \half \half}^{(2)}(x,z)} + F_{\half \half \half \half}^{(2)}(x,z) \overline{F_{\half \half \half \half}^{(1)}(x,z)}
\ee
and this reduces to the correct non-chiral correlator in $c=-2$ (\ref{eqn:cminus2singval}).

We now consider the hamiltonian reduction of the operators $\Theta^{\pm}(x,z)$. We have already calculated the conformal blocks (\ref{eqn:spinone}). Taking the limit as $x \rightarrow z$ we find:
\bea
F_{1111}^{(1)}(x,z)&\rightarrow &1 \nonumber \\
F_{1111}^{(2)}(x,z)&\rightarrow &\ln(z)  \\
F_{1111}^{(3)}(x,z)&\rightarrow &\ln(1-z) \nonumber 
\eea
which are precisely the conformal blocks of the $h_{1,3}=0$ fields in $c=-2$ (\ref{eqn:h13cminus2}) .The $\Theta^{\pm}(x,z)$ operator in $SU(2)_0$ reduces to the $\Theta^{\pm}(z)$ operator in $c=-2$ in complete agreement with the free field approach. In \cite{Kausch:2000fu} the $2N$-point function of these fields was calculated in $c=-2$. It would be interesting to calculate similar correlators in $SU(2)_0$ as they should give us a lot more information on the indecomposable representations $K,C$ and $D$ occurring in the OPE of the $\Theta^{\pm}(x,z)$ operators (\ref{eqn:OPEthetatheta}). There is however a slight difference in that the two point function of $\Theta^{\alpha}(x,z)$ vanishes in $SU(2)_0$ (\ref{eqn:ThetaThetaequalszero}) however in $c=-2$ the operator $\Theta^{\alpha}(z)$ has non-vanishing (actually constant) two point function. This is again related to the issues regarding overall normalisation of the correlators which we shall not discuss here. 

The $h_{1,4}=\f{3}{8}$ operator $\nu^{\pm}$ in the $c=-2$ triplet model has a doublet nature and therefore we expect that the $j=\f{3}{2}$ operator in $SU(2)_0$ should also be a doublet. Rather than discuss this in detail it is sufficient to note than in $c=-2$ we have:
\bea
\Psi^{\pm}(z) \mu(w,\bar{w}) \sim (z-w)^{-3/4} \nu^{\pm}(w,\bar{w})
\eea
where $\mu$ is the $h=-\f{1}{8}$ twist field. All the multiplet structure comes from the integer dimension fields and therefore it is sufficient to consider these.
\subsection{Rational correlators}
We shall now discuss the hamiltonian reduction of the rational solutions of the integer spin operators.
Now performing such a reduction on the $F_{1111}$ correlator (\ref{eqnjone}) we find (up to a numerical factor):
\bea
F_{1111}(x,z) \rightarrow 1
\eea
Such a simple correlator is due to the fact that the $j=1$ field reduces to an $h=0$ field. This is just the identity field $\Omega$ and, as such, does not generate a chiral algebra in the reduced theory. In fact in the $c=-2$ theory the identity is a zero-norm field \cite{Caux:1996nm}:
\bea
\left< \Omega(z_1) \Omega(z_2) \right> =0
\eea
and so in order to reproduce the conformal block one must actually insert the logarithmic partner of the identity $\omega$:
\bea
\left< \Omega(z_1) \omega(z_2,\bar{z}_2) \right> =1
\eea
We then have:
\bea
\left< \Omega(z_1) \Omega(z_2) \Omega(z_3) \omega(z_4,\bar{z}_4) \right>=1 
\eea
We see that the discussion of insertion of logarithmic partners to obtain the conformal blocks in $SU(2)_0$ parallels that of $c=-2$. However there is an important difference: in $c=-2$ this indecomposable representation is at the $h=0$ level only whereas for $SU(2)_0$ it is believed that the indecomposable representations occurs between the $j=0$ and $j=1$ states \cite{Caux:1997kq}. From (\ref{eqn:reduced}) we see that both $j=0,1$ fields reduce to $h=0$ fields.

For the $F_{2222}$ correlators (\ref{eqn:jtwo}) we get:
\bea
F_{2222}^{(1)}(x,z) &\rightarrow& 1- \f{1}{z^2} \\
F_{2222}^{(2)}(x,z) &\rightarrow& 1- \f{1}{(1-z)^2} \nonumber 
\eea
These reproduce precisely the conformal blocks of the $h_{2,1}=1$ operators (\ref{eqn:cminus2doublet}) in the $c=-2$ theory as claimed. We have deliberately named these fields in $SU(2)_0$ with the same letters but in case of confusion we shall write explicitly the $SU(2)_0$ fields as functions of $x$ and $z$ to indicate the additional affine Kac-Moody structure. The doublet nature is unaffected by the reduction. Thus the Hamiltonian reduction of the multiplets $\Psi^{\pm}(x,z)$ (\ref{eqn:jequaltwo}) gives precisely the fields $\Psi^{\pm}(z)$ which are the chiral fermionic doublet in $c=-2$ (\ref{eqn:symplecticfermion}).

When dealing with chiral algebras of any kind one has to make sure that the structure of the fields as defined by the OPEs is actually associative. In terms of the mode expansions this is expressed by the Jacobi identity whereas for the fields it is the crossing symmetry of the four point functions. In general the Jacobi identity may be satisfied only up to certain null fields and for consistency these must vanish in all correlation functions.

Extending the Virasoro algebra by the symplectic fermion fields $\Psi^{\pm}(z)$ leads to an algebra that is is automatically associative; in other words the Jacobi identity is immediately satisfied. However for the $j=2$ algebra in $SU(2)_0$ we found that the associativity constraint was only satisfied if certain null vectors decouple. For example:
\bea \label{eqn:nullvectorj2}
{\mathcal N} =4 J^3 \lb \Psi^+ \rb^{++}+J^{+} \lb \Psi^+ \rb^{+}-2 \p \lb \Psi^+ \rb^{++}
\eea
This null vector actually follows from the \KZ null vector:
\bea
{\mathcal N}=-2 \left[ L_{-1} - \half \lb J^a_{-1} J^a_0 \rb \right] \lb \Psi^+ \rb^{++}
\eea
It is not clear if all such null vectors are a direct consequence of the \KZ null vector and it would be interesting to study this in more detail.

Reduction of the $F_{3333}$ correlators (\ref{eqn:jthree}) gives:
\bea \label{eqn:cminus2triplet}
F_{3333}^{(1)}(x,z) &\rightarrow& \f{1}{(z-1)^6} z^4 \left(6-6z+z^2 \right) \nonumber \\
F_{3333}^{(2)}(x,z) &\rightarrow& \f{1}{z^6(z-1)^6} \left( 2-12z+12z^2+50z^3-225z^4+468z^5-588z^6+468z^7-225z^8 \right.\nonumber\\
&&\left.+50z^9+12z^{10}-12z^{11}+2z^{12} \right)\\
F_{3333}^{(3)}(x,z) &\rightarrow& \f{1}{z^6} \left( 1-9z^2+16z^3-9z^4+z^6 \right) \nonumber
\eea
These again, as expected, reproduce the conformal blocks for $h_{3,1}=3$ operators in $c=-2$. By analysing the pole structure and symmetries we find that these are related (up to normalisation) in the following way to the standard fields \cite{Kausch:1995py}:
\bea
F_{3333}^{(1)} &\rightarrow& \left< W^+(0) W^+(z) W^-(1) W^-(\infty) \right> \nonumber \\
F_{3333}^{(2)}  &\rightarrow& \left< W^3(0) W^3(z) W^3(1) W^3(\infty) \right> \\
F_{3333}^{(3)} &\rightarrow& \left< W^+(0) W^-(z) W^-(1) W^+(\infty) \right>\nonumber
\eea
The first of these is the unique solution that has no singular terms as $z \rightarrow 0$. The third is similar but with no singular terms as $z \rightarrow 1$. The second is the unique solution invariant under all crossing symmetries. Again we are being slightly sloppy because, as we have discussed before, we must insert a logarithmic zero mode into these correlators to reproduce the conformal blocks.

Analysing these is sufficient to fully reconstruct the triplet algebra (\ref{eqn:tripletalgebra}) generated by the $W^a$ fields \cite{Kausch:1995py}. The diagonal generator $W^3$ generates an automatically associative W-algebra that is precisely the Zamolodichikov $W(2,3)$ algebra \cite{Zamolodchikov:1985wn} at $c=-2$ whose correlator we have already given (\ref{eqn:ZamW23}). Therefore the triplet algebra \textbf{can} come from a hamiltonian reduction of an $SU(2)$ structure. This is in complete contrast to the normal case of Hamiltonian reduction of $\widehat{SU(2)}$ which leads to the Virasoro algebra only. It is the reduction of the extra chiral fields which produces the more complicated structure.
\section{General $SU(2)_k$ structure}
We shall now comment briefly on the structure that we expect in $SU(2)_k$ for $k \in N$. We have in general an affine Lie algebra vacuum null vector (\ref{eqn:introvacnullvect}) that transforms in the $j=k+1$ representation. Imposing this as a null vector on states gives the normal rational affine Lie algebra theory with $0 \le j \le \f{k}{2}$. Using the expression for the conformal weights (\ref{eqn:reduced}) we see that the reduction of this null vector gives an $h=0$ state in the $c_{k+2,1}$ theory. We thus see that extending the model by adding a logarithmic partner which prevents the affine Lie algebra null vector decoupling automatically leads to a logarithmic partner for the vacuum in the $c_{p,1}$ models. For $SU(2)_0$ this null vector was precisely the affine current $J^a$ and the rational set was trivial. There is indeed a single rational solution for correlators of $j=k+1$ fields \cite{Hadjiivanov:2001kr} generalising our discussion for $j=1$ in $SU(2)_0$. 

All the $c_{p,1}$ models have a triplet algebra generated by the $h_{3,1}=2p-1$ fields \cite{Kausch:1991vg}. These come from the hamiltonian reduction of $j=2k+3$ fields in $SU(2)_k$. We therefore conjecture that there will also be a triplet algebra generated by these fields and correspondingly, in general, three rational solutions to the KZ equation. We have confirmed this by explicit calculation in many cases but rather than comment further on this we shall discuss the more general appearance of rational correlation functions that we found in the $\widehat{SU(2)}$ and $c_{p,q}$ models.
\section{Correlation functions in $c_{p,q}$ models}
We shall make use of the quantum hamiltonian reduction of $SU(2)_k$ WZNW models, at rational level $k$, which gives a very efficient procedure to directly calculate differential equations for the $h_{1,s}$ fields in the $c_{p,q}$ models. By examining several examples we shall show that there is a very simple, and elegant, structure for a certain subset of $h_{1,s}$ operators.

We find that there is a single rational solution generated by the \mbox{$h_{1,2p-1}=(p-1)(q-1)$} field corresponding to the vacuum null vector of the irreducible theory. It is well known that decoupling such a null vector gives us a complete description of the `minimal' $c_{p,q}$ model \cite{Feigin:1992wv}. However at this conformal weight we find, in addition, two other primary \emph{fermionic} non-chiral operators. This extended index structure permeates the model.

We found that there are triplets of chiral bosonic fields at $h_{1,4p-1}=(2p-1)(2q-1)$. These are a natural generalisation of an algebra, generated by the $h_{1,3}=2q-1$ fields, that appears in the $c_{1,q}$ models. It has been previously conjectured by M. Flohr \cite{Flohr:1997vc} that these extended $c_{p,q}$ models should be formally considered as $c_{3p,3q}$ and we conjecture that the algebra of such $h_{1,4p-1}$ fields may yield rational extended $c_{p,q}$ models.  We also observed extra doublet structure in the $c_{p,q}$ model generated by the $h_{1,3p-1}=(\f{3p}{2}-1)(\f{3q}{2}-1)$ fields.

Much of what we shall say about $c_{p,q}$ models could presumably also be reinterpreted in the $\widehat{SU(2)}$ theory, as solutions for $F_0(z)$ lift up to solutions for the full $F(x,z)$, but we do not attempt this here. Recently a particular $SU(2)_k$ theory at rational level, namely $k=-\f{4}{3}$, was studied \cite{Gaberdiel:2001ny}. It was found that indecomposable, and even continuous, representations are created in the fusion of admissable representations and the theory was certainly \emph{not} rational. On hamiltonian reduction the discrete representations of $SU(2)$ with $2j \in Z$, which are different to the admissable representations, produce $h_{1,s}$ fields in the $c_{2,3}=0$ model. It would be interesting to see if the type of extended algebras studied here could be used to construct rational models of $\widehat{SU(2)}$ at fractional level.
\subsection{Vacuum null vector and its fermionic partners}
In this section we shall discuss the vacuum null vector. We shall also find new fermionic partner fields.
\subsubsection{Vacuum null vector}
It is known that by studying the vacuum null vector we can learn everything about the `minimal' sector of the $c_{p,q}$ models with weights $h_{r,s}$ given by (\ref{eqn:CpqKactableweights}) in the region $1 \le r \le q-1,~1 \le s \le p-1$ and with the identifications $h_{r,s}=h_{q-r,p-s}$\cite{Feigin:1992wv}.

For example the Ising model at $c_{3,4}=\half$ has a vacuum null vector given by:
\bea \label{eqn:nullIsing}
{\mathcal N}= 9 \p^4 T +264 ((\p^2 T) T) -186 (\p T \p T)) -128 (T(TT))
\eea
One can verify by using the Virasoro algebra (\ref{eqn:TTOPE}) and the normal ordering prescription:
\bea
(AB)(w) = \f{1}{2 \pi i} \oint_w \f{dz}{z-w} A(z) B(w)
\eea
that this null vector is indeed a primary field of conformal weight $6$:
\bea
T(z) {\mathcal N}(w) \sim \f{6 {\mathcal N}(w)}{(z-w)^2}+ \f{\p  {\mathcal N}(w)}{z-w}
\eea
In the irreducible theory this null vector should be set to zero in all correlation functions. In particular the zero mode of this must vanish when applied to Virasoro primary states $\left. | h \right>$. We know:
\bea
L_n \left. | h \right> &=&0 \quad \quad n \ge 1 \nonumber\\
L_0  \left. | h \right> &=& h \left. | h \right>
\eea
Now using the equivalent version of the normal ordering in terms of modes:
\bea
\lb A B \rb_m = \sum_{n \le -h_A} A_n B_{m-n} + \sum_{n > -h_A} B_{m-n} A_n
\eea
one finds:
\bea
{\mathcal N}_0 \left. | h \right> = -4h(2h-1)(16h-1) \left. | h \right> =0 
\eea
From this one finds the solutions $h=0,\f{1}{2}, \f{1}{16}$ which are the conformal weights of the operators present in the Ising model. In general imposing the zero modes of the $h=(p-1)(q-1)$ vacuum null vector gives us a polynomial of rank $r=\half (p-1)(q-1)$. Solving this gives us precisely the $r$ primary operators in the `minimal' $c_{p,q}$ model \cite{Feigin:1992wv}. Furthermore all fusion rules in this theory can, in principle, be found from such a null vector. More details can be found in \cite{Gaberdiel:2001tr} and references therein.

In particular if we wish to go beyond the minimal $c_{p,q}$ model and consider fields outside the region with $1 \le r \le q-1,~1 \le s \le p-1$ we must not decouple this vacuum null vector. In order to achieve this we would have to introduce a logarithmic partner for this field. In the case of the well studied $c_{p,1}$ models the vacuum null vector is at $h=0$ implying, as is well known, that all these extended models must have a logarithmic partner for the vacuum itself.

Continuing with the example of the Ising model we can calculate the correlator with four $h_{1,2p-1}=6$ operators and we find:
\bea \label{eqn:Ising4pt}
F(z)&=&\f{1}{z^6(1-z)^6} \left(2090z^6-6270z^5+10869z^4-11288z^3 \right. \nonumber \\
 && \hspace{2in} \left. +10869z^2-6270z+2090 \right)
\eea
This conformal block is easily seen to lead to a well behaved correlator invariant under all crossing symmetries. By analysing the leading singularity as $z \rightarrow 0$ we deduce that the two point function of these fields must vanish. To see that this must be true in general consider the OPE of two vacuum null vector fields of the irreducible theory having conformal weight $h$. This must have the form (up to normalisation):
\bea
{\mathcal N}(z) {\mathcal N}(w) \sim \f{{\mathcal N}(w)}{(z-w)^h} + \cdots
\eea
where $\cdots$ stands for other less singular terms. There cannot be other operators in the more singular terms as these would also be vacuum null vectors, of lower conformal weight, contradicting the fact that we are considering the vacuum null vector of the \emph{irreducible} theory. 

This is of course confirmed by explicitly calculating the OPE of (\ref{eqn:nullIsing}) with itself. However the vanishing of the two point function of ${\mathcal N}$ immediately implies that the four point function must also vanish. In order to make the four point function non-zero and realise the conformal block (\ref{eqn:Ising4pt}) we must have an insertion of a logarithmic partner field. We found in all cases (we checked $c_{p,q}$ with $p,q \le 6$) that there is indeed a single rational solution generated by the $h_{1,2p-1}=(p-1)(q-1)$ field as we expect. However, as we shall see in the next section, there were always two extra \emph{non-chiral} states as well.
\subsubsection{Non-chiral fermionic partners}
In general we found that the differential equation with four $h_{1,2p-1}=(p-1)(q-1)$ operators always admitted solutions of the form:
\bea \label{eqn:vacnullsolns}
F^{(1)}(z)&=&R_1(z) \nonumber\\ 
F^{(2)}(z)&=&R_1(z) \ln z + R_2(z) \\
F^{(3)}(z)&=&F^{(2)}(1-z)\nonumber
\eea
where $R_1(z)$ and $R_2(z)$ are $\emph{rational}$ functions. The first solution $F^{(1)}(z)$ is the conformal block of the four point function of the vacuum null vector, with the subtleties about insertions of a logarithmic partner, that we have just discussed. We have already commented that as this is a bosonic field we expect it to be invariant under all crossing symmetries:
\bea
R_1(z)=R_1(1-z) \quad \quad z^{-2h} R_1\left( \f{1}{z} \right) =R_1(z)
\eea
The set (\ref{eqn:vacnullsolns}) is clearly closed under monodromy transformations however in order to be closed under crossing symmetries we must have:
\bea
z^{-2h} R_2\left( \f{1}{z} \right) =-R_2(z) + \alpha R_1(z)
\eea
The constant $\alpha$ is arbitrary but we shall always redefine $F^{(2)}(z)$ by addition of $F^{(1)}(z)$ to set $\alpha$ to zero. 

The other solutions, as we shall presently see, correspond to extra non-chiral \emph{fermionic} operators. To see this it is interesting to consider the explicit example of the $c_{2,3}=0$ model. This is of great importance in the field of percolation and polymers \cite{Saleur:1992hk,Cardy:1992cm,Watts:1996yh}. The vacuum null vector in this case is the stress tensor $T$ itself and imposing the vanishing of this in correlators gives us just the `minimal' topological sector. Considering fields beyond this sector we must create a logarithmic partner for the stress tensor\cite{Gurarie:1999yx}.

We consider the four point function with four $h_{1,3}=2$ fields in this model. The differential equation for $F(z)$ is:
\bea
&&z^3(z-1)^3 \f{d^3 F}{d z^3}+6z^2(2z-1)(z-1)^2\f{d^2 F}{d z^2} \nonumber \\
&&\hspace{1in} +6z(z-1)(4z^2-4z-1)\f{d F}{d z}-24(2z-1)F(z)=0
\eea
and one can easily solve this to obtain the conformal blocks:
\bea \label{eqn:solsczero}
F^{(1)}(z)&=& \f{z^2-z+1}{z^2 (z-1)^2} \nonumber\\
F^{(2)}(z)&=& F^{(1)}(z) \ln(z) - \f{(5z^5-5z^4+12z^3+12z^2-5z+5)}{24 (z-1) z^4} \\
F^{(3)}(z)&=& F^{(2)}(1-z)\nonumber
\eea
These are indeed of the form (\ref{eqn:vacnullsolns}) as claimed. Before continuing to discuss these solutions we should comment on what occurs if one instead studies the correlators of the $h_{5,1}=2$ field. Then of course the equation to be solved is of fifth order but one finds as a subset of solutions the \emph{same} rational block $F^{(1)}(z)$. However the solution for $R_2(z)$ in (\ref{eqn:vacnullsolns}) is now slightly different:
\bea
R_2(z)=\f{(5z^5-5z^4-16z^3-16z^2-5z+5)}{32 (z-1) z^4}
\eea
There are also two other solutions having no simple form. This seems to be a universal feature and the rational functions, in this case just $F^{(1)}(z)$, always appear as a subset of solutions to both equations.

The rational solution $F^{(1)}(z)$ forms a well behaved chiral correlator on its own and corresponds to the vacuum null vector $T$. It is easy to see that this is the only primary $(2,0)$ operator in the theory as the other solutions in (\ref{eqn:solsczero}) on their own do not lead to single-valued correlators. It is also possible to have local $(2,2)$ operators in the theory. To see what these are we combine these conformal blocks with their anti-holomorphic components into the full correlator:
\be
G(z,\bar{z})=\sum_{a,b=1}^{3}{U_{a,b} F^{(a)}(z) \overline{F^{(b)}(z)}}
\ee
To make this single-valued everywhere we find:
\bea
G(z,\bar{z})&=&U_{1,1} F^{(1)}(z) \overline{F^{(1)}(z)} 
+ U_{1,2} \Bigl[ F^{(1)}(z) \overline{F^{(2)}(z)} + F^{(2)}(z) \overline{F^{(1)}(z)} \Bigr] \nonumber \\
& &+ U_{1,3} \Bigl[ F^{(1)}(z) \overline{F^{(3)}(z)} + F^{(3)}(z) \overline{F^{(1)}(z)} \Bigr]
\eea
As well as the solution corresponding to the stress tensor $F^{(1)}$ we also have two other solutions which, as we have logarithms present, do not have a diagonal form.

Now consider the effect of crossing symmetries on these solutions. Under $1 \leftrightarrow 3$ we have $z \rightarrow 1-z$ and:
\bea
F^{(1)} \rightarrow F^{(1)} \quad F^{(2)} \rightarrow F^{(3)} \quad F^{(3)} \rightarrow F^{(2)}
\eea
Under $1 \leftrightarrow 4$ we have $z \rightarrow \f{1}{z}$:
\bea
F^{(1)} \rightarrow z^4 F^{(1)} \quad F^{(2)} \rightarrow -z^4 F^{(2)} \quad F^{(3)} \rightarrow z^4 \left(-i \pi F^{(1)} - F^{(2)} + F^{(3)} \right)
\eea
We immediately see that the other two solutions are \emph{not} invariant under all crossing symmetries. This is exactly the same situation as we have already encountered in the $SU(2)_0$ case. We can therefore use the same set of arguments as in Section \ref{sec:Tripletsolutions} and deduce the existence of \emph{non-chiral fermionic} operators $\Theta^{\pm}(z,\bar{z})$ in the $c_{2,3}=0$ model. To get the correct crossing symmetries we must have:
\bea
\langle \Theta^+(0,0)  \Theta^-(z,\bar{z}) \Theta^-(1,1) \Theta^+(\infty,\infty)\rangle&=& F^{(1)}(z) \overline{F^{(2)}(z)} + F^{(2)}(z) \overline{F^{(1)}(z)}  \nonumber \\
\langle \Theta^+(0,0)  \Theta^+(z,\bar{z}) \Theta^-(1,1) \Theta^-(\infty,\infty)\rangle&=& F^{(1)}(z) \overline{F^{(3)}(z)} + F^{(3)}(z) \overline{F^{(1)}(z)}  \nonumber 
\eea
By expanding these we see:
\bea
\left< \Theta^{\alpha}(z_1,\bar{z_1})  \Theta^{\beta}(z_2,\bar{z_2}) \right> =0 \quad \quad \alpha,\beta=\pm
\eea
It appears to be a general fact that all the irreducible fields in $c=0$ models beyond the minimal sector have vanishing two point functions. It is the non-vanishing of the four point functions that gives us a non-trivial theory. It should be remembered however that we made an assumption regarding the normalisability of the vacuum state. This is clearly different from the situation in $c=-2$ in which the two point function of the twist field $\mu$ is non-zero and the vacuum has zero-norm. Clearly there remain some major questions regarding this model and $c=0$ theories in general.

It has been conjectured that fermionic partners to the stress tensor in $c=0$ generate a super-algebra with $U(1|1)$ symmetry \cite{Gurarie:1999yx}. As we have seen these fields are non-chiral and so cannot be generators of an affine algebra.
\subsection{Triplet solutions}
In general once one considers fusion of fields from outside the minimal region of the $c_{p,q}$ models we start to generate an infinite number of Virasoro primary states. However in the $c_{p,1}$ models this infinite number of fields become rearranged into a finite number of states with respect to a larger algebra 
generated by a triplet of $h_{3,1}=2p-1$ fields. Although these were originally found by different methods \cite{Kausch:1991vg} we have already seen earlier how they arise from the rational solutions for the conformal blocks. For example in the $c_{1,2}=-2$ model the $h_{1,3}$ or the $h_{7,1}$ fields, both of which have $h=3$, have exactly three rational four point functions (\ref{eqn:cminus2triplet}). However the arguments leading us to this triplet structure rely only on the existence of the three rational solutions with the stated pole structure and behaviour under crossing symmetry and one might suspect that such triplets of rational solutions can be found in more general $c_{p,q}$ models. We indeed found that this was true and it is useful to note that in $c_{p,1}$ we have $h_{3,1}=h_{1,4p-1}$. We found that it is the $h_{1,4p-1}=(2p-1)(2q-1)$ fields that become the triplet fields in the general $c_{p,q}$ models.
\subsubsection{Correlators in the $c_{1,1}$ model}
The $c_{1,1}=1$ model is a rather peculiar case and so we shall discuss it separately in this section. The $h_{1,s}$ fields have weights:
\bea
h_{1,2j+1}=h_{2j+1,1}=j(j+1)-j=j^2
\eea
In this case we find the first few fields have dimensions: $0,\f{1}{4},1, \cdots$. The $j \in Z$ fields have integer dimensions and all conformal blocks of these fields that we studied were found to be rational functions. In particular we found that the $h_{1,3}=1$ fields have three rational solutions behaving exactly as before and so we deduce a triplet algebra with correlators:
\bea \label{eqn:c11triplet}
\left< W^+(0) W^+(z) W^-(1) W^-(\infty) \right>&=& \f{z^2}{(1-z)^2} \nonumber\\
\left< W^3(0) W^3(z) W^3(1) W^3(\infty) \right>&=& \f{(1-z+z^2)^2}{z^2(1-z)^2}  \\
\left< W^+(0) W^-(z) W^-(1) W^+(\infty) \right>&=& \f{(1-z)^2}{z^2} \nonumber
\eea
The $c_{1,1}$ model comes from a hamiltonian reduction of the $SU(2)_{-1}$ theory. These $h_{1,3}$ correlators come from reduction of a set of three rational solutions for the $j=1$ fields. This theory is sufficiently simple that we can again write a free field representation for the triplet fields. We use as before the Wakimoto representation (\ref{eqn:Wakimoto}), this time at level $k=-1$:
\bea
J^+&=&\beta \nonumber \\
J^3&=& \f{i}{\sqrt{2}} \p \phi + \gamma \beta \\
J^-&=&- i \sqrt{2} \p \phi \gamma + \p \gamma - \beta \gamma^2 \nonumber
\eea
These obey the $SU(2)_{-1}$ algebra with Sugawara stress tensor:
\bea
T=-\beta \p \gamma - \f{1}{2} \p \phi \p \phi - \f{i}{\sqrt{2}} \p^2 \phi
\eea
with central charge $c=\f{3k}{k+2}=-3$. The $(\beta,\gamma)$ bosonic ghost system has weights $(1,0)$ and $c=2$ as usual. For the $\phi$ system with $c=-5$ we do not have a representation in terms of symplectic fermions as we did for $c=-2$. In this respect this illustrates the more generic situation. There are however mutually local integer dimension fields namely the vertex operators: $e^{i \sqrt{2} \phi},e^{-i \sqrt{2} \phi}$ with weights $2,0$ respectively. In a similar way to $SU(2)_0$ we found the following expressions for the $j=1$ multiplets:
\bea \label{eqn:levelminusonetriplets}
\left( W^+ \right)^+&=& e^{i \sqrt{2} \phi} \nonumber \\
\left( W^+ \right)^3&=& -2 \gamma e^{i \sqrt{2} \phi} \nonumber \\
\left( W^+ \right)^-&=& 2 \gamma^2 e^{i \sqrt{2} \phi} \nonumber \\
\nonumber \\
\left( W^3 \right)^+&=& \beta \p \phi + \f{i}{\sqrt{2}} \p \beta \nonumber \\
\left( W^3 \right)^3&=& -2 \beta \gamma \p \phi - i \sqrt{2} \p \phi \p \phi - i \sqrt{2} \p \beta \gamma + \p^2 \phi \\
\left( W^3 \right)^-&=& 2 \beta \gamma \gamma \p \phi + 2 i \sqrt{2} \gamma \p \phi \p \phi - 2 \gamma \p^2 \phi + i \sqrt{2} \p \beta \gamma^2 \nonumber \\
\nonumber \\
\left( W^- \right)^+&=& \beta \beta e^{-i \sqrt{2} \phi} \nonumber \\
\left( W^- \right)^3&=& -2 \beta \beta \gamma e^{-i \sqrt{2} \phi} -2 i \sqrt{2} \beta \p \phi e^{-i \sqrt{2} \phi} -2 \p \beta e^{-i \sqrt{2} \phi} \nonumber \\
\left( W^- \right)^-&=& 2 \beta^2 \gamma^2 e^{-i \sqrt{2} \phi} +4 i \sqrt{2} \beta \gamma \p \phi e^{-i \sqrt{2} \phi} -4 \p \phi \p \phi e^{-i \sqrt{2} \phi} \nonumber \\
&& + 4 \p \beta \gamma e^{-i \sqrt{2} \phi} + 2 i \sqrt{2} \p^2 \phi e^{-i \sqrt{2} \phi} \nonumber
\eea
Using the general expressions in Section \ref{sec:Freefieldapproach} we find the stress tensor in the reduced theory:
\bea
T= - \f{1}{2} \p \phi \p \phi - \f{i}{\sqrt{2}} \p^2 \phi + \p \left( \f{i}{\sqrt{2}} \p \phi \right) =-\f{1}{2} \p \phi \p \phi 
\eea
This is exactly the expected result for the $c_{1,1}=1$ theory. The fields $e^{\pm i \sqrt{2} \phi}$ now both have dimension $1$ in this theory.

In the same way as for $SU(2)_0$ in the reduced theory only the top component of the multiplets, given in (\ref{eqn:levelminusonetriplets}), is physical. Therefore we find the reduced currents are:
\bea
W^+&=&  e^{i \sqrt{2} \phi} \nonumber \\
W^3&=& \p \phi  \\
W^-&=& e^{-i \sqrt{2} \phi} \nonumber 
\eea
These are better known as the currents corresponding to an $SU(2)_1$ algebra! The correlators (\ref{eqn:c11triplet}) are exactly those corresponding to four point functions of these fields in the extended $c_{1,1}=1$ model.

We began with the $SU(2)_{-1}$ theory with three Kac-Moody currents and the triplet of $j=1$ fields. Note in this case there is potential confusion as the extended fields are triplets of the $SU(2)_{-1}$ algebra (they have $j=1$) \emph{and} also have an extended triplet index. After hamiltonian reduction the $SU(2)_{-1}$ structure is lost but the extended one remains. What is remarkable, in this example, is that the extended structure after reduction is in fact \emph{itself} an $SU(2)$ affine Kac-Moody algebra, this time at level $k=1$. As the $SU(2)_1$ model is one of the very simplest rational CFTs one may hope by considering the extended triplet algebra in $SU(2)_{-1}$ that this model should be a relatively simple example of a rational non-unitary CFT. It is not clear if this theory involves indecomposable representations or not. The four point correlators for the irreducible representations were all rational functions but further fusions may yield other representations.
\subsubsection{Correlators in the $c_{p,q}$ models}
We found for every $c_{p,q}$ model (we tested $p \le 5, q \le 5$) that there was always exactly three rational solutions for the $h_{1,4p-1}=(2p-1)(2q-1)$ fields. Rather more non-trivially if one exchanges $p$ and $q$ the differential equations are of a different order but the same set of three rational solutions solves both of them. In all the cases we studied these three rational solutions always had the same behaviour under crossing symmetry giving us the same triplet structure.

As we have already discussed the $c_{2,1}=c_{1,2}=-2$ case we shall begin with the first new example: the $c_{2,3}=0$ theory. In the $c_{2,3}=0$ model the solutions for the $h_{1,7}=15$ fields are given by:
\bea
F^{(1)}(z)&=& \f{z^{10}}{(1-z)^{28}} \left(357106464-2856851712 z+10509841628 z^2-23573986436 z^3 \right. \nonumber \\
&&+36044249670 z^4-39790427248 z^5+32773983814 z^6-20529517008 z^7 \nonumber \\
&&+9880147186 z^8-3667147120 z^9+1048374600 z^{10}-229634210 z^{11} \nonumber\\
&&\left.+38248769 z^{12}-4810728 z^{13}+452625 z^{14}-30294 z^{15}+1122 z^{16} \right)  \nonumber\\
F^{(2)}(z)&=& \f{1}{z^{28} (1-z)^{28} } \left( 2244-60588 z+905250 z^2-9621456 z^3+76497538 z^4 \right. \nonumber\\
&&-459268420 z^5+2096749200 z^6-7334294240 z^7+19760294372 z^8 \nonumber\\
&&-41059034016 z^9 +65547967628 z^{10}-79580854496 z^{11}+72088499340 z^{12} \nonumber\\
&&-36330724836 z^{13}-200733901482 z^{14}+2212292459088 z^{15}-14422439940116 z^{16} \nonumber\\
&&+68562493363130 z^{17}-254028569259777 z^{18}+763908934818536 z^{19}\\
&&-1917517271406737 z^{20}+4101816418782654 z^{21}-7599053781520630 z^{22}\nonumber\\
&&+12352604911298080 z^{23}-17809256023135980 z^{24}+22972890487011504 z^{25}\nonumber\\
&&-26689578674273868 z^{26}+28044134317298400 z^{27}-26689578674273868 z^{28}\nonumber\\
&&+22972890487011504 z^{29}-17809256023135980 z^{30}+12352604911298080 z^{31}\nonumber\\
&&-7599053781520630 z^{32}+4101816418782654 z^{33}-1917517271406737 z^{34}\nonumber\\
&&+763908934818536 z^{35}-254028569259777 z^{36}+68562493363130 z^{37}\nonumber\\
&&-14422439940116 z^{38}+2212292459088 z^{39}-200733901482 z^{40}\nonumber\\
&&-36330724836 z^{41}+72088499340 z^{42}-79580854496 z^{43}+65547967628 z^{44}\nonumber\\
&&-41059034016 z^{45}+19760294372 z^{46}-7334294240 z^{47}+2096749200 z^{48}\nonumber\\
&&\left. -459268420 z^{49}+76497538 z^{50}-9621456 z^{51}+905250 z^{52}-60588 z^{53}+2244 z^{54} \right)\nonumber \\
F^{(3)}(z)&=&  \f{(1-z)^{10}}{z^{28}} \left( 1122+12342 z+132855 z^2+1026528 z^3+5156450 z^4 \right. \nonumber\\
&&+17580680 z^5+42038555 z^6+70854550 z^7+83500300 z^8+70854550 z^9 \nonumber \\
&&+42038555 z^{10}+17580680 z^{11}+5156450 z^{12}+1026528 z^{13}\nonumber\\
&&\left.+132855 z^{14}+12342 z^{15}+1122 z^{16} \right) \nonumber
\eea
where we have again used the same conventions as before in the labelling of the $F^{(i)}$. Although the detailed form of the solutions is extremely complicated their structure is very simple. In an exactly analogous way to the arguments used for the extra indicial nature of the triplet fields in $c=-2$ we deduce that the triplet of dimension $h_{1,7}=15$ fields in $c_{2,3}=0$ behave as:
\bea
\left< W^+(0) W^+(z) W^-(1) W^-(\infty) \right> &=& F^{(1)}(z) \nonumber \\
\left< W^3(0) W^3(z) W^3(1) W^3(\infty) \right> &=& F^{(2)}(z) \\
\left< W^+(0) W^-(z) W^-(1) W^+(\infty) \right>&=&F^{(3)}(z) \nonumber
\eea
It is extremely difficult to read off the operator content from the rational correlation functions. However from some simple observations one may deduce that the algebra of the $W^a$ fields is closed. For example in $c_{2,3}=0$ the $h_{1,7}=15$ fields have the following BPZ \cite{Belavin:1984vu} fusion rule:
\bea
h_{1,7} \otimes h_{1,7} = h_{1,1}+h_{1,3}+ h_{1,5}+\cdots + h_{1,13}
\eea
As we know from their rational correlation functions the $h_{1,7}=15$ fields are chiral. Therefore we should only have contributions from the singular terms which requires operators with $h < 30$. Using the above fusion rules this leaves only \mbox{$h_{1,1}=0$},\mbox{$h_{1,3}=2$},\mbox{$h_{1,5}=7$},\mbox{$ h_{1,7}=15$},\mbox{$h_{1,9}=26$}. We have seen from its correlator that the $h_{1,3}=2$ field is just the stress tensor $T$ which is just a Virasoro descendent of the $h_{1,1}=0$ identity field. We shall also see in the next section that the $h_{1,5}=7$ field is fermionic due to the behaviour of its $2$ rational conformal blocks under crossing symmetry. In a similar way the $h_{1,9}=26$ operator has $4$ rational blocks and is also of fermionic form. Clearly the fusion of two bosonic operators cannot produce fermionic operators and therefore we have the OPE:
\bea
h_{1,7} \otimes h_{1,7}= h_{1,1} + h_{1,7}
\eea
and we conclude that the $W$-algebra of the $h_{1,7}=15$ fields in $c_{2,3}=0$ is in fact closed.  
\subsection{Doublet fields}
In addition to the triplet algebra we also found that if $p$ and $q$ were not both odd then there was also a doublet of rational solutions in the $c_{p,q}$ model generated by the $h_{1,3p-1}=(\f{3p}{2}-1)(\f{3q}{2}-1)$ fields. In the cases where these fields have integer dimension they behave fermionically and for half integer they are bosonic. If $p$ and $q$ are both odd then one still finds a simple doublet of solutions but they now have square root branch cuts and the fields are parafermions \cite{Fateev:1985mm}.

The simplest example is the $h_{2,1}=1$ and $h_{1,5}=1$ fields in the $c_{2,1}=-2$ model. As we have already commented on these in great detail we shall discuss the next case in the $c_{p,1}$ series namely $c_{3,1}=-7$ with $h_{2,1}=\f{7}{4}$. We then have conformal blocks:
\bea
F^{(1)}(z)&=&\f{1}{\sqrt{z(1-z)}} \f{(z-1)^2}{z^3} \lb 2 z^2+3z+2 \rb \\
F^{(2)}(z)&=& F^{(1)}(1-z) \nonumber
\eea
We see that both solutions have branch cuts in the complex plane. They are therefore not generated by a chiral algebra but instead by a parafermionic one \cite{Fateev:1985mm}. They still have the same form as before and so the same arguments can be made to deduce that this field has in fact a doublet nature:
\bea
\left< \Psi^+(0) \Psi^-(z) \Psi^-(1) \Psi^+(\infty) \right> &=& F^{(1)}(z)\\
\left< \Psi^+(0) \Psi^+(z) \Psi^-(1) \Psi^-(\infty) \right> &=& F^{(2)}(z) \nonumber
\eea
In $c_{2,3}=0$ we also find similar solutions for $h_{1,5}=7$ operators:
\bea
F^{(1)}(z)&=& \f{(1-z)}{z^{12}} \left(-22 z^9-44 z^8-323 z^7-859 z^6-1302 z^5-1302 z^4-859 z^3 \right. \nonumber \\
&& \left.-323 z^2-44 z-22 \right) \\
F^{(2)}(z)&=& F^{(1)}(1-z)\nonumber
\eea
We have checked many other $c_{p,q}$ models and always found similar doublet nature for the $h_{1,3p-1}=(\f{3p}{2}-1)(\f{3q}{2}-1)$ fields.
\subsubsection{Role of doublet sub-structure}
We know from standard Virasoro fusion rules \cite{Belavin:1984vu} that:
\bea \label{eqn:BPZfuse}
h_{2,1} \otimes h_{2,1} = \left[ h_{1,1} \right] + \left[ h_{3,1} \right] 
\eea
Now, as we have already explained, to create a rational $c_{p,1}$ triplet model we extend the chiral algebra by the $h_{3,1}=2p-1$ fields. Therefore with respect to the full chiral algebra all the terms on the RHS of (\ref{eqn:BPZfuse}) are descendents of the unit operator $h_{1,1}$. In the $c=-2$ case the $h_{2,1}=1$ field is normally denoted $\nu_1$ and one can see in the fusion rules (\ref{eqn:oldfusionrules}) we do indeed have a closed sub-algebra:
\bea
\nu_1 \otimes \nu_1 &=& \nu_0
\eea
where $\nu_0$ is the identity representation. We can now consider splitting the total space into equivalence classes under the action of $\nu_1$ in the following way: if there exists a non-negative integer $n$ (actually $n=0,1$ is enough) such that:
\bea
\nu_1^n \otimes X=Y 
\eea
where $\nu_1^n$ denotes the $n^{th}$ fusion product $\nu_1 \otimes \cdots \otimes \nu_1$ then we put $X$ and $Y$ into the same equivalence class i.e.~  $X  \cong Y $. In the $c=-2$ case using the fusion rules (\ref{eqn:oldfusionrules}) we find:
\bea
\CR_0  &\cong& \CR_1  \nonumber\\
\nu_0  &\cong&  \nu_1  \\
\nu_{-1/8}  &\cong&  \nu_{3/8} \nonumber
\eea
We can also easily obtain the fusion rules for the equivalence classes:
\bea
 \nu_0  \otimes  X  &=&  X  \quad \quad  {\rm for~ all ~ X } \nonumber\\
 \nu_{-1/8}  \otimes  \nu_{-1/8}  &=&  \CR_0  \\
 \nu_{-1/8}  \otimes  \CR_0  &=& 4  \nu_{-1/8}  \nonumber\\
 \CR_0  \otimes  \CR_0  &=& 4  \CR_0 \nonumber
\eea
We stress that these are fusion rules for the chiral theory and not the local one. They are considerably simpler than those of the full triplet model and are essentially the fusion rules of the symplectic fermion model \cite{Kausch:2000fu} as the representation $\CR_0$ is generated from the state $\omega$ (See Figure \ref{fig:indecomprepn}). The use of additional chiral algebras to simplify fusion rules is well known in normal CFT \cite{Moore:1989ss}. In general it seems that use of the simple fusion rules of the $h_{2,1}$ field allows one to effectively reduce the number of basic fields present in the triplet model from those of the $c_{3p,3}$ model to $c_{2p,2}$. It would be interesting to study this in more detail in other examples where we found similar doublet algebras.
\subsection{General structure}
The structure of rational solutions and Bose/Fermi assignments to operators is very suggestive. It seems that when the operators had integer conformal weights an odd number of rational solutions corresponded to bosonic operators and and an even number to fermionic ones. The cases we studied all fit into the sequence $h_{1,2np-1}=(np-1)(nq-1)$ having $2n-1$ rational solutions where $n=1,2,3,\cdots$ for bosonic fields and $n=\f{3}{2},\f{5}{2},\cdots$ for fermionic fields.

If these are indeed the only chiral $h_{1,s}$ operators in the theory then we immediately see in the singular terms of the OPE of two $h_{1,4p-1}$ triplet fields we can only create $h_{1,2p-1}$ and $h_{1,4p-1}$ fields. The next possible bosonic field is at $h_{1,6p-1} > 2 h_{1,4p-1}$ and so lies beyond the singular terms in the chiral OPE. Therefore the triplet algebra should close as a $W$-algebra with a schematic OPE:
\bea
h^a_{1,4p-1} \otimes h^b_{1,4p-1} = g^{ab} \left[h_{1,2p-1} \right] + f^{ab}_c \left[h^c_{1,4p-1} \right]
\eea
where $g^{ab}$ and $f^{ab}_c$ are the metric and structure constants of $SU(2)$ and $[h]$ denotes an operator and all its Virasoro descendents. Recall that the $h_{1,2p-1}$ field is the vacuum null vector of the irreducible theory and so $[h_{1,2p-1}]=[1]$.

The appearance of extended chiral algebras generated by the integer dimension primary fields also occurs within the integrable sector of certain ordinary $SU(2)_k$ models for $k \in N$. The extended algebras and their rational correlation functions are in exact coincidence \cite{MST,ST} with the A-D-E classification of $\widehat{SU(2)}$ modular invariants \cite{Cappelli:1987xt}. These lead to a similar classification for the minimal models \cite{Cappelli:1987hf}. Considering fields beyond the minimal sector leads, as we have seen, to other, more complicated, extended algebras and one may suspect that there is a similar classification of modular invariants for LCFTs. This would be very interesting to investigate further.
\chapter{LCFT as a limit}

\section{The moduli space of CFTs}
It has often been commented in the literature that logarithmic theories can be approached in a smooth way from non-logarithmic theories. This statement is somewhat misleading and we hope to clarify some aspects of this.

We shall analyse the approach to local logarithmic CFTs in two particular cases. The first is the approach to the $c=-2$ triplet model where we shall discuss certain subtelties in the limiting procedure. The second is the case of $c=0$ theories and the appearance of a logarithmic partner for the stress tensor.
\subsection{Approaching the $c=-2$ triplet model}
We shall first analyse the appearance of an indecomposable representation and then examine a situation in which operators may have extended indices. We shall work with the well known $c_{2,1}=-2$ model as the operator content is particularly well known.

The first correlator that we shall analyse is the original one studied by Gurarie \cite{Gurarie:1993xq} for the $h_{1,2}=-\f{1}{8}$ operators:
\bea
\left< \mu(z_1,\bar{z}_1) \mu(z_2,\bar{z}_2) \mu(z_3,\bar{z}_3) \mu(z_4,\bar{z}_4) \right> = |z_{13} z_{24} |^{1/2} |z(1-z)|^{1/2} G(z,\bar{z}) 
\eea
As we have already seen in (\ref{eqn:cminus2singval}) the unique single-valued combination is:
\bea \label{eqn:cminus2singvalagain}
G(z,\bar{z})= K(z) \overline{K(1-z)} + K(1-z) \overline{K(z)} 
\eea
where $K(z)$ is the hypergeometric function: $_2F_1\left( \half,\half;1;z \right)$. This correlator is invariant under all exchanges of operators. We shall now discuss how the logarithmic singularities emerge as $c \rightarrow -2$. We shall examine the $c_{k+2,1}$ model in the limit as $k$ approaches zero. The central charge of $c_{k+2,1}$ model is given by (\ref{eqn:CpqKactable}):
\bea
c=13-6 \left( k+2 + \f{1}{k+2} \right) = -2 -\f{9 k}{2} +O(k^2)
\eea
The first few operators in the Kac-table have dimensions:
\bea
h_{1,1}&=&0 \nonumber\\
h_{1,2}&=&\f{3}{4(k+2)}-\half = -\f{1}{8} -\f{3 k}{16} + O(k^2) \\
h_{1,3}&=&\f{2}{k+2}-1= -\f{k}{2} + O(k^2)\nonumber
\eea
At the point $k=0$ we have $h_{1,3}=h_{1,1}=0$ but for generic values of $k$ there is no degeneracy in the levels. Now:

\bea
\left< h_{1,2} (z_1,\bar{z}_1) h_{1,2} (z_2,\bar{z}_2) h_{1,2} (z_3,\bar{z}_3) h_{1,2} (z_4,\bar{z}_4) \right> = |z_{13} z_{24} |^{-4 h} |z|^{\f{2k+1}{k+2}} |1-z|^{\f{1}{k+2}} G(z,\bar{z}) \nonumber
\eea
where:
\bea
G(z,\bar{z})=\sum_{i,j=1}^{2} U_{i,j} F_i(z) \overline{F_j(z)} 
\eea 
and the conformal blocks $F_i(z)$ are found by solving differential equations or via the Coulomb gas approach. They are:
\bea
F_1(z)&=&  \f{\Gamma \left(\f{k+1}{k+2}\right) \Gamma \left(\f{k+1}{k+2}\right)}{\Gamma \left(\f{2k+2}{k+2}\right) }   {}_2F_1 \left( \f{1}{k+2}, \f{k+1}{k+2};\f{2k+2}{k+2};z \right) \\
F_2(z)&=& z^{\f{-k}{k+2}} \f{\Gamma \left(\f{1-k}{k+2}\right) \Gamma \left(\f{k+1}{k+2}\right)}{\Gamma \left(\f{2}{k+2}\right) }   {}_2F_1 \left( \f{1}{k+2},\f{1-k}{k+2};\f{2}{k+2};z \right) \nonumber 
\eea
We have included the normalisations so that we can use standard results. The solutions $F_1(z)$ and $F_2(z)$ are the conformal blocks for the contributions from the $h_{1,1}$ and $h_{1,3}$ operators respectively as can be seen from the leading behaviour as $z$ approaches zero. However we also see in the limit as $k \rightarrow 0$ that these two solutions become identical.

The full correlator must of course be single-valued everywhere. Monodromy around $z=0$ leads to the requirement that:
\bea \label{eqn:monod}
U_{1,2}e^{2 \pi i \f{k}{k+2}}= U_{1,2} \\
U_{2,1}e^{-2 \pi i \f{k}{k+2}}= U_{2,1}\nonumber
\eea
Now for the case of generic values of $k$ we have $\f{k}{k+2} \notin Z$ and we conclude $U_{1,2}=U_{2,1}=0$ and so the correlator must be diagonal:
\bea
G(z,\bar{z})=U_{1,1} |F_1(z)|^2 + U_{2,2} |F_2(z)|^2
\eea
Now imposing the monodromy around $z=1$ leads to the condition \cite{Dotsenko:1984nm}:
\bea
\f{U_{1,1}}{U_{2,2}}=\f{\sin \pi(a+b+c) \sin \pi b }{\sin \pi a \sin \pi c}
\eea
where $a=\f{-2k-1}{k+2}~,~~b=c=\f{-1}{k+2}$.
Expanding this in the limit $k \rightarrow 0$ we get:
\bea
\f{U_{1,1}}{U_{2,2}}=-1 + O(k^2) 
\eea
Therefore:
\bea \label{eqn:mumumumu}
G(z,\bar{z})=U_{1,1} \left( |F_1(z)|^2 - |F_2(z)|^2 \right)
\eea
The minus sign is absolutely crucial. It signifies that we have negative norm states close to $c=-2$. Logarithms can occur when these are cancelled to leading order by the positive norm states. Expanding $F_1$ and $F_2$ gives:
\bea
F_1(z) = 2 K(z) + k C(z)  \\
F_2(z) = 2 K(z) +k D(z)
\eea
where:
\bea
C-D=- \pi K(1-z)
\eea
In order to make the full correlator non-vanishing in the limit $k \rightarrow 0$ we have to choose the overall rescaling of the four point function (\ref{eqn:mumumumu}) to be $U_{1,1} \sim \f{1}{k}$. With this choice we find that we have a smooth limit as $k \rightarrow 0$.
\bea
G(z,\bar{z})&=&\f{1}{k} \left( (2 K +kC)\overline{(2 K + kC )} - (2 K+kD)\overline{( 2 K +k D )} \right) \\
&&\rightarrow -2 \pi \left[ K(z) \overline{K(1-z)} + K(1-z) \overline{K(z)} \right] \nonumber
\eea
This is, up to normalisation, the same result as was obtained at the limiting $c=-2$ point (\ref{eqn:cminus2singvalagain}). In this case we were able to get a smooth approach to a logarithmic correlator from a non-logarithmic one. One might be therefore tempted to think that LCFT is merely some continuous limit of ordinary CFT. However we shall soon see that this is \emph{not} always the case.

To illustrate this we shall examine the correlator:
\bea \label{eqn:correl2233}
\left< h_{1,2} (z_1,\bar{z}_1) h_{1,2} (z_2,\bar{z}_2) h_{1,3} (z_3,\bar{z}_3) h_{1,3} (z_4,\bar{z}_4) \right> &=&  |z_{34}|^{4h_{1,2}-4h_{1,3}} |z_{24}|^{-4h_{1,2}}|z_{13}|^{-4h_{1,2}} \nonumber\\
&& ~~~ |z|^{\f{2k+1}{k+2}} |1-z|^{\f{2}{k+2}} G(z,\bar{z}) 
\eea
Evaluating this correlator for the $c=-2$ theory (i.e $k=0$) we get two solutions:
\bea
\CF_1(z)&=& (1-z)^{-1/2} \\
\CF_2(z)&=& (1-z)^{-1/2} \arctan( \sqrt{z-1})\nonumber
\eea
These can also be obtained from the hamiltonian reduction of the solutions $F_{\half \half 1 1}(x,z)$ for $SU(2)_0$ given in (\ref{eqn:halfhalf11}). The function $\arctan( \sqrt{z-1})$ has the following behaviour near $z=0$ \footnote{This is most easily seen using:
\bea
\arctan(\sqrt{z-1}) = \int \f{1}{2z\sqrt{z-1}} ~ dz &=& \f{1}{2i} \int \left[ \f{1}{z} + \f{1}{2} + \f{3}{8}z + \cdots \right] dz \nonumber \\
&=& \f{1}{2i} \ln z + {\rm regular} \nonumber 
\eea}:
\bea
\arctan(\sqrt{z-1}) \sim  - \f{i}{2} \ln z + {\rm regular}
\eea
The full single-valued correlator (\ref{eqn:correl2233}) must be single valued and we find \emph{two} possible solutions:
\bea
G(z,\bar{z})=U_{1,1}\CF_1(z) \overline{\CF_1(z)} + U_{1,2} \left( \CF_1(z) \overline{\CF_2(z)} + \CF_2 \overline{\CF_1(z)} \right)
\eea
The solution with $U_{1,2}=0$ corresponds to the correlator:
\bea
\left< \mu (z_1,\bar{z}_1) \mu (z_2,\bar{z}_2) \Omega (z_3,\bar{z}_3) \Omega (z_4,\bar{z}_4) \right> =  |z_{12}|^{1/2} \nonumber
\eea
The other solution with logarithmic terms corresponds to the correlator:
\bea
\left< \mu (z_1,\bar{z}_1) \mu (z_2,\bar{z}_2) \Theta^+ (z_3,\bar{z}_3) \Theta^- (z_4,\bar{z}_4) \right> =  |z_{12}|^{1/2} \left( \arctan( \sqrt{z-1}) + \overline{\arctan( \sqrt{z-1})} \right)  \nonumber
\eea
where $\Omega(z,\bar{z})$ is the normal vacuum and $\Theta^{\pm}(z,\bar{z})$ are the non-chiral fermionic $h=0$ operators that we have already discussed.

We shall now analyse the correlator (\ref{eqn:correl2233}) in the $c_{k+2,1}$ theory. For any value of $k$ we can again solve to find the conformal blocks. They are:
\bea
F_1(z)&=& \f{\Gamma \left(\f{k+1}{k+2}\right) \Gamma \left( \f{k+1}{k+2}\right)}{\Gamma \left(\f{2k+2}{k+2}\right) }  {}_2F_1 \left( \f{2}{k+2}, \f{k+1}{k+2};\f{2k+2}{k+2};z \right) \\
F_2(z)&=& z^{\f{-k}{k+2}} \f{\Gamma \left(\f{2-k}{k+2}\right) \Gamma \left(\f{k}{k+2}\right)}{\Gamma \left(\f{2}{k+2}\right) }  {}_2F_1 \left( \f{1}{k+2},\f{2-k}{k+2};\f{2}{k+2};z \right)
\eea
These now have the leading forms:
\bea
F_1(z) &=& \pi \CF_1 + O(k) \\
F_2(z) &=& \f{2 \CF_1}{k} + O(1)  \nonumber 
\eea
Again for generic values of $k$ we must have the diagonal correlator:
\bea \label{eqn:newcorrel}
G(z,\bar{z})= U_{2,2} \left\{ \f{U_{1,1}}{U_{2,2}} |F_1|^2 + |F_2|^2 \right\}
\eea
Now imposing monodromy around $z=1$ we find:
\bea
\f{U_{1,1}}{U_{2,2}}= - \f{2}{ \pi^2 k^2} -\f{2}{\pi^2 k} + O(1) 
\eea
We therefore see that, in order to have a well defined limit in (\ref{eqn:newcorrel}), we must take $U_{2,2} \sim k^2$ but then we find:
\bea
G(z,\bar{z}) \rightarrow |\CF_1(z)|^2\nonumber
\eea
Therefore we see that in the limit of the correlator we do \emph{not} find the second solution $\CF_2(z)$ corresponding to operators $\Theta^{\pm}(z,\bar{z})$. We now have a rather interesting puzzle. For $k \ne 0$ we have no degeneracy and get a unique correlator. However at the point $k=0$ we have a \emph{choice} of two different correlators coming from the extra $h=0$ operators $\Theta^{\pm}(z,\bar{z})$. The fundamental reason for this is that the moduli space of solutions to the monodromy constraints: 
\bea
U_{1,2}e^{2 \pi i \f{k}{k+2}}= U_{1,2}\nonumber
\eea
is not smooth as a function of $k$. We see that the condition is trivial if $\f{k}{k+2} \in Z \Leftrightarrow h_{1,3}-h_{1,1} \in Z$. It is exactly in the cases in which conformal dimensions differ by integers, and we may get logarithms, that the monodromy constraints break down.

This conclusion is applicable to any conformal field theory in which one has an extended multiplet structure at a certain point. The limit of the correlators is not necessarily the same as solving the theory at the limiting  point. It would be particularly interesting to analyse this in the context of disordered systems which can be studied in the replica limit or using the super-symmetric approach \cite{Bhaseen:2000mi}.
\section{$c=0$ Logarithmic CFT}
We shall now analyse another situation in which logarithmic CFT naturally emerges - namely in $c=0$ CFTs. This has been previously discussed in \cite{Cardy,Gurarie:1999yx,CardyTalk}. As our discussion will be limited to the two and three point functions it will naturally be less complete than the previous section. However we hope that it may nevertheless describe some generic features.

The general problems encountered in $c=0$ models are very similar to those of $SU(2)_0$. The stress tensor is a Virasoro primary and therefore a potential null vector. Of course if we decouple this then all states in the theory would obey:
\bea
L_{0} \state{\phi}=0 \\
L_{-1} \state{\phi} =0 \nonumber 
\eea
and we would be left with a purely topological sector. To go beyond this trivial sector one must introduce (or produce in fusion) a field, which we denote as $t$, which stops $T$ decoupling. 

It is easy to see that in $c=0$ theories if $t$ is a descendent of any fields $\left.|A\right>,\left.|B\right>,\left.|C\right>,...$ then it cannot prevent $T$ from decoupling. For general $c$ we have:
\bea
\left<T|t\right>&=&\left<0|L_{2}|\right. \left\{ L_{-1}\left.|A\right> + L_{-2}\left.|B\right> + L_{-3}\left.|C\right> + \cdots \right\} \nonumber \\
&=&\left<0|\right. \left\{ 3L_1 \left.|A\right> + \left(4L_{0}+\f{c}{2} \right) \left.|B\right>+ 5L_{-1}\left.|C\right> + \cdots \right\} \\
&=& \f{c}{2} \left< 0|B \right> \nonumber
\eea
where we do not use any properties of the states  $\left.|A\right>,\left.|B\right>,\left.|C\right>$ only the Virasoro algebra (\ref{eqn:Virasoroalg}) and the conjugate of the vacuum (\ref{eqn:vacuumdefn}) namely:
\bea
\left<0|\right. L_{n} =0 ~~ n \le 1 
\eea
We thus conclude that in the $c=0$ theories there must be a field of dimension $(2,0)$, which is not a pure descendent, which is responsible for the non-decoupling of $T$. Of course there may also be a part of the $t$ field which is a descendent but this is not essential\footnote{Note that in LCFT, as we have indecomposable representations, a non-descendent field is not necessarily a primary in the sense of (\ref{eqn:primary}).}. This is in contrast to the situation, for general $c$, discussed in \cite{Moghimi-Araghi:2000qn,Moghimi-Araghi:2000dt} in which $t$ was assumed to be a pure descendent. 

In the following we shall see how this field $t$ emerges in the limit as $c \rightarrow 0$. We begin by considering the two point function of a primary field of conformal dimension $(h(c),\bar{h}(c))$. By conformal invariance one knows this must be of the form:
\bea
\left< V(z_1,\bar{z}_1) V(z_2,\bar{z}_2) \right> = \f{A(c)}{z_{12}^{2h}\bar{z}_{12}^{2 \bar{h}}}
\eea
Then  we consider the correlator with an insertion of the stress tensor:
\bea \label{eqn:TVV}
\left< T(z) V(z_1,\bar{z}_1) V(z_2,\bar{z}_2) \right> = \f{A(c)~h}{(z-z_1)^2(z-z_2)^2 z_{12}^{2h-2}\bar{z}_{12}^{2 \bar{h}}}
\eea
The coefficient of the three point function is uniquely fixed by considering the limit $z \rightarrow z_1$ and using the property of a primary field:
\bea
T(z) V(z_1,\bar{z}_1) \sim \f{h V(z_1,\bar{z}_1)}{(z-z_1)^2}+\f{\p V(z_1,\bar{z}_1)}{z-z_1} + \cdots
\eea
Of course we have similar results following from insertions of $\bar{T}(\bar{z})$ in the correlator and in the following we take $h=\bar{h}$ for simplicity. We shall also make the same simplifying assumption as for $SU(2)_0$ namely that the identity field is \emph{not} a zero-norm state and so one can normalise $\la \Omega(z) \ra=1$. We now have:
\bea
\la T(z) T(0) \ra &=& \f{c}{2 z^4}
\eea
We shall also assume $c=\bar{c}$. If $T$ is the only $(2,0)$ field present in our model then from (\ref{eqn:TVV}) we deduce:
\bea \label{eqn:OPEVV}
V(z,\bar{z}) V(0,0) \sim \f{A(c)}{z^{2h}\bar{z}^{2 h}} \left[ 1 +  \f{2h}{c} z^2 T(0) + \f{2 \bar{h}}{c} \bar{z}^2 \bar{T}(0) +  \cdots \right]
\eea
Clearly for $c=0$ the above OPE may not be well defined. There are several remedies to this:
\begin{itemize}
\item $h(c) \rightarrow 0$
\item $A(c) \rightarrow 0$ 
\item More $(2,0)$ operators exist
\end{itemize}
For non-trivial theories some operators would have $h \ne 0$ and therefore we shall concentrate our attention on the case in which other operators with dimension $(2,0)$ emerge and comment on other possibilites later.

There are clearly many possible ways to create other operators of dimension $(2,0)$ at the limiting point and for simplicity we shall discuss only limiting cases of primary operators. Suppose that near to $c=0$ there are other local primary fields $X(z,\bar{z})$ and $\bar{X}(z,\bar{z})$, of dimensions $(2+\alpha(c),\alpha(c))$ and $(\alpha(c),2+\alpha(c))\,$ respectively, that converge in the limit to $(2,0)$ and $(0,2)$ fields. Then for $c \ne 0$ we must modify the previous expression (\ref{eqn:OPEVV}):
\bea \label{eqn:OPE}
V(z,\bar{z}) V(0,0) \sim \f{A(c)}{z^{2h}} \left[ 1 +  \f{2h}{c} z^2 T(0)
+ 2 X(0,0) z^{2+\alpha(c)} \bar{z}^{\alpha(c)} + \cdots \right]
\eea
It is simplest to analyse this by insisting that the correlation functions are well defined in the limit as we approach $c=0$. The only non-trivial 2-point correlators (up to relation by conjugation) are:
\bea \label{eqn:XX}
\left< T(z_1) T(z_2) \right> &=& \f{c}{2 z_{12}^4} \\
\left< X(z_1,\bar{z}_1) X(z_2,\bar{z}_2) \right>&=& \f{1}{c}\,\f{B(c)}{z_{12}^{4+2\alpha(c)}\bar{z}_{12}^{2 \alpha(c)}}
\eea
and $\left<T(z_1) X(z_2,\bar{z}_2) \right>$ vanishes as they have different dimensions. We define the new fields $t$, $\bar{t}$ via:
\bea
t= \f{b}{c}T+\f{b}{h} X, ~
 \hskip 1 cm \bar{t}= \f{b}{c} \bar{T}+\f{b}{h} \bar{X}
\eea
The parameter $b$ is defined through:
\bea \label{eqn:bdef}
b^{-1} \equiv -\lim_{c \rightarrow 0} \f{\alpha(c)}{c}=-\alpha'(0)
\eea
We can now calculate the two point function:
\bea
\left< T(z_1) t(z_2,\bar{z}_2) \right> &=& 
\left< T(z_1) \left[ \f{b}{c} T + \f{b}{h}  X \right](z_2,\bar{z}_2)
\right>  
 = \f{b}{c} \left< T(z_1) T(z_2) \right> \nonumber \\
&=& \f{b}{2}\f{1}{ z_{12}^4}
\eea
We assume that $b \ne 0$ as otherwise we have not succeeded in constructing a field which prevents $T$ from decoupling and we would have to have more fields with dimensions $(2,0)$.
\bea
\left< t(z_1,\bar{z}_1) t(z_2,\bar{z}_2) \right> &=& 
 \left< \left[ \f{b}{c} T + \f{b}{h}  X \right](z_1,\bar{z}_1) \left[ \f{b}{c} T + \f{b}{h}  X \right](z_2,\bar{z}_2) \right> \nonumber \\
&=& \f{b^2}{c^2} \left< T(z_1) T(z_2) \right> +
 \f{b^2}{h^2} \left< X(z_1,\bar{z}_1) X(z_2,\bar{z}_2) \right> \nonumber \\
&=&\f{b^2}{2c}\f{1}{z_{12}^4}+\f{b^2 B(c)}{h^2 c} \f{1}{z_{12}^{4+2\alpha(c)} \bar{z}_{12}^{2 \alpha(c)}}  \\
&=&\f{1}{z_{12}^4} \left\{ \left( \f{b^2}{2c}+\f{b^2 B(c)}{h^2 c} \right) - \f{2b^2 B(c)\alpha(c)}{h^2c}\ln |z_{12}|^2 + \cdots \right\} \nonumber
\eea
As this is to be well defined we see that we must have $B(c)\!=\!-\f{1}{2} h^2\!+\!B_1c\! +\!O(c^2) $. Now using (\ref{eqn:bdef}) we get the standard OPEs for a logarithmic pair:
\bea
\left< T(z_1) T(z_2) \right>&=&0 \nonumber \\
\left< T(z_1) t(z_2,\bar{z}_2) \right>&=& \f{b}{2 z_{12}^4} \\
\left< t(z_1,\bar{z}_1) t(z_2,\bar{z}_2) \right>&=& 
\f{B_1 -b\ln |z_{12}|^2}{z_{12}^4} \nonumber
\eea
The constant $B_1$ can  be removed by a redefinition of $t$ and we shall assume that this has been done. Note that although $t$ is a $(2,0)$ field it is not chiral as $\bar{\p} t \ne 0$. We have similar expressions for the correlation functions related to these by conjugation. Also:
\bea
\left< T(z_1) \bar{t}(z_2,\bar{z}_2) \right>&=& \left< T(z_1) \left[ \bar{T} \f{b}{c} +\f{b}{h} \bar{X} \right] (z_2,\bar{z}_2) \right>=0 \\
\left<  t(z_1,\bar{z}_1) \bar{t}(z_2,\bar{z}_2)\right>&=& \left<  \left[ T \f{b}{c} +\f{b}{h} X \right] (z_1,\bar{z}_1) \left[ \bar{T} \f{b}{c} +\f{b}{h} \bar{X} \right] (z_2,\bar{z}_2) \right>=0
\eea
The OPE (\ref{eqn:OPE}) now becomes:
\bea 
V(z,\bar{z}) V(0,0) \sim \f{A(0)}{z^{2h}} \left[ 1\! + \!
 \f{2h}{b} z^2 \left( T  \ln |z|\!+\! t \right)\! + \! \f{2h}{b} \bar{z}^2 
\left( \bar{T} \ln |z|\!+\! \bar{t} \right)\! +\! \cdots \right]
\label{logOPE}
\eea
which now only involves quantities that are perfectly well defined in the limit as $c \rightarrow 0$. We can now continue and insist that $t$ is also well defined in the three point functions (See Appendix \ref{chap:3ptfns}). These now involve more arbitrary constants. Assuming that the algebra closes these are then sufficient to determine the OPEs. These are:
\bea
&& T(z_1) t(z_2,\bar{z}_2) 
\sim \f{b}{2 z_{12}^4} + \f{2 t(z_2,\bar{z}_2) +
T(z_2)}{z_{12}^2} + \f{\p t(z_2,\bar{z}_2) }{z_{12}}+ \cdots \\
%
&& T(z_1) \bar{t}(z_2,\bar{z}_2)  \sim
 \f{\bar{T}(\bar{z}_2)}{z_{12}^2} +
\f{\p \bar{t}(z_2,\bar{z}_2)}{z_{12}}+ \cdots \\
%
&& t(z_1,\bar{z}_1) t(z_2,\bar{z}_2) \sim   \f{-b \ln|z_{12}|^2}{z_{12}^4} + \f{1}{z_{12}^2} 
\Bigg[ \left( 1-4 \ln |z_{12}|^2 \right) t(z_2,\bar{z}_2)\nonumber\\ 
&&  
\hspace*{2.5cm} +\left( \f{2a}{b} - \ln|z_{12}|^2 -2\ln^2 |z_{12}|^2 \right) T(z_2)
\Bigg]   \\
&&\hspace*{2.5cm}+ \,\f{\bar{z}_{12}^2}{z_{12}^4} \left[ \bar{t}(z_2,\bar{z}_2) + \left( \f{2f}{b} + \ln|z_{12}|^2 \right) \bar{T}(\bar{z}_2) \right] + 
\cdots  \nonumber\\
%
&&t(z_1,\bar{z}_1) \bar{t}(z_2,\bar{z}_2) \sim \f{1}{\bar{z}_{12}^2} \left[ \left( \f{2f}{b} - \ln|z_{12}|^2 \right) T(z_2) + \cdots \right] \\
&&\hspace*{2.5cm}+\,\f{1}{z_{12}^2} \left[ \left( \f{2f}{b}-\ln|z_{12}|^2 \right) \bar{T}(\bar{z}_2) + \cdots \right]  \nonumber 
\eea
The appearance of a state $\left.|t\right> = t\left.|0\right> $ in this way is equivalent to postulating a logarithmic partner for the null vector $T$ \cite{Rohsiepe:1996qj}. This prevents $T$ from decoupling despite the fact that it is a zero-norm state.
Note that once one fixes:
\bea
L_{0}\left.|t\right>=2\left.|t\right>+\left.|T\right>\, \hspace{20mm}
\bar{L}_0 \left.|t\right>=\left.|T\right>
\eea
then the parameter $b$ cannot be removed by rescaling and thus different values of $b$ correspond to inequivalent representations. The parameter $b$ in our notation is different by a factor of two from a definition given in \cite{Gurarie:1999yx}. As the $t(z,\bar{z})$ operator is non-chiral the singular terms are not sufficient to reconstruct the OPEs and we require more than one parameter (infinitely many in general) to describe them fully.

As we commented earlier there is another way in which one may avoid the $c=0$ catastrophe. If $A(c)$ in (\ref{eqn:OPE}) also vanishes in the limit then this may cancel the divergence. This is exactly what occurs in the Kac-Moody theories at level zero. For clarity we shall only discuss here $SU(2)_0$ although arguments are identical for general Lie groups. We have the OPE:
\bea
J^a(z) J^b(w) \sim \f{k g^{ab}}{(z-w)^2} + \f{i f^{ab}_c J^c(w)}{z-w}
\eea
and the central charge is given by:
\bea
c=\f{3k}{k+2}
\eea
Thus for small $c$ we have $k=A(c)=2c/3$. The divergent term $1/c$ in front of $T$ in (\ref{eqn:OPE}) is exactly cancelled leaving a finite result. This is expected because the Sugawara stress tensor is still perfectly well defined in these theories.

A similar argument to that given previously shows that for $SU(N)_0$ there must be a non-descendent $(1,0)$ field to prevent $J^a$ decoupling. For $SU(2)_0$ this is precisely the field $N^a$ that we discussed earlier (\ref{eqn:JNOPE}):
\bea
J^a(z) N^b(w) \sim \f{\delta^{ab}}{(z-w)^2}+\f{i f^{abc}N^c(w)}{z-w}
\eea
However in general there will be other operators whose two point functions are non-vanishing for example $K(x,z)$ in (\ref{eqn:KKOPE}). Then by the previous arguments we must have:
\bea
K^a(z) K^a(0) \sim \f{1}{z^2} \left[ 1 +  \f{2}{b} z^2 (\ln |z| T(0) + t) + \cdots \right]
\eea
We see that it is possible by examining the $O(1)$ and $O(\ln |z|)$ terms to read off the operators $T$ and $t(z,\bar{z})$. It then remains to compute their OPE and find the value of $b$ that is realised in this system. Using so-called `pre-logarithmic' currents to generate partners of the stress tensor was further investigated in \cite{Kogan:2001ku}.
\subsection{$c=0$ and separability}
There is a third rather trivial way out of the paradox at $c=0$. It is simply that the full theory is constructed from two separate parts $T=T_1 \oplus T_2,~c=c_1+c_2=0$ both having $c_i \ne 0$. Operators in the full theory are just the direct product $V=V_1 \otimes V_2$. Therefore in the OPE of two fields from one part we will obviously only see the stress tensor for that part rather than the full one:
\bea
&&V(z) V(0) = V_1(z) V_1(0) \otimes V_2(z) V_2(0)  \\
&&\sim  \f{1}{z^{2h_1}} \left( 1+ z^2 \f{2h_1}{c_1}T_1(0) + \cdots
\right) 
 \otimes \f{1}{z^{2h_2}} \left( 1+ z^2 \f{2h_2}{c_2}T_2(0) + \cdots \right)
\nonumber
 \\[1mm]
&&\sim \f{1}{z^{2h}} \left[ 1+ z^2 \left( \f{2h_1 T_1}{c_1} 
+ \f{2h_2 T_2}{c_2} \right) +\cdots \right] \nonumber 
\eea
This expression is now perfectly well defined as $c_1,c_2 \ne 0$. Of course this is as expected as the two decoupled theories are perfectly regular.

In critical string theory the ghost and matter sectors are normally assumed to be non-interacting. However this is not the most general if we allow not just positive but also zero norm states in our final theory \cite{SUSY30}.

\chapter{Conclusion}

In this thesis we have analysed particular examples of logarithmic conformal field theories. We have seen that by careful study of the correlation functions one may deduce the presence of extended algebras and extra quantum numbers carried by the fields. We believe that the concepts and arguments presented here are very general.

In chapter 2 we analysed correlation functions of discrete representations of $SU(2)$ using of the \KZ equation. In the $SU(2)_0$ example we saw the first appearance of indecomposable representations when fusing two $j=\half$ operators. At $j=1$ we observed the first instance of a rational solution corresponding to the $\widehat{SU(2)}$ chiral algebra itself. This closed subset, on which one may perform the conformal bootstrap, was related to factorisation of the \KZ equations and null vectors. We found that the other solutions corresponded to non-chiral fermionic operators.

In chapter 3 we studied further the rational solutions in $SU(2)_0$ and saw that these were in exact correspondence with the possible extensions of the chiral algebra. They had a very simple multiplet structure and Bose/Fermi assignments that were all deducible from the correlation functions. As further evidence we presented an explicit free field construction allowing direct calculation of the OPEs.

In chapter 4 we used the powerful technique of hamiltonian reduction, in both the free field representation and at the level of correlation functions, to relate $\widehat{SU(2)}$ and $c_{p,q}$ models. This procedure gave us a practical approach to study certain fields, namely the $h_{1,s}$ ones, in the extended $c_{p,q}$ models. We used this explicitly in the $SU(2)_0$ example where it reduced to the well known $c=-2$ model. The multiplet and extended chiral structure is similar in both models.  In the $c_{1,1}=1$ case we were again able to directly construct the extended algebra in the free field approach and hence verify the deductions made from the correlation functions. We found that there is a very simple, and generic, structure to many of the extended algebras in $\widehat{SU(2)}$ and $c_{p,q}$ models which would be interesting to understand more fully.

In chapter 5 we discussed some of the subtleties involved in the approach to LCFTs. We saw that although some correlators in LCFT can be obtained in a smooth manner from non-logarithmic theories there are others which could not. This was related to the appearance of rational solutions and extended multiplet structure in the theory. We also discussed non-trivial $c=0$ theories and presented some general arguments that require the existence of a logarithmic partner to the stress tensor $T$. We then studied some of the properties of such a field.

A major aspect which we have not touched upon is the constraints placed on these theories on higher genus Riemann surfaces and in particular the torus. This is an essential task as the action of the modular group on $\widehat{SU(2)}$ LCFTs will almost certainly force us to consider representations other than the finite dimensional primary ones considered here. The full desciption of these and the indecomposable representations remains the main outstanding issue.

\appendix
\chapter{Rational correlation functions for $j=3$ in $SU(2)_0$}
%
The correlation functions for $j_1=j_2=j_3=j_4=3$ in $SU(2)_0$ are:
{\tiny
\bea \label{eqn:jthree}
F_{3333}^{(1)}(x,z)&=&\f{1}{462(z-1)^{11}}z^3 \left\{ (-252z+756z^2-910z^3+560z^4-190z^5+36z^6-3z^7)\right. \nonumber \\ 
&&+(-2520+9576z-14532z^2+11340z^3-4860z^4+1140z^5-138z^6+6z^7)x\nonumber\\
&&+(6300-21000z+27510z^2-18000z^3+6150z^4-1050z^5+75z^6)x^2\nonumber\\
&&+(-5600+16240z-17840z^2+9200z^3-2200z^4+200z^5)x^3\nonumber\\
&&+(2100-5010z+4200z^2-1425z^3+150z^4)x^4\nonumber \\
&&\left.+(-300+516z-258z^2+30z^3)x^5+(10-8z+z^2)x^6 \right\}\nonumber
\\
F_{3333}^{(2)}(x,z)&=&\f{1}{231z^{11} (z-1)^{11}} \nonumber\\
&&\left\{ (-z^5+8z^6-10z^7-70z^8+455z^9-1456z^{10}+ 3003z^{11}-4290z^{12}+4290z^{13} \right.\nonumber\\
&&-2860z^{14}+875z^{15}+560z^{16}-910z^{17}+560z^{18}-190z^{19}+36z^{20}-3z^{21})\nonumber\\
 &&+ (-30z^4+258z^5-516z^6-1380z^7+11970z^8-40950z^9+88998z^{10}\nonumber\\
&&-134706z^{11}+145860z^{12}-111540z^{13}+54510z^{14}-7770z^{15}-12180z^{16}\nonumber\\
&&+11340z^{17}-4860z^{18}+1140z^{19}-138z^{20}+6z^{21})x\nonumber\\
&&+(-150z^3+1425z^4-4200z^5-870z^6+46200z^7-184275z^8+436800z^9\nonumber\\
&&-718575z^{10}+858000z^{11}-750750z^{12}+471900z^{13}-194775z^{14}+30450z^{15}\nonumber\\
&&+21630z^{16}-18000z^{17}+6150z^{18}-1050z^{19}+75z^{20})x^2\nonumber\\
&&+(-200z^2+2200z^3-9200z^4+13920z^5+26880z^6-197400z^7+564200z^8\nonumber\\
&&-1058200z^9+1430000z^{10}-1430000z^{11}+1058200z^{12}-564200z^{13}\nonumber\\
&&+197400z^{14}-26880z^{15}-13920z^{16}+9200z^{17}-2200z^{18}+200z^{19})x^3\\
&&+(-75z+1050z^2-6150z^3+18000z^4-21630z^5-30450z^6+194775z^7-471900z^8\nonumber\\
&&+750750z^9-858000z^{10}+718575z^{11}-436800z^{12}+184275z^{13}-46200z^{14}\nonumber\\
&&+870z^{15}+4200z^{16}-1425z^{17}+150z^{18})x^4\nonumber\\
&&+ (-6+138z-1140z^2+4860z^3-11340z^4+12180z^5+7770z^6-54510z^7\nonumber\\
&&+111540z^8-145860z^9+134706z^{10}-88998z^{11}+40950z^{12}-11970z^{13}\nonumber\\
&&+1380z^{14}+516z^{15}-258z^{16}+30z^{17})x^5\nonumber\\
&&+(3-36z+190z^2-560z^3+910z^4-560z^5-875z^6+2860z^7\nonumber\\
&&\left.-4290z^8+4290z^9-3003z^{10}+1456z^{11}-455z^{12}+70z^{13}+10z^{14}-8z^{15}+z^{16})x^6 \right\}\nonumber
\\
F_{3333}^{(3)}(x,z)&=&\f{1}{462 z^{11}}(z-1)^3 \left\{ (-z^5-6z^6-3z^7)+(-30z^4-162z^5-54z^6+6z^7)x \right.\nonumber\\
&&+(-150z^3-675z^4+75z^6)x^2+(-200z^2-600z^3+600z^4+200z^5)x^3\nonumber\\
&&\left.+(-75z+675z^3+150z^4)x^4+(-6+54z+162z^2+30z^3)x^5+(3+6z+z^2)x^6 \right\}\nonumber
\eea}
\chapter{Three point functions}
\label{chap:3ptfns}
As in the main text we suppose that near to $c=0$ there are other local primary fields $X(z,\bar{z})$ and $\bar{X}(z,\bar{z})$, of dimensions $(2+\alpha(c),\alpha(c))$ and $(\alpha(c),2+\alpha(c))\,$ respectively, that converge in the limit to $(2,0)$ and $(0,2)$ fields. We now consider the three point functions of such fields:
\bea 
\label{eqn:OPEs}
\left< T(z_1) T(z_2) T(z_3) \right> &=& \f{c}{z_{12}^2 z_{13}^2 z_{23}^2 } \nonumber\\
\left< T(z_1) X(z_2,\bar{z}_2) X(z_3,\bar{z}_3) \right>
&=&\f{1}{c}\,\f{C(c)}{z_{12}^2 z_{13}^2 z_{23}^{2+2\alpha(c)}
\bar{z}_{23}^{2\alpha(c)} }\nonumber \\
\left< X(z_1,\bar{z}_1) X(z_2,\bar{z}_2) X(z_3,\bar{z}_3) \right> &=&
\f{1}{c^2}\f{D(c)}{z_{12}^{2+\alpha(c)} z_{13}^{2+\alpha(c)} 
z_{23}^{2+\alpha(c)} \bar{z}_{12}^{\alpha(c)}
 \bar{z}_{13}^{\alpha(c)} \bar{z}_{23}^{\alpha(c)} } \\
\left< T(z_1) \bar{X}(z_2,\bar{z}_2) \bar{X}(z_3,\bar{z}_3) \right> &=&
\f{E(c)}{z_{12}^2 z_{13}^2 z_{23}^{2 \alpha(c)-2}
\bar{z}_{23}^{4+2\alpha(c)}}\nonumber \\
\left< X(z_1,\bar{z}_1) X(z_2,\bar{z}_2) \bar{X}(z_3,\bar{z}_3) \right>
&=& \f{1}{c}\f{F(c)}{z_{12}^{4+\alpha(c)} z_{13}^{\alpha(c)}
z_{23}^{\alpha(c)} \bar{z}_{12}^{\alpha(c)-2} \bar{z}_{13}^{2+\alpha(c)} 
\bar{z}_{23}^{2+\alpha(c)} } \nonumber
\eea
We note that all correlators are single-valued for any $\alpha(c)$ and therefore must also be at the critical $c=0$ point. This is important as logarithmic terms should only emerge in the form $\ln|z|$.

We have already fixed the normalisation of the two point functions (\ref{eqn:XX}). Then by expanding the three point functions we see that:
\bea \label{eqn:known}
C(c)&=&(2+\alpha(c)) B(c)=(2+\alpha(c))\left(-\f{1}{2} h^2+B_2c^2 +\cdots\,\right) \\
E(c)&=&\f{\alpha(c)}{c} B(c)= \f{\alpha(c)}{c}  (-\f{1}{2} h^2+B_2c^2 
+\cdots ) 
\eea
As we wish to have well defined operators $T,~t,~\bar{T},~\bar{t}$ they must have regular 3-point functions. This will be enough to determine the leading behaviour of the functions above. Consider:
\bea
\left< T(z_1) T(z_2) t(z_3,\bar{z}_3) \right> = \left< T(z_1) T(z_2) \left[ \f{b}{c}T+\f{b}{h} X \right] (z_3,\bar{z}_3) \right> 
= \f{b}{z_{12}^2 z_{13}^2 z_{23}^2 }
\eea
%
%
\vspace{-3mm}
\bea
\hspace{-1cm}\left< T(z_1) t(z_2,\bar{z}_2) t(z_3,\bar{z}_3) \right>  
&=& \left< T(z_1) 
 \left[ \f{b}{c}T+\f{b}{h} X \right] (z_2,\bar{z}_2) \left[ \f{b}{c}T+\f{b}{h} X \right] (z_3,\bar{z}_3) \right> \nonumber\\
&=& 
\f{b^2}{c^2}\f{c}{z_{12}^2 z_{13}^2 z_{23}^2 } +
\f{b^2}{h^2 c}\f{C(c)}{z_{12}^2 z_{13}^2 z_{23}^{2+2\alpha(c)}
\bar{z}_{23}^{2\alpha(c)}  }
\\
&=& \f{b^2}{z_{12}^2 z_{13}^2 z_{23}^2 } \left[ \f{1}{c} + \f{C(c)}{h^2 c}\left(1-2\alpha(c) \ln |z_{23}|^2 +\cdots \right) \right] \nonumber
\eea
Now using the form of $C(c)$ from (\ref{eqn:known}) we get:
\bea
\left< T(z_1) t(z_2,\bar{z}_2) t(z_3,\bar{z}_3) \right>
 =\f{-2b\ln |z_{23}|^2+\f{b}{2}}{z_{12}^2 z_{13}^2 z_{23}^2 }
\eea
and:
\bea
&&\left< t(z_1,\bar{z}_1) t(z_2,\bar{z}_2) t(z_3,\bar{z}_3) \right>  
\nonumber \\[1mm]
&& =\left<  \left[ \f{b}{c}T+\f{b}{h} X \right] (z_1,\bar{z}_1)   \left[ \f{b}{c}T+\f{b}{h} X \right] (z_2,\bar{z}_2) \left[ \f{b}{c}T+\f{b}{h} X \right] (z_3,\bar{z}_3) \right>  \nonumber  \\[1mm]
&&= \f{b^3}{c^3} \left< T(z_1) T(z_2) T(z_3) \right> + \f{b^3}{h^2 c} \left( \left< X(z_1,\bar{z}_1) X(z_2,\bar{z}_2) T(z_3) \right> \right.  \nonumber  \\[1mm]
&&\left. + \left< X(z_1,\bar{z}_1) T(z_2) X(z_3,\bar{z}_3) \right> + \left< T(z_1) X(z_2,\bar{z}_2) X(z_3,\bar{z}_3) \right> \right) \\[1mm]
&&+\f{b^3}{h^3} \left< X(z_1,\bar{z}_1) X(z_2,\bar{z}_2) X(z_3,\bar{z}_3) \right>  \nonumber  \\[1mm]
&&=\f{b^3}{c^2} \f{1}{z_{12}^2 z_{13}^2 z_{23}^2 }
 + \f{b^3}{h^3 c^2} \f{D(c)}{z_{12}^2 z_{13}^2 z_{23}^2 }
z_{12}^{-\alpha(c)} z_{13}^{-\alpha(c)} z_{23}^{-\alpha(c)} \nonumber   \\[1mm]
&&+ \f{b^3}{h^2 c^2} \f{C(c)}{z_{12}^2 z_{13}^2 z_{23}^2 } \left
[ z_{12}^{-2\alpha(c)}\bar{z}_{12}^{-2\alpha(c)} +
z_{13}^{-2\alpha(c)}\bar{z}_{13}^{-2\alpha(c)} +
z_{23}^{-2\alpha(c)}\bar{z}_{23}^{-2\alpha(c)} \right] \nonumber 
\eea
Now expanding this and using (\ref{eqn:bdef}):
\bea
&&\left< t(z_1) t(z_2) t(z_3) \right>  = \f{b^3}{h^3 c^2}\left(-2h^3+D_0\right) \\
&&+ \f{b^2}{h^3 c} \left[ \left(D_0-2h^3)(\ln |z_{12}|^2+ \ln |z_{13}|^2+ \ln |z_{23}|^2 \right) + bD_1+\f{3}{2}h^3 \right] +O(1) \nonumber 
\eea
Thus if this is to be regular in the limit we must have:
\bea
D_0=2h^3, ~~~~~ D_1=-\f{3h^3}{2b}
\eea
Then from the $O(1)$ terms we get:
\vspace{-3mm}
\bea
&&\left< t(z_1,\bar{z}_1) t(z_2,\bar{z}_2) t(z_3,\bar{z}_3) \right>  = 
\f{1}{z_{12}^2 z_{13}^2 z_{23}^2 } 
\Biggl\{ -b\left( \ln^2 |z_{12}|^2 +  \ln^2 |z_{13}|^2 +  \ln^2
|z_{23}|^2 \right) \nonumber   \\ 
&&+ 2b \left(\ln |z_{12}|^2 \ln |z_{13}|^2 + \ln |z_{12}|^2 \ln |z_{23}|^2 + \ln |z_{13}|^2 \ln |z_{23}|^2 \right)   \\
&& - \f{b}{2} \left( \ln |z_{12}|^2 +  \ln |z_{13}|^2 +  \ln |z_{23}|^2  \right) + a \Biggr\}  \nonumber
\eea
where we have defined the constant $a$ by:
\bea
a \equiv -\f{b^3}{2h^3}\left( -2 D_2 -12hB_2 +\f{3}{2}h^3 \alpha''(0) \right)
\eea
Now consider correlators involving the $\bar{T},\bar{X}$ fields as well. For instance:
\bea
&&\left< T(z_1) \bar{T}(\bar{z}_2) t(z_3,\bar{z}_3) \right> = \left< T(z_1) \bar{T}(\bar{z}_2) \left[ \f{b}{c}T + \f{b}{h} X \right] (z_3,\bar{z}_3) \right> =0  \nonumber\\
&&\left< T(z_1) t(z_2,\bar{z}_2) \bar{t}(z_3,\bar{z}_3) \right> =
 \left< T(z_1) \left[ \f{b}{c} T + \f{b}{h} X \right] (z_2,\bar{z}_2) \left[ \f{b}{c}\bar{T} + \f{b}{h} \bar{X} \right] (z_3,\bar{z}_3) \right> =0  \nonumber
\eea
More non-trivially:
\bea
&&\left< T(z_1) \bar{t}(z_2,\bar{z}_2) \bar{t}(z_3,\bar{z}_3) \right> =
 \left< T(z_1) \left[ \f{b}{c} \bar{T} + 
\f{b}{h} \bar{X} \right] (z_2,\bar{z}_2) \left[ \f{b}{c}\bar{T} + 
\f{b}{h} \bar{X} \right] (z_3,\bar{z}_3) \right>   \nonumber\\
&&\hspace*{3cm}\;= \f{b^2}{h^2} \,\f{E(c)}{z_{12}^2 z_{13}^2 z_{23}^{2 \alpha(c)-2} \bar{z}_{23}^{4+2\alpha(c)}}
\eea
Inserting the known expression for of $E(c)$ we get:
\bea
\left< T(z_1) \bar{t}(z_2,\bar{z}_2) \bar{t}(z_3,\bar{z}_3) \right> =  
\f{\f{b}{2}}{z_{12}^2 z_{13}^2 z_{23}^{-2} \bar{z}_{23}^{4}}
\eea
The last correlator we have to consider is the following:
\bea
&&\left< t(z_1,\bar{z}_1) t(z_2,\bar{z}_2) \bar{t}(z_3,\bar{z}_3) \right> 
\\[1mm]
&&=\left< \left[ \f{b}{c} T +\f{b}{h} X \right](z_1,\bar{z}_1) \left
[ \f{b}{c} T +\f{b}{h} X \right](z_2,\bar{z}_2) \left[ \f{b}{c} \bar{T}
+\f{b}{h} 
\bar{X} \right](z_3,\bar{z}_3) \right>  \nonumber \\
&&=\f{b^3}{c h^2} \left< X(z_1,\bar{z}_1) X(z_2,\bar{z}_2) 
\bar{T}(z_3,\bar{z}_3) \right> + \f{b^3}{h^3} \left< X(z_1,\bar{z}_1) 
X(z_2,\bar{z}_2) \bar{X}(z_3,\bar{z}_3) \right> \nonumber  \\
&&=\f{b^3 E(c)}{c h^2} \f{1}{z_{12}^{4+2 \alpha(c)} \bar{z}_{12}^{2
\alpha(c)-2} \bar{z}_{13}^2 \bar{z}_{23}^2 } \nonumber \\
&&+ \,
\f{b^3 F(c)}{c h^3}\f{1}{z_{12}^{4+\alpha(c)} z_{13}^{\alpha(c)}
z_{23}^{\alpha(c)} \bar{z}_{12}^{\alpha(c)-2} \bar{z}_{13}^{2+\alpha(c)}
\bar{z}_{23}^{2+\alpha(c)} } \nonumber
\eea
Thus we find:
\vspace{-2mm}
\bea
F(c)=-\f{h^3}{2b}+F_1 c + O(c^2)
\eea
Finally we get
\bea
&&\left< t(z_1,\bar{z}_1) t(z_2,\bar{z}_2) \bar{t}(z_3,\bar{z}_3) \right>
=\f{ \f{b}{2} \left( \ln|z_{12}|^2-\ln|z_{13}|^2-\ln|z_{23}|^2 \right) + f}{z_{12}^4 \bar{z}_{12}^{-2} \bar{z}_{13}^2 \bar{z}_{23}^2 }
\eea
where the coefficient $f$ is given by:
\bea
f=-\f{b^3(-\f{1}{2}h^3\alpha''(0) -2F_1)}{2h^3}
\eea
In summary we have found the following correlators which yield the OPEs given in the text:
\bea
\left< T(z_1) t(z_2,\bar{z}_2) \right>&=& \f{b}{2 z_{12}^4} \nonumber\\
\left< t(z_1,\bar{z}_1) t(z_2,\bar{z}_2) \right> &=&
\f{-b\ln |z_{12}|^2}{z_{12}^4} \nonumber\\
\left< T(z_1) T(z_2) t(z_3,\bar{z}_3) \right>&=&
\f{b}{z_{12}^2 z_{13}^2 z_{23}^2 } \nonumber\\
\left< T(z_1) t(z_2,\bar{z}_2) t(z_3,\bar{z}_3) \right> &=&
\f{-2b\ln |z_{23}|^2+\f{b}{2}}{z_{12}^2 z_{13}^2 z_{23}^2 } 
\eea
\vspace{-5mm}
\bea
 && \hspace{-8mm} \left< t(z_1,\bar{z}_1) t(z_2,\bar{z}_2) t(z_3,\bar{z}_3) \right>  =
\f{1}{z_{12}^2 z_{13}^2 z_{23}^2 } \Biggl\{ -b\left( \ln^2 |z_{12}|^2 + 
 \ln^2 |z_{13}|^2 +  \ln^2 |z_{23}|^2 \right)  \nonumber\\
&&\hspace{20mm} + 2b \left(\ln |z_{12}|^2 \ln |z_{13}|^2 +
 \ln |z_{12}|^2 \ln |z_{23}|^2 + \ln |z_{13}|^2 \ln |z_{23}|^2 \right) \nonumber  \\
&& \hspace{20mm}- \f{b}{2} \left( \ln |z_{12}|^2 +  \ln |z_{13}|^2 +  \ln |z_{23}|^2  
\right) + a \Biggr\}  \nonumber \\
&&\hspace{-8mm} \left< T(z_1) \bar{t}(z_2,\bar{z}_2) \bar{t}(z_3,\bar{z}_3) \right> =
 \f{\f{b}{2}}{z_{12}^2 z_{13}^2 z_{23}^{-2} \bar{z}_{23}^{4}} \nonumber\\
&&\hspace{-8mm}\left< t(z_1,\bar{z}_1) t(z_2,\bar{z}_2) \bar{t}(z_3,\bar{z}_3) \right>
=  \f{\f{b}{2} \left( \ln|z_{12}|^2-\ln|z_{13}|^2-\ln|z_{23}|^2
\right) + 
f }{z_{12}^4 \bar{z}_{12}^{-2} \bar{z}_{13}^2 \bar{z}_{23}^2 } \nonumber
\eea
%

\singlespacing
\addcontentsline{toc}{chapter}
                 {\protect\numberline{Bibliography\hspace{-96pt}}} 






\end{document}